\documentclass[Afour,sageh,times]{sagej_arxiv}
 \usepackage{moreverb,url}
\usepackage{amssymb,amsmath,amsfonts}
\usepackage{subfig}
\usepackage{graphicx}
\usepackage{rotating}
\usepackage{geometry}
\usepackage{setspace}
\usepackage{lipsum}
\newcommand{\verbatimfont}[1]{\renewcommand{\verbatim@font}{\ttfamily#1}}
\usepackage{epstopdf}
\usepackage{alltt}
\usepackage{color}
\usepackage{multirow}
\usepackage{listings}

\usepackage{fancyvrb}
\fvset{fontsize=\scriptsize}
\RecustomVerbatimEnvironment{verbatim}{Verbatim}{}

\usepackage{floatrow}
\usepackage{algpseudocode,algorithm,algorithmicx}
\usepackage{algcompatible}
\usepackage{todonotes}

\def\cH{{\cal H}}
\def\cJ{{\cal J}}

\def\scriptO{{{\it O}\kern -.42em {\it `}\kern + .20em}}
\def\RR{{{\rm l}\kern - .15em {\rm R} }}
\def\PP{{{\rm l}\kern - .15em {\rm P} }}
\def\L2{{{\sf L}^2}}
\def\H1{{{\sf H}^1}}

\def\PN2{{\PP_{N}-\PP_{N-2}}}

\def\complex{{{\rm C} \kern - .53em {\rm l} \kern + .38em}}

\def\a1{{ | \lambda_{\min} |}}

\def\l1{{   \lambda_{\min}  }}

\def\uut{{\tilde {\underline u}}}

\def\bu0{{\underline {\bf 0}}}

\def\bu{{\bf u}}

\def\bx{{\bf x}}

\def\Dt{{\tilde D}}
\def\It{{\tilde I}}

\def\ih{{\hat \imath}}
\def\Ih{{\hat I}}
\def\jh{{\hat \jmath}}
\def\eh{{\hat e}}
\def\kh{{\hat k}}

\def\Dh{{\hat D}}

\def\Oh{{\hat \Omega}}

\def\Jh{{\hat J}}

\def\ub{{\underline b}}

\def\ur{{\underline r}}

\def\uu{{\underline u}}

\def\uv{{\underline v}}
\def\uw{{\underline w}}

\def\u0{{\underline 0}}
\def\n12{{n_{\frac{1}{2}}}}
\def\t12{{t_{\frac 1 2}}}
\def\n12{{n_{0.8}}}
\def\t12{{t_{0.8}}}

\def\nl{{n_{\mbox{\em \tiny local}}}}

\newcommand{\pp}[2]{\frac{\partial #1}{\partial #2} }

\definecolor{code}{rgb}{0, 0, 0}
\newcommand{\code}[1]{\texttt{\small\color{code} #1}}

\begin{document}

 \title{Scalability of High-Performance PDE Solvers}



\author{Paul Fischer,\affilnum{1,2,3} 
        Misun Min,\affilnum{1} 
        Thilina Rathnayake,\affilnum{2}
        Som Dutta,\affilnum{9}
        Tzanio Kolev,\affilnum{4}
        Veselin Dobrev,\affilnum{4}
        Jean-Sylvain Camier,\affilnum{4}
        Martin Kronbichler,\affilnum{5}
        Tim Warburton,\affilnum{8} 
        Kasia {\'S}wirydowicz,\affilnum{7}
        Jed Brown\affilnum{6}} 

\affiliation{\affilnum{1}Mathematics and Computer Science, Argonne National Laboratory, Lemont, IL 60439\\
\affilnum{2}Department of Computer Science, University of Illinois at Urbana-Champaign, Urbana, IL 61801\\
\affilnum{3}Department of Mechanical Science and Engineering, University of Illinois at Urbana-Champaign, Urbana, IL 61801\\
\affilnum{4}Center for Applied Scientific Computing, Lawrence Livermore National Laboratory, Livermore, CA 94550\\
\affilnum{5}Institute for Computational Mechanics, Technical University of Munich, 85748 Garching b. Muenchen, Germany\\
\affilnum{6}Department of Computer Science, University of Colorado, Boulder, CO 80309\\ 
\affilnum{7}National Renewable Energy Laboratory, Lakewood, CO, 80401\\
\affilnum{8}Department of Mathematics, Virginia Tech, Blacksburg, VA 24061\\
\affilnum{9}Mechanical \& Aerospace Engineering, Utah State University, UT 84322}

\corrauth{Misun Min,
Mathematics and Computer Science Division, Argonne National Laboratory, Lemont, IL 60439}
\email{mmin@mcs.anl.gov}

 \begin{abstract}

Performance tests and analyses are critical to effective HPC software development and are central components 
in the design and implementation of computational algorithms for achieving faster simulations on existing and
future computing architectures for large-scale application problems. In this paper, we explore performance and 
space-time trade-offs for important compute-intensive kernels of large-scale numerical solvers for PDEs that 
govern a wide range of physical applications. We consider a sequence of PDE-motivated bake-off problems designed 
to establish best practices for efficient high-order simulations across a variety of codes and platforms. 
We measure peak performance (degrees of freedom per second) on a fixed number of nodes
and identify effective code optimization strategies for each architecture. 
In addition to peak performance, we identify
the minimum time to solution at 80\% parallel efficiency. 
The performance analysis is based on spectral and $p$-type finite elements 
but is equally applicable to a broad spectrum of numerical PDE discretizations,
including finite difference, finite volume, and $h$-type finite elements.


\end{abstract}


\maketitle


 \section{Introduction}

One characteristic common to many science and engineering applications in
high-performance computing (HPC) is their enormous range of spatial and temporal
scales, which can manifest as billions of degrees of freedom (DOFs) to be updated
over tens of thousands of timesteps.
Typically, the large number of spatial scales is addressed through
parallelism---processing all elements of the spatial problem
simultaneously---while temporal complexity is most
efficiently addressed sequentially, particularly in the case of highly
nonlinear problems such as fluid flow, which defy linear transformations that
might expose additional parallelism (e.g., through frequency-domain analysis).
Simulation campaigns for these problems can require weeks or months of
wall-clock time on the world's fastest supercomputers.  One of the
principal objectives of HPC is to reduce these runtimes to manageable levels.

In this paper, we explore  performance and space-time trade-offs for
computational model problems that typify the important compute-intensive
kernels of large-scale numerical solvers for partial differential equations
(PDEs) that govern a wide range of physical applications.
We are interested in peak performance (degrees of freedom
per second) on a fixed number of nodes and in minimum time to solution at
reasonable parallel efficiencies, as would be experienced by computational
scientists in practice.  While we consider matrix-free implementations of
$p$-type finite and spectral element methods as the principal vehicle for
our study, the performance analysis presented here is relevant to a broad
spectrum of numerical PDE solvers, including finite difference, finite volume,
and $h$-type finite elements, and thus is widely applicable.

Performance tests and analyses are critical to effective HPC software
development and are central components of the recently formed U.S. Department
of Energy Center for Efficient Exascale Discretizations 
(CEED)\footnote{\cite{CEED}: https://ceed.exascaleproject.org}.  One of the
foundational components of CEED is a sequence of PDE-motivated bake-off
problems designed to establish best practices for efficient high-order methods
across a variety of platforms.  The idea is to pool the efforts of several
high-order development groups to identify effective code optimization
strategies for each architecture.  Our first round of tests features
comparisons from the software development projects
Nek5000\footnote{https://github.com/Nek5000/Nek5000},
MFEM\footnote{https://github.com/mfem/mfem},
deal.II\footnote{https://github.com/dealii/dealii},
and libParaNumal\footnote{https://github.com/paranumal/libparanumal}.
Principal findings are that high-order operator evaluation and 
solvers can in fact be {\em less expensive} per degree-of-freedom
than their low-order counterparts and that strong scaling of CPU-only
platforms can yield lower time per iteration than even highly-tuned
kernels on GPU nodes.  We note, however, that other important metrics, 
such as energy and capital costs, are not considered in this evaluation.

The rest of this paper is organized as follows.
Section 2 provides an overview of the performance metrics that we use
throughout the paper and also gives motivation for the suite of bake-off problems.
Section 3 describes the bake-off problem (BP) specifications.
Sections 4 and 5 discuss the detailed mathematical formulations.
Sections 6 and 7 demonstrate performance results and analysis for the three production 
codes on IBM's BG/Q at Argonne Leadership Computing Facility (ALCF).
In Section 8, we present results using NVIDIA V100s on the Summit at
Oak Ridge Leadership Computing Facility (OLCF), provided with
discussion and further analysis of these results.
We summarize our findings in Section 9.
Details regarding the code implementations are provided in the Appendix.

 \section{Performance Metrics}

Scalability is an important metric when assessing performance of parallel
computing applications.  An open question remains, however, of how one should
quantify scalability.  Is it best to study strong scaling, in which the problem
size $n$ (say, number of grid points) is fixed and the number of cores $P_c$ is
increased?  Or is weak scaling, in which $n/P_c$ is fixed while $P_c$ increases, better?
The concern with each of these options is that they often fail to reveal performance
at the level of granularity that is of paramount interest to HPC users, namely,
the point where parallel efficiency starts to drop below a tolerable fraction
$\eta \in [0,1]$.  We refer to this tolerable limit as the $\eta$-performance or
{\em strong-scale} limit, with typical values of $\eta=0.5$ or 0.8 corresponding
respectively to 50\% and 80\% parallel efficiency.

\begin{figure*}[t]
 \centering
 \subfloat[BP5: time vs. $P_c$]{{\includegraphics[width=0.42\textwidth]{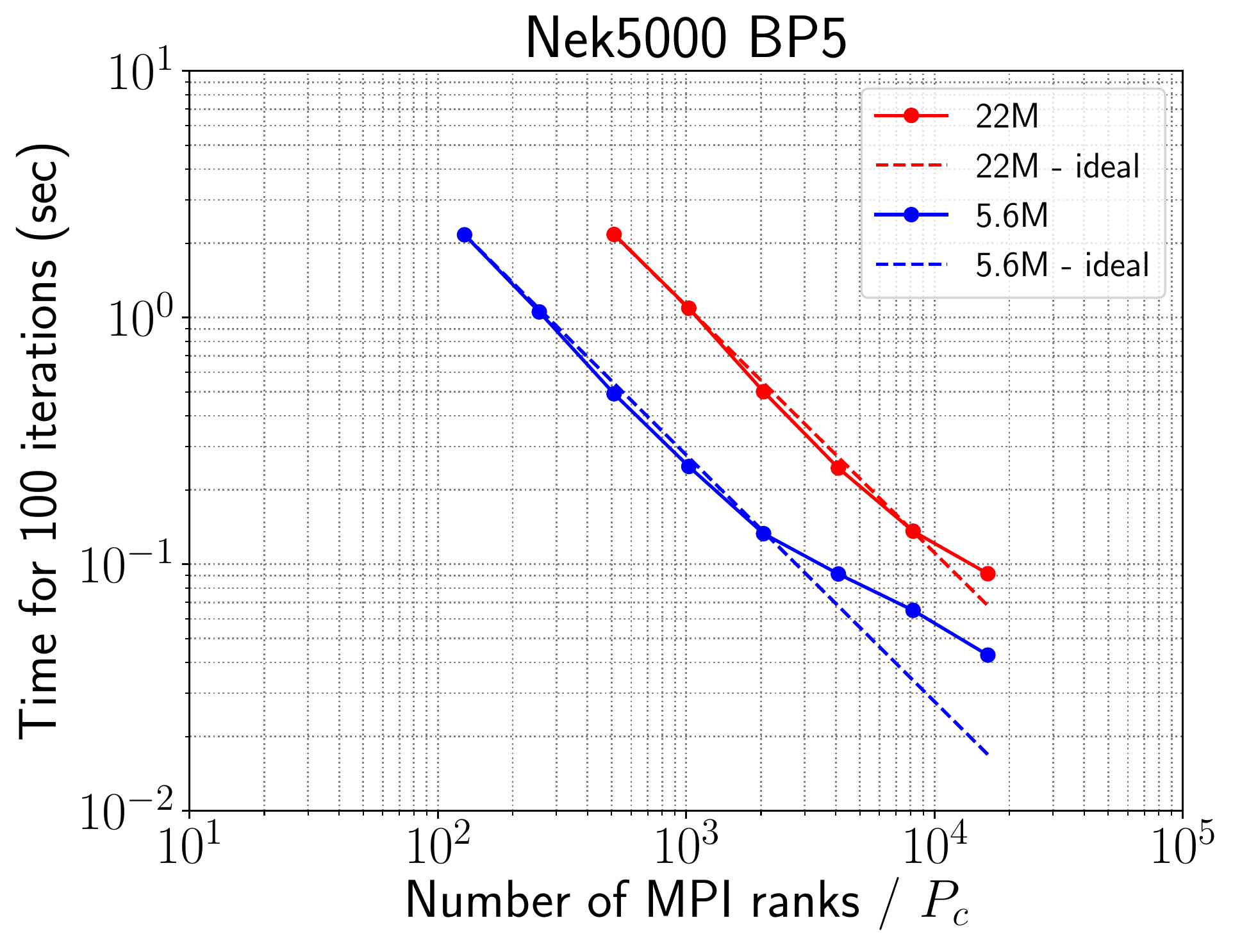}}}
 \hskip.1in
 \subfloat[BP5: time vs. $n/P_c$]{{\includegraphics[width=0.42\textwidth]{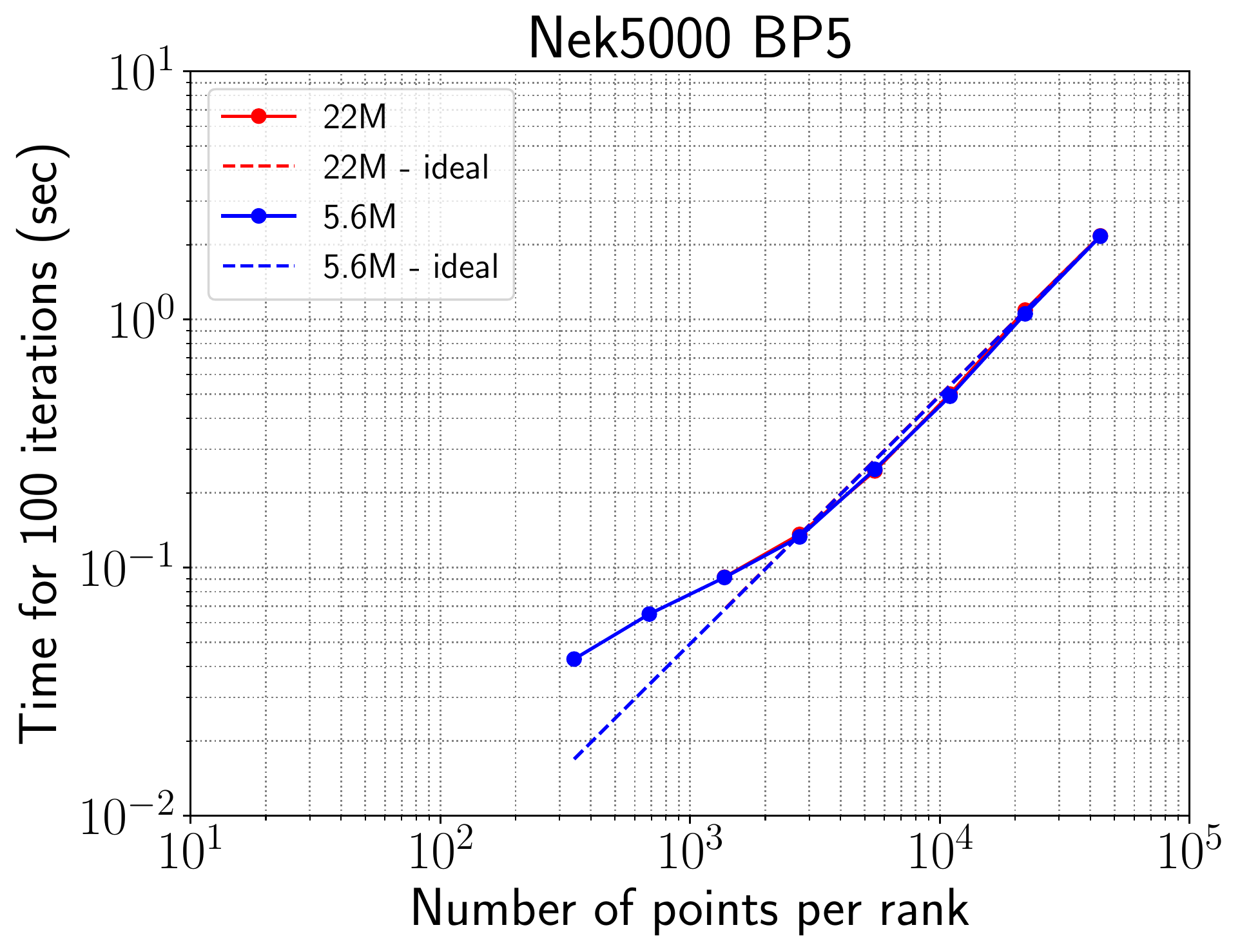}}}
 \caption{\label{fig:strong1}Strong-scale study for BP5-Nek5000 with $n$= 22M and 5.6M.
 $n/P_c$ is the problem size per core, and strong-scale limit is observed at $n/P_c=2744$.}
\end{figure*}

\begin{figure*}[t]
 \centering
 \subfloat[BP5: efficiency vs. $n/P_c$]
  {{\includegraphics[width=0.42\textwidth]{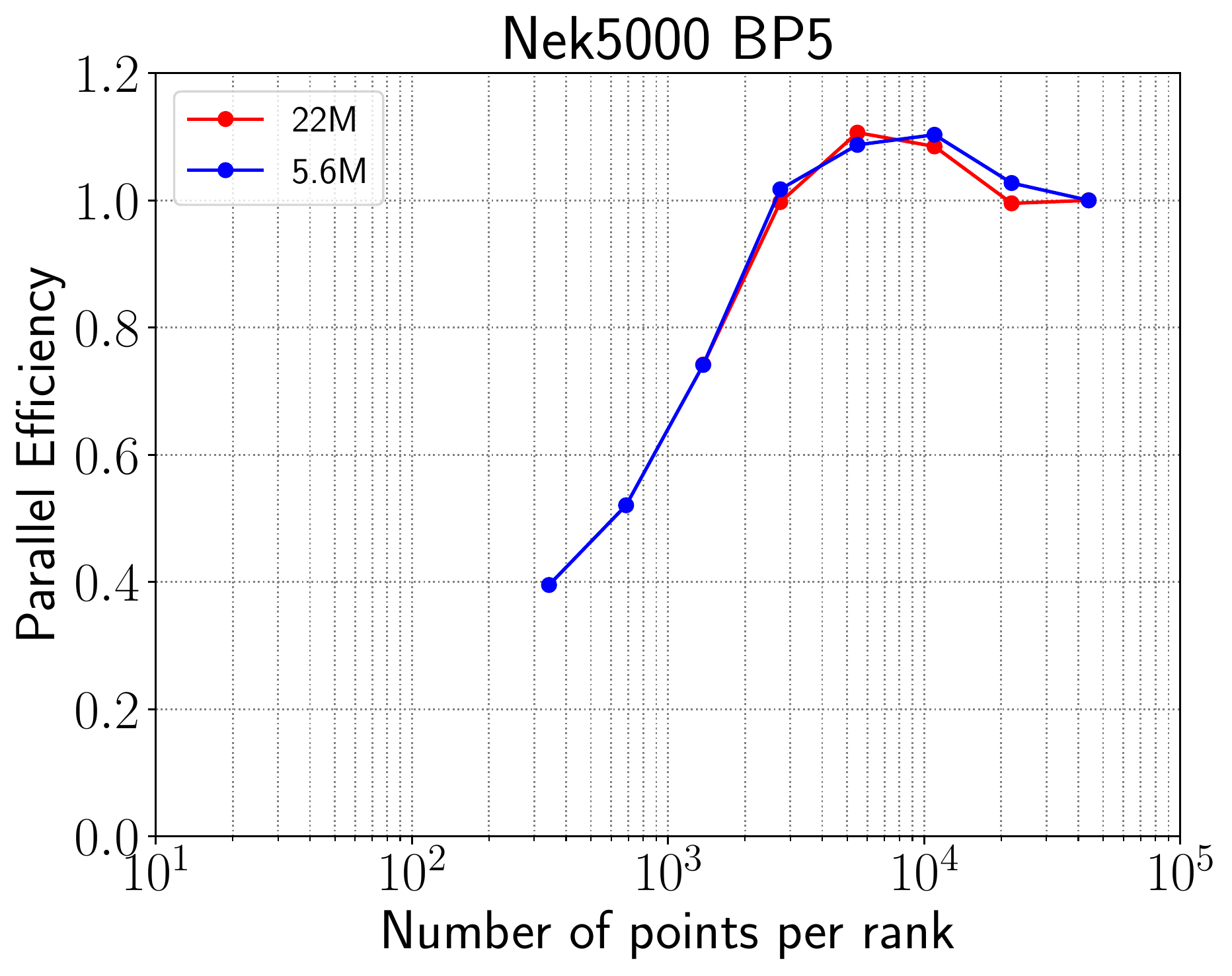}}}
 \hskip.1in
 \subfloat[BP5: DOFs vs. $n/P_c$]
  {{\includegraphics[width=0.42\textwidth]{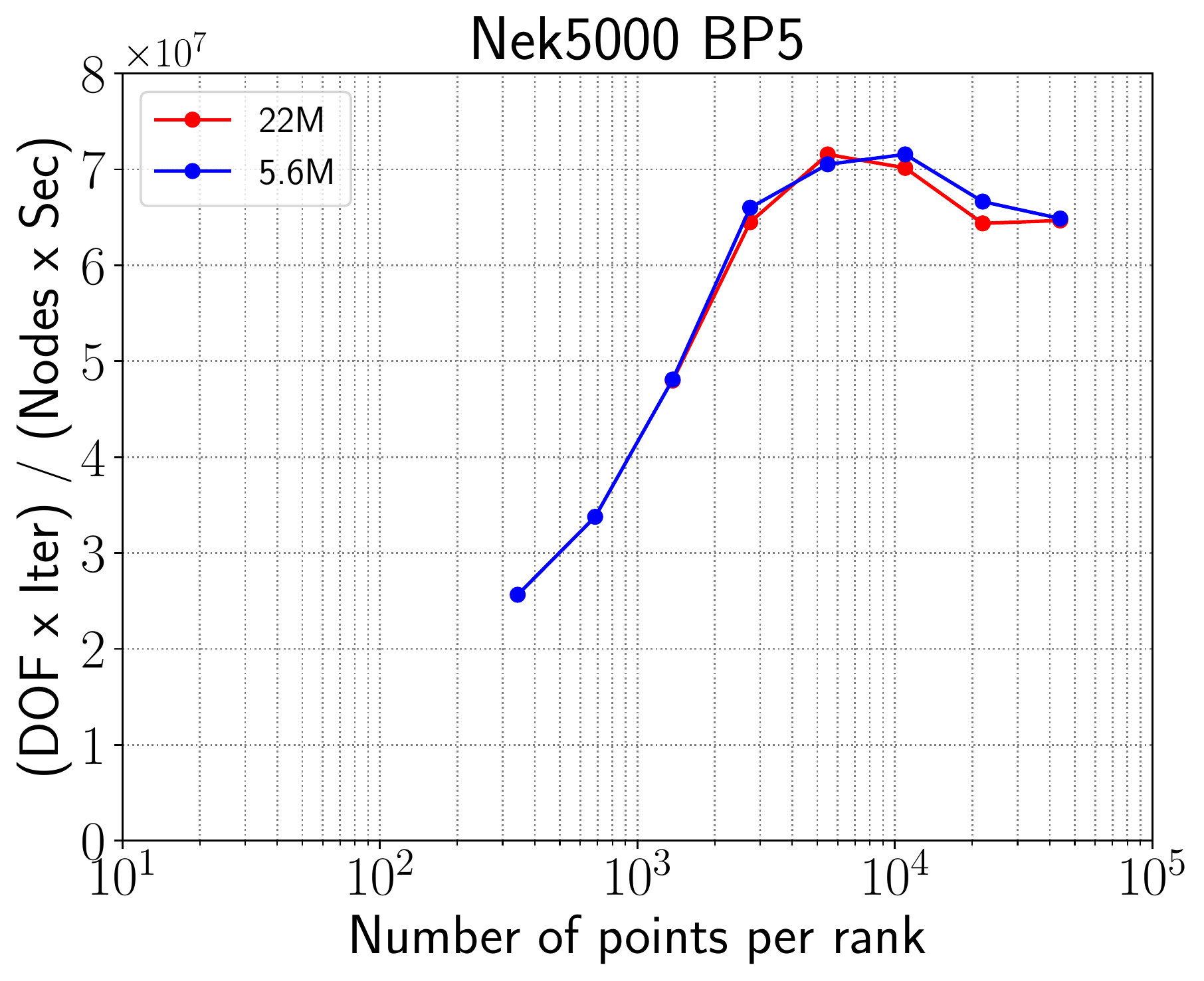}}}
 \caption{\label{fig:strong2}Strong-scale study for BP5-Nek5000.
  $n/P_c$ is the problem size per core. 
  Order unity parallel efficiency can be achieved for $n/P_c \geq 2744$. }
\end{figure*}

As an example, we show in Figures~\ref{fig:strong1} and \ref{fig:strong2}
strong-scale performance of Nek5000 using 32 MPI ranks per node on Cetus, the
development-level IBM BG/Q at ALCF. 
The model problem is a 3D Poisson equation solved at different spectral element
resolutions, which are described in Section~\ref{sec:bps} as bake-off problem
number 5 (BP5). With polynomial order $p=7$ and number of elements $E=2^{14}$
and $E=2^{16}$, the number of points are $n=$5,619,712 and 22,478,848,
respectively. Note that we define the number of points by $n=p^3E$ throughout
the paper, without redundancy and without boundary condition effects. In other
words, $n$ is the total number of parallel unique degrees of freedom, including
all boundary unknowns. The polynomial order $p$ is fixed for each direction in 
the reference element $\Oh=[-1,1]^3$.

Figure~\ref{fig:strong1}(a) shows the time versus the number of MPI ranks, $P_c$,
for each case of fixed $n=22M$ and $n=5.6M$,
along with ideal linear scaling plots scaling as $P_c^{-1}$.
The lower curve, corresponding to the smaller problem, exhibits linear scaling
out to $P_c=2048$, beyond which the time decreases more slowly.  In contrast,
scaling for the larger problem is linear out to $P_c=8192$ before the performance
drop-off is observed.  In isolation, one might conclude that the $n=22,478,848$
performance is indicative of good strong-scale behavior.   By contrast, the
$n=5,619,712$ graph might be viewed with skepticism regarding the ability of the
code to strong-scale beyond $P_c=2048$.

In fact, if we change the $x$-axis in Figure~\ref{fig:strong1}(a)
to be $n/P_c$, we see that the large and small problem results collapse to a
single curve, as evident in Figure~\ref{fig:strong1}(b).
In this case, for $n/P_c
\geq 2744$ the solution time is directly proportional to the amount of work on
each rank, namely,
\begin{eqnarray}
 t_{\mbox{\footnotesize{wall}}} \sim n/(S \, P_c),
\end{eqnarray}
where $S$ is a constant that scales as the rate of work per rank in this
work-saturated limit.
From this result,  we can infer that we have order unity parallel
efficiency for $n/P_c \geq 2744$, meaning that {\em for any case with $n/P_c > 2744$
one can effectively increase the number of processors with a
corresponding decrease in time to solution, all at fixed energy and cost},
assuming that the machine has enough processors to do so.  This is the
important promise of
distributed-memory parallel computing.  We often refer to such a break point
(e.g., $n/P_c=2744$ in Figure~\ref{fig:strong1}(b)) as the strong-scale limit towards
which users will naturally gravitate. Running with $n/P_c$ significantly
exceeding this limiting value implies unnecessarily long runtimes.  In
practice, users might choose a value of $n/P_c$ below this break point if they
are willing to tolerate a certain level of inefficiency.

To this end, we define parallel efficiency,
\begin{eqnarray} \label{eq:eff}
  \eta(P) &:=&  \frac{T_{P_{\min}} \times P_{\min}}{T_P \times P},
\end{eqnarray}
where $P_{\min}$ is the minimum number of processors used in the strong-scale
study (usually, the memory-saturation-bound limit) and $T_P$ is the wall-clock
time for $P$ processors.  (Here, processors means cores, nodes, GPUs, etc.,
according to the sensible metric for the given platform.)
Figure~\ref{fig:strong2}(a) shows the corresponding efficiencies for the pair
of strong-scale studies from Figure~\ref{fig:strong1}.  We see that $\eta
\approx 0.8$ for $n/P_c=1372$ and $\eta \approx 0.5$ for $n/P_c=686$.  Here, we
denote $n_{0.8}=1372$ and $n_{0.5}=686$ as the values of $n/P_c$ that
correspond to 80\%  and 50\% efficiency, respectively.

We note that in leadership-class computing environments the granularity of the
simulation, $n/P$, is a runtime parameter selectable by the user unless $P$ is
bounded by the total number of processors accessible in the system.  To
minimize solution time, users are generally interested in operating as far to
the left as possible in Figure~\ref{fig:strong2}(a) while staying within a
tolerable parallel efficiency, $\eta$.
Beyond the strong-scale limit $n/P_c=2744$ discussed previously, one could have
a 1.6$\times$ speedup by doubling the number of processors (which might be
reasonable for long simulations) or a 2$\times$ speedup by quadrupling the
number of processors (which likely is not acceptable to most users).  

For cross-code comparisons, one must be able to compare
rates of work between differing implementations.
Here, we retain the independent variable, $n/P$, but change the dependent variable
to be
\begin{eqnarray} \label{eq:rate}
\frac{\mbox{\em work}}{\mbox{\em resources}} =
\frac{\mbox{\em iterations $\times$ DOFs}}{\mbox{\em nodes $\times$ seconds}}
\; =: \; \mbox{DOFs},
\end{eqnarray}
where
{\em iterations} is the number of conjugate gradient (CG) iterations
(see the BP definitions in Section~\ref{sec:bps});
{\em DOFs} is the number of degrees of freedom (here, grid points);
{\em nodes} is the number of BG/Q nodes;
and
{\em seconds} refers to the amount of wall-clock time required to perform
the given number of CG iterations.
(For the bake-off comparisons, the independent variable, $n/P$, is taken to be the
number of points per BG/Q node.)

With this background, we have developed a sequence of bake-off problems
to generate performance figures similar to Figure \ref{fig:strong2}(b) for
a variety of polynomial orders, element counts, operators, and codes
(over 2,000 simulations in total).
The principal objective is to identify the fastest implementations for each
operator, at the given resolution (e.g., polynomial order $p$).
Equally important, however, is to identify the peak performance value and the
strong-scale limit
where performance efficiency drops to a fraction $\eta=0.8$ of
this peak value.  For the bake-off problems, the number of MPI ranks is fixed
to be a large value (generally, $P=512$ nodes, for a total of 16,384 ranks
with -c32 mode on BG/Q, using 2 MPI processes per core and 16 cores per node);
and the independent variable is the number of degrees of
freedom per node, $n/P=(p^3E)/P$.

\section{Bake-Off Specifications}
\label{sec:bps}

\begin{table*} \centering
\footnotesize
{\caption{\label{tab:bps}Bake-Off Kernel/Problem Summary}}
\begin{tabular}{ |l|ll|l|l|l|l|} \hline
              &\multicolumn{2}{c|}{System}& Form    &  BCs  & Quadrature Points & Nodal Points \\ \hline
{\bf BK1/BP1}  &  $B\uu  $&\hspace*{-.15in}$=\ur  $ & scalar  &  homogeneous Neumann   & ($p+2$) GL  & ($p+1$) GLL\\ \hline
{\bf BK2/BP2}  &  $B\uu_i$&\hspace*{-.15in}$=\ur_i$ & vector  &  homogeneous Neumann   & ($p+2$) GL  & ($p+1$) GLL\\ \hline
{\bf BK3/BP3}  &  $A\uu  $&\hspace*{-.15in}$=\ur  $ & scalar  &  homogeneous Dirichlet & ($p+2$) GL  & ($p+1$) GLL\\ \hline
{\bf BK4/BP4}  &  $A\uu_i$&\hspace*{-.15in}$=\ur_i$ & vector  &  homogeneous Dirichlet & ($p+2$) GL  & ($p+1$) GLL\\ \hline
{\bf BK5/BP5}  &  $A\uu  $&\hspace*{-.15in}$=\ur  $ & scalar  &  homogeneous Dirichlet & ($p+1$) GLL & ($p+1$) GLL\\ \hline
{\bf BK6/BP6}  &  $A\uu_i$&\hspace*{-.15in}$=\ur_i$ & vector  &  homogeneous Dirichlet & ($p+1$) GLL & ($p+1$) GLL\\ \hline
\end{tabular}
\end{table*}

The first suite of problems is focused on runtime performance, measured in
DOFs for bake-off kernels (BKs) and
bake-off problems (BPs).  The {\bf BK}s are
defined as the application of a local (unassembled) finite-element operator to
a scalar or vector field, without the communication overhead associated with
assembly.  These tests essentially demonstrate the vectorization performance of
the PDE operator evaluation (i.e., matrix-free matrix-vector products, or matvecs).
As they represent the most computationally-intensive operations in bake-off
problems, the BKs provide an upper bound on realizable floating-point
performance (DOFs or MFLOPS) for an application based on such implementations.
The {\bf BP}s are mock-up solvers that use the BKs inside a
diagonally preconditioned conjugate gradient (PCG) iteration.  The point of
testing the BKs inside PCG is to establish realistic memory-access patterns, to
account for communication overhead of assembly (i.e., nearest-neighbor
exchange), and to include some form of global communication (i.e., dot
products) as a simple surrogate for global coarse-grid solves that are
requisite for large-scale linear system solves.

To meet these goals in a measurable way, we decided to run the initial BPs on 16,384
MPI ranks (i.e., 512 nodes in -c32 mode)
of the IBM BG/Q {\em Cetus} at the ALCF,
which, by virtue of its convex network partitions and provision for a 17th
``OS'' core on each node, realizes timings that are repeatable to 3 digits from
one job submission to the next.  The number of MPI ranks was deemed large enough to
reveal obvious strong-scale deficiencies yet small enough to allow minimal
queue times for all the cases that needed to be submitted.  The range of
problem sizes was chosen to span from the performance-saturated limit (a lot of
work per node) to beyond the strong-scale limit (very little work per node).

To date, there are six BKs and six BPs.  Kernels BK1, BK3, and BK5 operate on
scalar fields, while BK2, BK4, and BK6 are corresponding operators applied to
vector fields having three components each.  Each BP$j$, $j=1,\dots,6$
corresponds to PCG applied to the assembled BK$j$ system, using a matrix-free
implementation if that is faster.  The even-numbered (vector-oriented) kernels
allow amortization of matrix-component memory references across multiple
operands and provide a realistic optimization for vector-based applications
such as fluid dynamics (three fields) and electromagnetics (six fields).  The
vector-based implementations can also benefit from coalesced messaging, thus
reducing the overhead of message latency, which is important when running in
the strong-scale (fast) limit.

The BPs include solution of the mass matrix, $B \uu = \ur$ (BP1--BP2), and the
stiffness matrix, $A\uu=\ur$ (BP3--BP6).  Approximation orders are
$p=1,\dots,15$.  For each problem, $q$-point quadrature is used in each
direction within the reference element (for a total of $q^3E$ quadrature points
throughout the domain).
For BK1--BK4, Gauss-Legendre (GL) quadrature is used with $q=p+2$.
For BK5--BK6, Gauss-Lobatto-Legendre (GLL) quadrature is used with
$q=p+1$ quadrature points corresponding to the nodal points of the
underlying Lagrangian bases.   The nodal and quadrature point choices as well
as the boundary conditions (BCs) are
summarized in Table \ref{tab:bps}.  The number of elements is $E=2^k$, for
$k=14, \ldots, 21$ ($2^{14}=16,384$ and $2^{21}=2,097,152$) 
so that at least one element per MPI rank is ensured. 
The elements are arranged in a tensor-product array with the number of elements in
each direction differing by at most a factor of 2.  Since the tests are designed
to mimic real-world applications, the benchmark codes assume that the
elements are full {\em curvilinear} elements, approximated by the same order
of polynomial, $p$, with iso-parametric mappings and are not allowed to exploit
the global tensor-product structure of the element layout.
We also note that the geometric matrices are all precomputed in the codes.

 \section{Mathematical Formulation}
We can cast the BP specifications in the context of
the following scalar Helmholtz equation,
 \begin{eqnarray} \label{eq:hm1}
  - \; \nabla \cdot \mu \nabla u\;+\; \beta \, u \;  = f \;\; \mbox{in $\Omega$},
  \end{eqnarray}
 where $\beta$ and $\mu$ may be nonnegative functions of $\bx \in \Omega$.
With $X^p$ a finite-dimensional subset of $\cH^1$, the Sobolev space comprising
square-integrable functions on $\Omega$ whose gradient is also square integrable,
the discrete variational formulation of (\ref{eq:hm1}) is
  {\em Find $u \in X^p$  such that}
  \begin{eqnarray} \label{eq:hm2}
  a(v,u) \;+\; (v,u)_{\beta}  &=& (v,f) \;\; \forall \; v \, \in \, X^p,
  \end{eqnarray}
  where, for sufficiently regular $v$, $u$, and $f$, we have
  \begin{eqnarray}
  \label{eq:ip1}
  a(v,u) &=& \int_{\Omega} \, \mu \nabla v \, \cdot \, \nabla u \, d\bx, \\
  \label{eq:ip2}
  (v,u)_{\beta} &=& \int_{\Omega} \, v \, \beta \, u \, d\bx, \\
  \label{eq:ip3}
  (v,f) &=& \int_{\Omega} \,   v \, f \, d\bx.
  \end{eqnarray}
  We approximate the scalar functions $u$ and $v$ by finite expansion of
  nodal (Lagrangian) basis functions
  \begin{eqnarray}
  \label{eq:u}
  u(\bx) &=&\sum_{l=1}^n u_l \, \phi_l(\bx), \\
  \label{eq:v}
  v(\bx) &=&\sum_{\hat l=1}^n v_{\hat l} \, \phi_{\hat l}(\bx).
  \end{eqnarray}
  Insertion of (\ref{eq:u})--(\ref{eq:v}) into (\ref{eq:ip1})--(\ref{eq:ip3})
  leads to the inner products defined as
  \begin{eqnarray} \label{eq:ip2a}
  a(v,u) &=& \uv^T A \uu , \\ \label{eq:ip2g}
  (v,u)_{\beta} &=& \uv^T B \uu, \\ \label{eq:ip2b}
  (v,f) &=&  \uv^T \ub.
  \end{eqnarray}
  Here, we have introduced the {\it stiffness matrix}, $A$,
  the (weighted) {\it mass matrix}, $B$,
  and the right-hand side, $\ub$,
  \begin{eqnarray}
   A_{\hat l l} &=&  a(\phi_{\hat l},\phi_l), \\
   B_{\hat l l} &=&  (\phi_{\hat l},\phi_l)_{\beta}, \\
   b_{\hat l}  &=&  (\phi_{\hat l},f),
  \end{eqnarray}
  where the index sets are $l,{\hat l} \; \in \; \{1,\dots,n\}$.
  The system to be solved is thus
  \begin{eqnarray} \label{eq:H}
  H \uu = \ub, \quad \mbox{where}\quad
  H := \, A \,+\, B.
  \end{eqnarray}
BPs 1--2 correspond to $\mu=0$ ($A=0$).
BPs 3--6 correspond to $\beta=0$ ($B=0$).

  The system (\ref{eq:H}) denotes a generic Galerkin discretization of
  (\ref{eq:hm2}).  The point of departure for the finite element/spectral element formulation is in the choice of basis
  functions, $\phi_i$, and choice of quadrature rules for the inner products
  (\ref{eq:ip1})--(\ref{eq:ip3}).
With the preceding definitions, we now consider the fast tensor-product
evaluations of the inner products introduced in
(\ref{eq:ip2a})--(\ref{eq:ip2g}).  To begin, we assume $\Omega = \cup_{e=1}^E
\Omega^e$, where the non-overlapping subdomains (elements) $\Omega^e$ are
images of the reference domain, $\Oh = [-1,1]^3$, given by
    \begin{eqnarray} \nonumber
    \left.  \bx \right|^{}_{\Omega^e}
    \!\!  &=&  \! \!
          \bx^e(r,s,t)\\ \label{eq:xijk}
    \!\!  &=& \! \!
    \sum_{k=0}^p\sum_{j=0}^p\sum_{i=0}^p
    \bx^e_{ijk} \,
    h_i(r) \,
    h_j(s) \,
    h_k(t).
    \;\;\; 
    \end{eqnarray}
Here, the $h_i$s are assumed to be the Lagrange interpolation polynomials based
on the Gauss-Lobatto-Legendre (GLL) quadrature points, $\xi_j \in [-1,1]$,
$j=0,\dots,p$.  This choice of points yields well-conditioned operators and
affords the option of direct GLL quadrature, if desired.

All functions are assumed to have expansions similar to (\ref{eq:xijk}).
For example, the solution $u(\bx)$ on $\Omega^e$ takes the form
\begin{eqnarray} \label{eq:uijk}
    \left.  u \right|^{}_{\Omega^e} =
    \sum_{k=0}^p\sum_{j=0}^p\sum_{i=0}^p
    u^e_{ijk} \,
    h_i(r) \,
    h_j(s) \,
    h_k(t),&&
\end{eqnarray}
for the $(p+1)^3$ unknown basis coefficients in $\Omega^e$.

To ensure interelement continuity ($u,v \in X^p \subset \cH^1$),
one must constrain coefficients at shared element interfaces
to be equal.   That is, for any given sets of index coefficients
$(i,j,k,e)$ and $(\ih,\jh,\kh,\eh)$,
\begin{eqnarray} \label{eq:dssum}
\bx_{ijk}^e \,=\, \bx_{\ih \jh \kh}^{\eh}  & \longrightarrow &
u_{ijk}^e \,=\, u_{\ih \jh \kh}^{\eh}.
\end{eqnarray}
The statement (\ref{eq:dssum}) leads to the standard finite element
processes of matrix assembly and assembly of the load and residual
vectors.  Fast matrix-free algorithms that use iterative solvers
require assembly only of vectors, since the matrices are never formed.
To implement the constraint
(\ref{eq:dssum}), we introduce a {\em global-to-local map}, formally expressed
as a sparse matrix-vector product, $\uu_L = Q \uu$, which takes (uniquely defined)
degrees of freedom $u_l$ from the index set $l \in \{1,\dots,n\}$ to their
(potentially multiply defined) local counterparts $u_{ijk}^e$.  We further
define $A_L =$block-diag$(A^e)$,
and arrive at the {\em assembled} stiffness matrix
\begin{eqnarray}
 A=Q^T A_L Q.
\end{eqnarray}
We refer to $A_L$ as the unassembled stiffness matrix
and its constituents $A^e$ as local stiffness matrices.  With this factored
form, a matrix-vector product can be evaluated as $\uw = Q^T A_L Q \uu$, which
allows parallel evaluation of the work-intensive step of applying $A^e$ to
basis coefficients in each element $\Omega^e$.  Application of the Boolean
matrices $Q$ and $Q^T$ represents the communication-intensive phases of the
process.\footnote{In the case of nonconforming elements, $Q$ is not Boolean but
can be factored into a Boolean matrix times a local interpolation matrix,
\cite{dfm02}.}

\section{Matrix-Free $Q_p$ Formulations}

Efficient matrix-free formulations of $Q_p$ tensor-product spaces date
back to Fourier-based spectral methods pioneered by Orszag and others
in the early 1970s.  In a landmark paper, \cite{sao80} showed that the
principal advantage of Fourier spectral methods in $\RR^3$ derives from the
tensor product forms (e.g., (\ref{eq:uijk})), which reduce the cost of applying
differential operators for $p^3$ degrees of freedom from $O(p^6)$ to $O(p^4)$.
Orszag further noted that if one exploits
symmetries in the one-dimensional operators, the constant can potentially be
improved; and, with enough symmetry, such as in the Fourier and Chebyshev cases,
the asymptotic complexity can be reduced to $O(p^3 \log p)$.
With the introduction of the spectral element method (SEM),
\cite{pat84} extended Orszag's work to a variational framework using Chebyshev
bases, and \cite{ronpa871} developed the nodal-bases approach using
Gauss-Lobatto-Legendre quadrature that is now commonplace in the
SEM and other high-order $Q_p$ implementations.

With the use of iterative solvers, the bake-off problems amount to implementing
fast matrix-vector products.  As noted earlier, matrix-free evaluations are
effected with a communication phase for assembly and a local work-intensive
phase to evaluate the physics.  For the stiffness matrix, $A$, the local matrix
vector products $\uw^e := A^e \uu^e$ can be implemented in the matrix-free form
outlined in 4.4.7 of \cite{dfm02},
\begin{eqnarray} \label{eq:bk5x}
\uw^e =
\left(\!  \! \begin{array}{c}
D_1 \\[.3ex]
D_2 \\[.3ex]
D_3 \end{array}\! \!  \right)^{\! \! \! \! \! T} \! \! \!
\left(\!  \begin{array}{ccc}
G^e_{11} & G^e_{12} & G^e_{13} \\[.3ex]
G^e_{21} & G^e_{22} & G^e_{23} \\[.3ex]
G^e_{31} & G^e_{32} & G^e_{33} \end{array} \! \right) \! \! \! \!
\left( \! \! \begin{array}{c}
D_1 \\[.3ex]
D_2 \\[.3ex]
D_3 \end{array}\! \!  \right) \uu^e.
\end{eqnarray}
Here, the derivative matrices $D_j$ involve tensor products of the
one-dimensional $(q \times p_1)$
interpolation ($\Jh$) and derivative ($\Dh$) matrices: $D_1= \Jh \otimes \Jh \otimes
\Dh$, $D_2= \Jh \otimes \Dh \otimes \Jh$, and $D_3= \Dh \otimes \Jh \otimes \Jh$,
where $p_1=p+1$ is the number of nodal points in each direction within an
element and $q$ is the corresponding number of quadrature points.
For BP5 and BP6, we use GLL quadrature on the $q=p_1$
nodal points, so $\Jh$ becomes the $q \times q$ identity matrix, which
yields considerable savings in computational cost.

Each of the  geometric factors is a  diagonal matrix of size $q^3 \times q^3$
(i.e., comprising only $q^3$ nontrivial entries).
For each quadrature point $(\eta_i,\eta_j,\eta_k)\in \Oh$,
\begin{eqnarray} \label{eq:gij}
\left[ G^e_{mm'}  \right]^{}_{ijk} =
\left[ \sum_{l=1}^3 \pp{r_m}{x^e_l}\pp{r_{m'}}{x^e_l} \right]^{}_{ijk}
\!\!\!
\cJ^e_{ijk} \rho_i \rho_j \rho_k,
\end{eqnarray}
where $\cJ^e_{ijk}$ is the geometric Jacobian evaluated at the quadrature points
and the $\rho_j$s are the GL quadrature weights for BP3--BP4
and GLL weights for BP5--BP6.
${\bf G}^e$ is a symmetric tensor, $G^e_{mm'}=G^e_{m'm}$, so only $6q^3$ memory
references per element are required for (\ref{eq:bk5x}), in addition to the
$p_1^3$ reads required to load $\uu^e$.  $\Jh$ and $\Dh$ require only $q
\times p_1$ reads, which are amortized over multiple elements and therefore
discounted in the analysis.

The majority of the computational effort in (\ref{eq:bk5x}) is in application
of the $D_j$s. Thus, it is worthwhile to explore the complexity for these tensor
contractions, which are central to fast matrix-free formulations.  We consider
two of the many possible approaches.
First, we need to evaluate for each element $\Omega^e$,
\begin{eqnarray} \label{eq:gradr}
  \uu^e_r &=& D_1 \uu^e \;=\; (\Jh \otimes \Jh \otimes \Dh) \uu^e \\ \label{eq:grads}
  \uu^e_s &=& D_2 \uu^e \;=\; (\Jh \otimes \Dh \otimes \Jh) \uu^e \\ \label{eq:gradt}
  \uu^e_t &=& D_3 \uu^e \;=\; (\Dh \otimes \Jh \otimes \Jh) \uu^e,
\end{eqnarray}
each of which is a map from $p_1^3$ nodes to $q^3=\gamma^3 p_1^3$
quadrature points, where $\gamma := q/p_1$.

While the differentiation matrices $D_j$ are full, with $\gamma^3 p_1^6$ nonzeros,
they can be applied in factored form as tensor contractions with only $O(p^3)$
storage and $O(p^4)$ work.  For example, (\ref{eq:gradr}) is typically
implemented as
\begin{eqnarray} \label{eq:tens1a}
\uu^e_r =
(\Jh \otimes \Ih \otimes \Ih) (\Ih \otimes \Jh \otimes \Ih) (\Ih \otimes \Ih \otimes \Dh) \uu^e,
\end{eqnarray}
with $\Ih$ an appropriately sized identity matrix.  More explicitly,
(\ref{eq:tens1a}) is evaluated for $i,j,k \, \in \, \{1,\dots,q\}^3$ as
\begin{eqnarray} \label{eq:tens1}
(\uu^e_r)_{ijk} =
\sum_{\kh=0}^{p} \Jh_{k \kh} \left(
\sum_{\jh=0}^{p} \Jh_{j \jh} \left(
\sum_{\ih=0}^{p} \Dh_{i \ih} u_{\ih \jh \kh} \right) \right) \uu^e,
\end{eqnarray}
which has computational complexity
\begin{equation} \label{eq:work}
\begin{split}
  \mbox{W} &= 2(qp_1^3 + q^2 p_1^2 + q^3p_1)  \\
    &= 2p_1^4(\gamma + \gamma^2 + \gamma^3)
\end{split}
\end{equation}
operations.
This cost is nominally repeated for (\ref{eq:grads})--(\ref{eq:gradt}), save
for the common term $(\Ih \otimes \Ih \otimes \Jh) \uu^e$, which does not need to be
re-evaluated.  We note that $[D_1^T \, D_2^T \, D_3^T]$ can be applied in a
similar fashion, so the total work for (\ref{eq:bk5x}) with this approach is
\begin{eqnarray} \label{eq:work2}
 \mbox{W}= 4p_1^4(3\gamma^3 + 3\gamma^2 + 2\gamma) + 15\gamma^3 p_1^3.
\end{eqnarray}
The $O(p^3)$ term arises from application of the $3 \times 3$ $G^e_{mm'}$
tensor.

An alternative approach to (\ref{eq:gradr})--(\ref{eq:gradt}) is to first
interpolate $u$ to the quadrature points,
$\uut^e = J \uu^e = (\Jh \otimes \Jh \otimes \Jh) \uu^e$, followed by differentiation
\begin{eqnarray} \label{eq:gradr2}
  \uu^e_r &=& (\It \otimes \It \otimes \Dt) \uut^e \\ \label{eq:grads2}
  \uu^e_s &=& (\It \otimes \Dt \otimes \It) \uut^e \\ \label{eq:gradt2}
  \uu^e_t &=& (\Dt \otimes \It \otimes \It) \uut^e,
\end{eqnarray}
where $\It$ and $\Dt$ are respectively $q \times q$ identity and derivative matrices
on the one-dimensional array of quadrature points.  This strategy leads to a
complexity of $\mbox{W}=2p_1^4(3 \gamma^4+\gamma^3+\gamma^2+\gamma)$
for the gradient and an overall complexity for (\ref{eq:bk5x}) of
\begin{eqnarray} \nonumber \label{eq:work3}
 \mbox{W} = 4p_1^4(3\gamma^4+\gamma^3 + \gamma^2 + \gamma) + 15\gamma^3 p_1^3.
\end{eqnarray}
For $\gamma=1$, the second approach yields a reduction in work of $\approx$25\%
over (\ref{eq:tens1}).
By contrast, for $\gamma=3/2$ (commonly used in evaluating nonlinear advection
terms), the second approach incurs roughly a 12\% increase in operation count.

The mass matrix has a similar tensor product form. Specifically,
\begin{eqnarray} \label{eq:mass}
\uw^e = B^e \uu^e =
J^T 
{\tilde B}^e
J\uu^e, 
\end{eqnarray}
where ${\tilde B}^e=\mbox{diag}(\rho_i \rho_j \rho_k \cJ^e_{ijk})$
is the diagonal mass matrix on the quadrature points and
$J = \Jh \otimes \Jh \otimes \Jh$ is the interpolation operator
that maps the basis functions to the quadrature points.  Application of
(\ref{eq:mass}) has a complexity of $4p_1^4(\gamma^3+\gamma^2+\gamma)$
operations and $p_1^3(1+\gamma^3)$ reads from memory.
(We remark that the spectral element method uses
a diagonal mass matrix on the GLL points, for overall
work and memory complexity of only $p_1^3$ per element
for either the forward or inverse mass matrix application.)

While the tensor contractions have complexity of $O(p^4)$, every other
operation is of order $O(p^3)$ or lower (e.g., $O(p^2)$ for surface
operations).  On traditional architectures, the tensor contractions are
effectively implemented as dense matrix-matrix products (e.g., \cite{dfm02}).
On highly threaded architectures, such as GPUs, other approaches that
better exploit the tensor structure are often more effective because the
straightforward BLAS3 implementations do not expose sufficient parallelism and
data reuse.  We revisit this point in Section \ref{sec:gpu}.

We further note that the operation count for all tensor contractions can be
roughly halved by exploiting the symmetry of the GLL and GL point
distributions, as noted by \cite{solomonoff92} and \cite{Kopriva09}.  Such an
approach is used in deal.II and in some of our libParanumal results for the
Nvidia V100, as detailed in the Appendix.  In addition to operation counts, one
must recognize that the kernel (BK) performance can be heavily influenced by
whether $q$ or $p+1$ match cache-line sizes or vector-lane widths, which are
typically 4 or 8 words wide.  Moreover, application of the work-intensive
operators, $D_i$ in (\ref{eq:bk5x}) and $J$ in (\ref{eq:mass}), is identical for
each element (of the same order, $p$) because these operators are applied in
the reference domain $\Oh$.  One can therefore vectorize over blocks of
elements, rather than just applying the operators to a single element at a
time.  If the number of elements per rank is a multiple of (say) 4 or 8, one
can easily arrange the data to exploit this level of SIMD parallelism, assuming
that there are enough elements per MPI rank.

\begin{figure}
   \begin{center}
    \includegraphics[width=1.0\textwidth,height=1.6in]{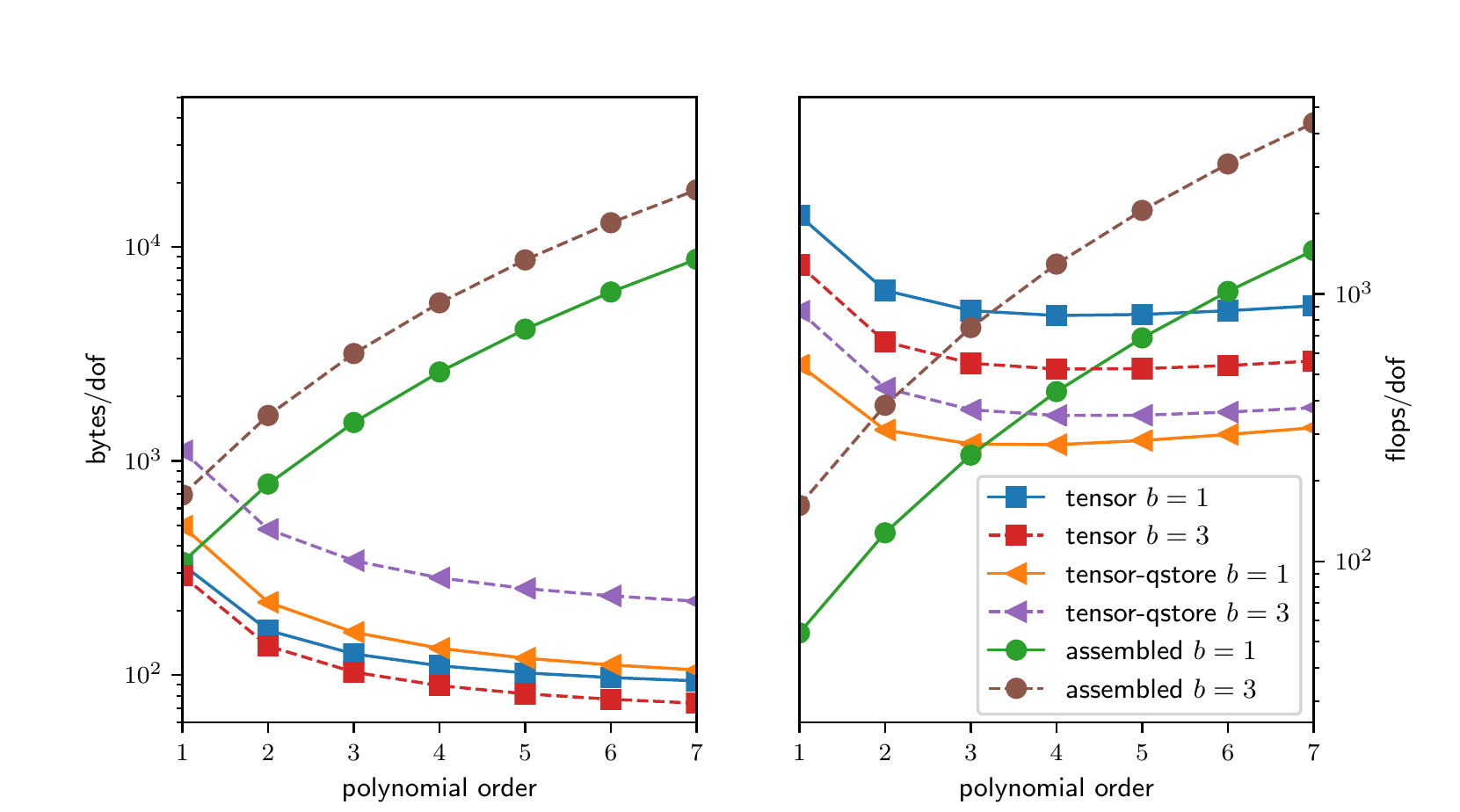}
    \caption{\label{fig:TensorVsAssembly}
      Memory transfer and floating-point operations per
      degree of freedom 
      for a PDE in 3D with $b$ components and variable coefficients.  
      The ``tensor'' computes metric terms on the fly and stores a compact
      representation of the coefficients at quadrature points;
      the ``tensor-qstore'' pulls the metric terms into the stored representation as 
      in (\ref{eq:gij}); the ``assembled'' uses a (block) CSR format.}
   \end{center}
  \end{figure}

We close this section by noting that, implemented properly, the
tensor-product-based matrix-free formulation requires the optimal amount of
memory transfers (with respect to the polynomial order) and near-optimal FLOPs
for operator evaluation.  The importance of these overhead costs is illustrated
in Figure~\ref{fig:TensorVsAssembly}, which contrasts the matrix-vector product
costs with fully assembled matrices (in block compressed sparse row format) with
matrix-free approaches.

 \begin{table*} \centering
\footnotesize
\caption{\label{tab:sys}System Configuration}
\begin{tabular}{l|l|l} \hline
              & ALCF BG/Q (Cetus \& Mira)  & OLCF Summit  \\ \hline
 Processor    & 16-core 1.6 GHz IBM PowerPC A2 &   IBM Power9$^{\rm TM}$ (2/node)    \\
 Nodes        &  2,048 (Cetus), 49,152 (Mira)  &   4,608                             \\
 CPU Cores/node   &  16                            &   42                                \\
 Total CPU Cores  &  32,768 (Cetus), 786,432 (Mira)&   193,536                           \\
 Total GPUs       &   --                           &   27,648 NVIDIA Volta V100s (6/node)\\
 Memory/node  &  16 GB RAM                     &   512 GB DDR4 + 96 GB HBM2         \\
 Peak Performancec & 10 PF (Mira)              &   42 TF                            \\
 Interconnect &  5D Torus Proprietary Network  &   Mellanox EDR 100F InfiniBand, Non-blocking Fat Tree \\ \hline
\end{tabular}
\end{table*}

\section{Bake-Off Performance on BG/Q}
\label{sec:bgq_results}

In this section, we present results for BP1--6 using Nek5000, MFEM, and deal.II.
Nek5000 is an F77 code developed at Argonne National Laboratory and originating
from the Nekton 2 spectral element code written at M.I.T.  by
\cite{ronquist88}, \cite{pff89}, and \cite{ho89}.  MFEM is a general-purpose
finite element library written in C++ at Lawrence Livermore National
Laboratory with the goal of enabling research and development of scalable finite
element discretizations \cite{mfem}. The deal.II library of~\cite{dealII85} is also
a C++ code that originally emerged from work at the Numerical Methods Group at
Universit\"{a}t Heidelberg, Germany. The main focus of deal.II is the
computational solution of partial differential equations using adaptive finite
elements. Further algorithmic details for each code are described in the Appendix.

All simulations were performed on the ALCF BG/Q Cetus in -c32 mode.
The BG/Q system configuration is shown in Table~\ref{tab:sys}.
The initial runs were based on the GNU-based compiler, or briefly  gcc.  On BG/Q, however, the
performance of gcc is much lower than that of the native IBM XL (xlc/xlf) compiler, or briefly xlc, 
so the battery of tests was rerun by using xlc, save for deal.II, which is unable to link
properly with xlc version of Cetus. We present both the gcc and xlc results.

We measured the rate of work in DOFs (\ref{eq:rate}).
The principal metrics of interest are
\begin{itemize}
\item $r_{\max}$, the peak rate of work per unit resource,
\item $\n12$, the local problem size $n/P$ on the node required to realize
      80\% of $r_{\max}$, and 
\item $t_{0.8}$, the time per iteration when running at
      80\% of $r_{\max}$.
\end{itemize}
Note that $\n12$ is defined in terms of points and that $r_{\max}$
(for any given $p$) is
the peak rate of work across all the implementations.
The importance of $\n12$, $r_{\max}$, and
the scaled ratio, $t_{0.8} = 1.25 \,\,\n12 / r_{\max}$, is discussed in
Section \ref{sec:cpu}.

All results are plotted as the rate of work per unit resource (\ref{eq:rate})
versus the number of points per compute node, $n/P$. We used a fixed iteration
of 100.  To simplify the notation,
we will refer to the performance variable---the $y$ axis---as DOFs (or MDOFs,
for millions of DOFs).  Choosing the number of points per
node as the independent variable on the $x$-axis, rather than number of DOFs
per node, leads to a data collapse in the case where components of vector
fields are computed independently:  one obtains a single curve for any number
of independent components.  When the component systems are solved
simultaneously, as in BP2, BP4, and BP6, benefits such as increased data reuse
or amortized messaging overhead should manifest as shifts up and to the left in
the performance curves.  We note that for the scalar problems (BP1, BP3, and
BP5), the number of DOFs is equal to the number of grid points.

\begin{figure*}
 \subfloat[Nek5000 gcc]{{\includegraphics[width=0.32\textwidth]{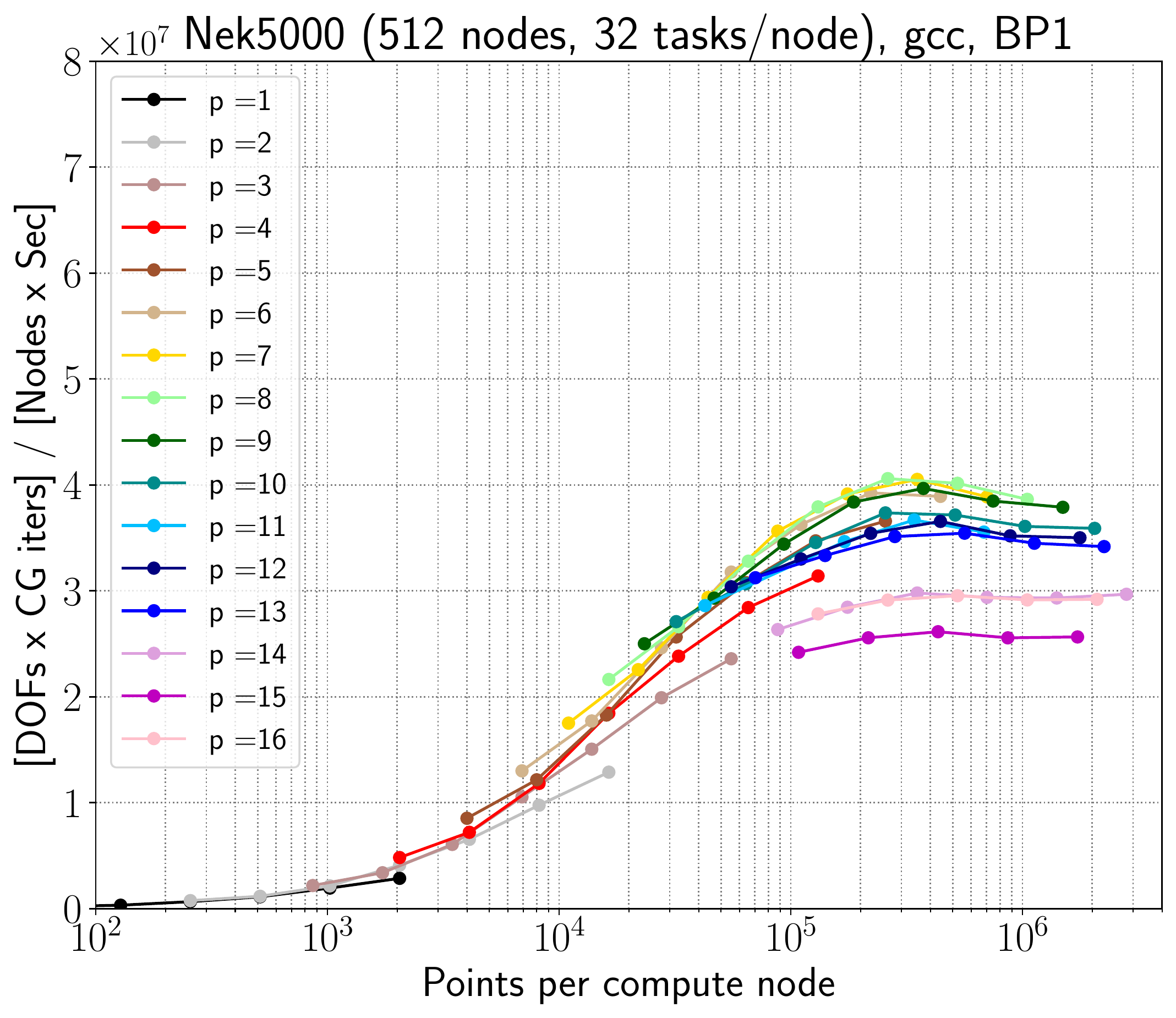}}}
 \hskip.1in
 \subfloat[MFEM    gcc]{{\includegraphics[width=0.32\textwidth]{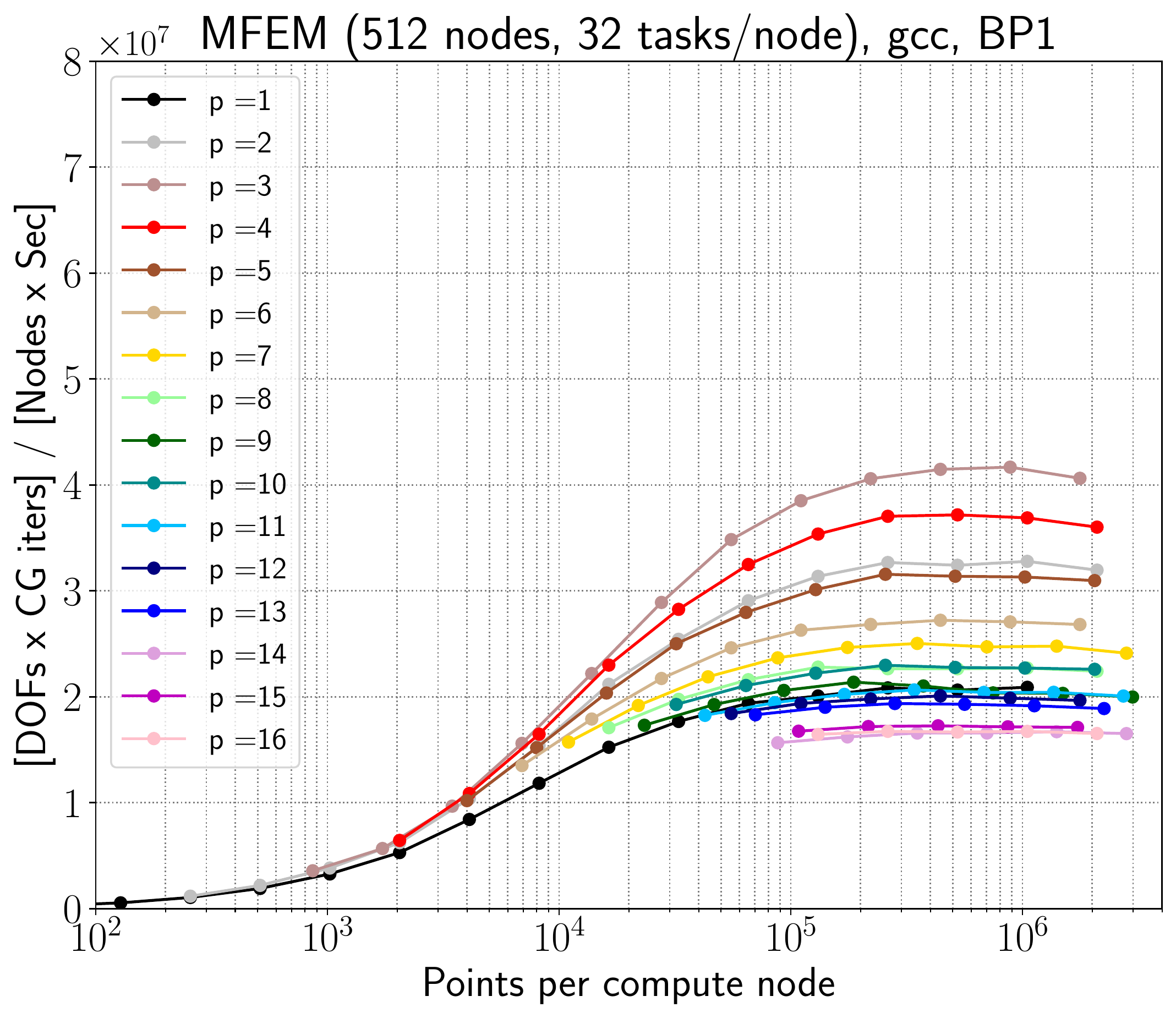}}}
 \hskip.1in
 \subfloat[deal.II gcc]{{\includegraphics[width=0.32\textwidth]{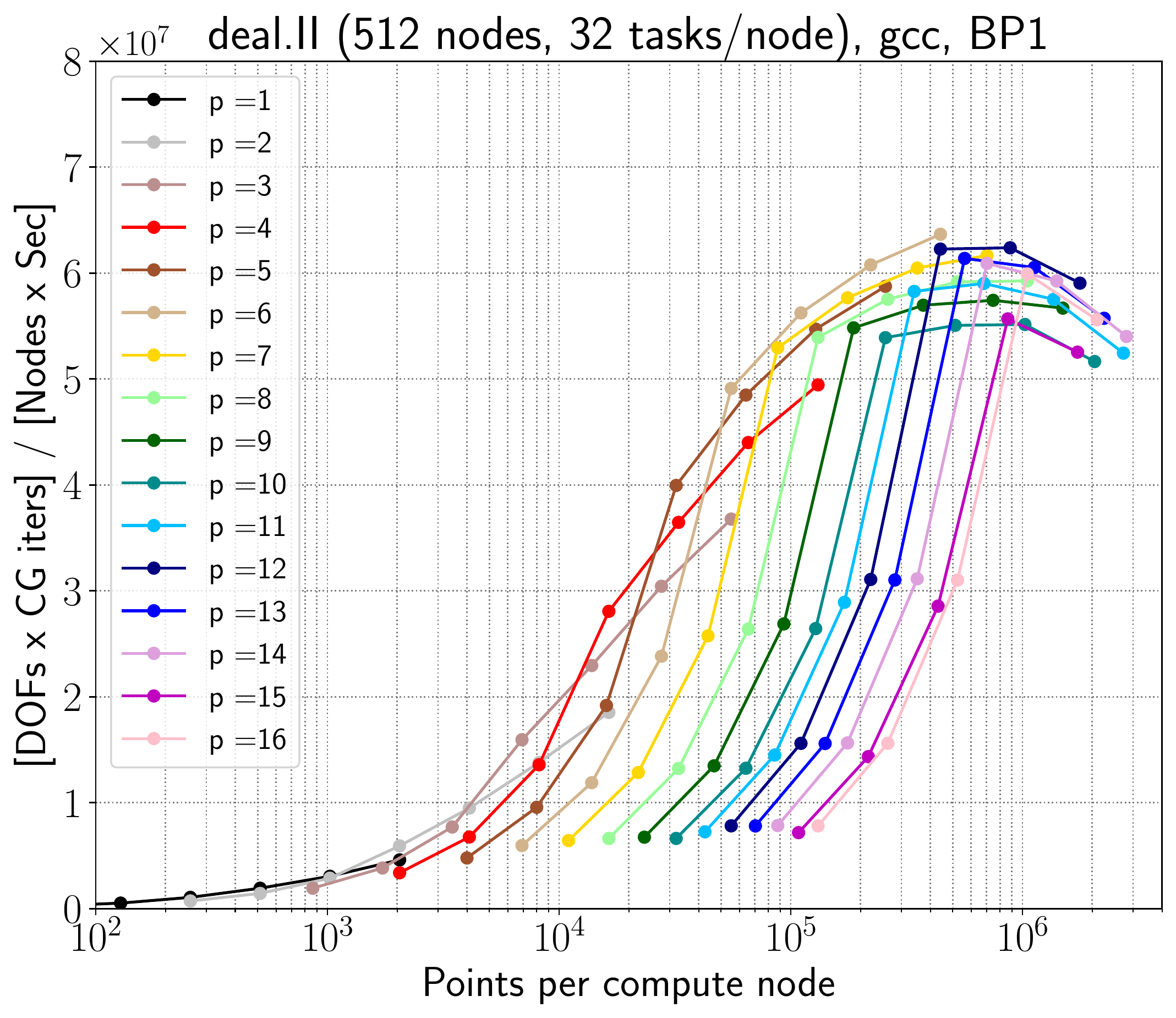}}}
 \caption{\label{fig:bp1_bgq_gcc}BP1 results with gcc compiler using 16,384 MPI ranks on 512 nodes of BG/Q;
          varying polynomial order ($p=1,...,16$) and quadrature points ($q=p+2$).}
 \end{figure*}

\begin{figure*} 
 \centering
 \subfloat[Nek5000 xlc]{{\includegraphics[width=0.32\textwidth]{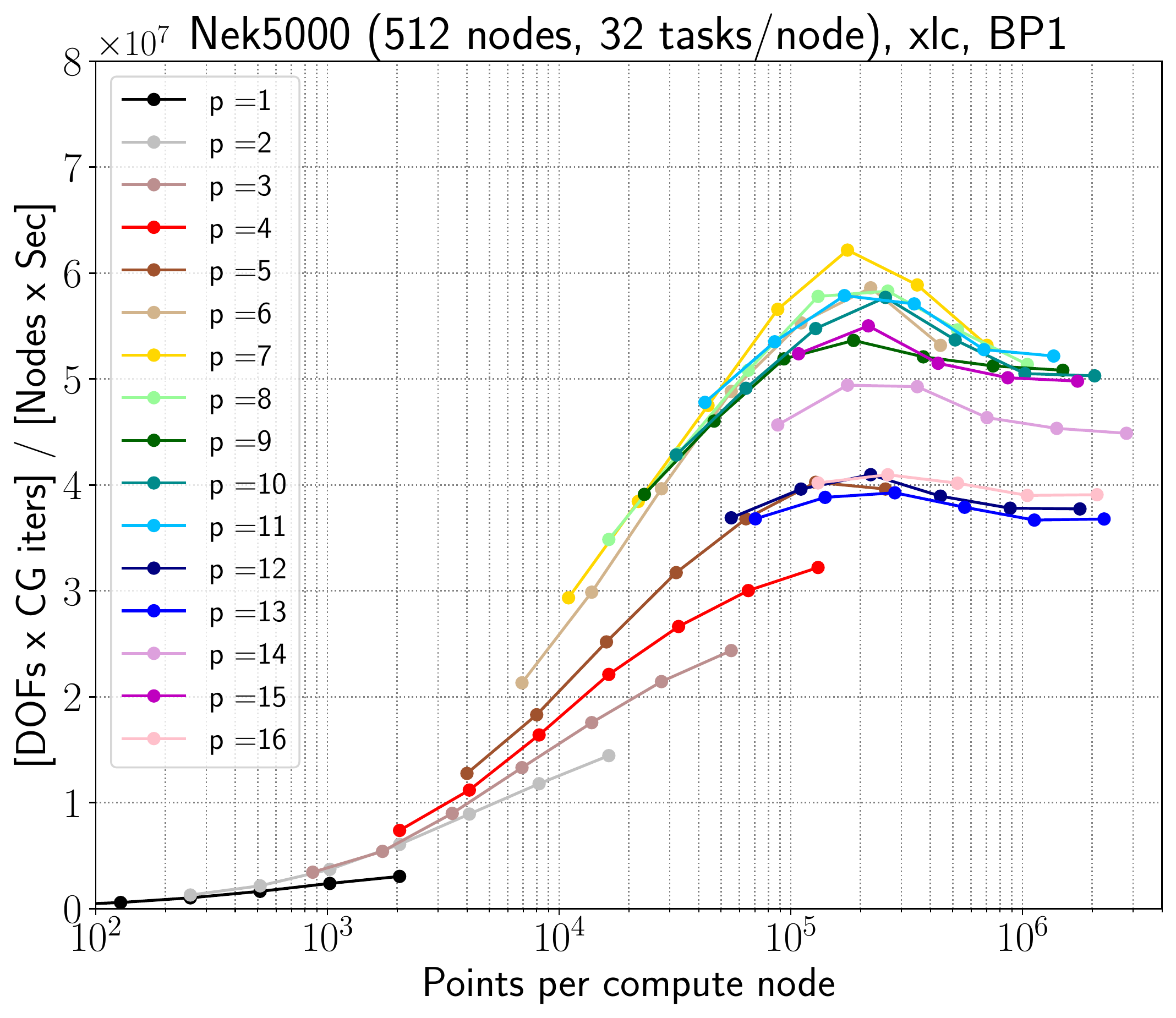}}}
 \hskip.1in
 \subfloat[MFEM    xlc]{{\includegraphics[width=0.32\textwidth]{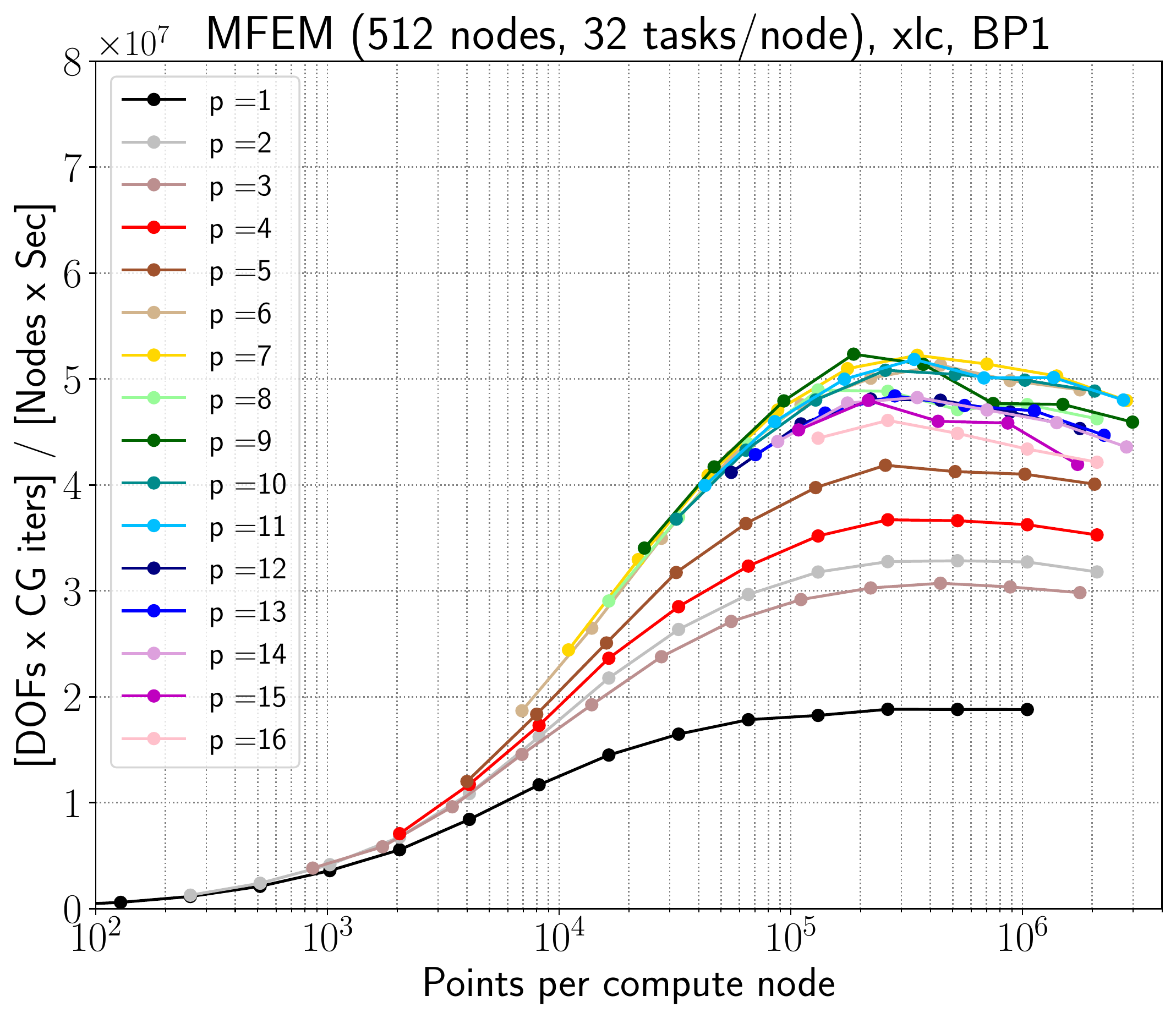}}}
 \hskip.1in
 \subfloat[MFEM    xlc/x86]{{\includegraphics[width=0.32\textwidth]{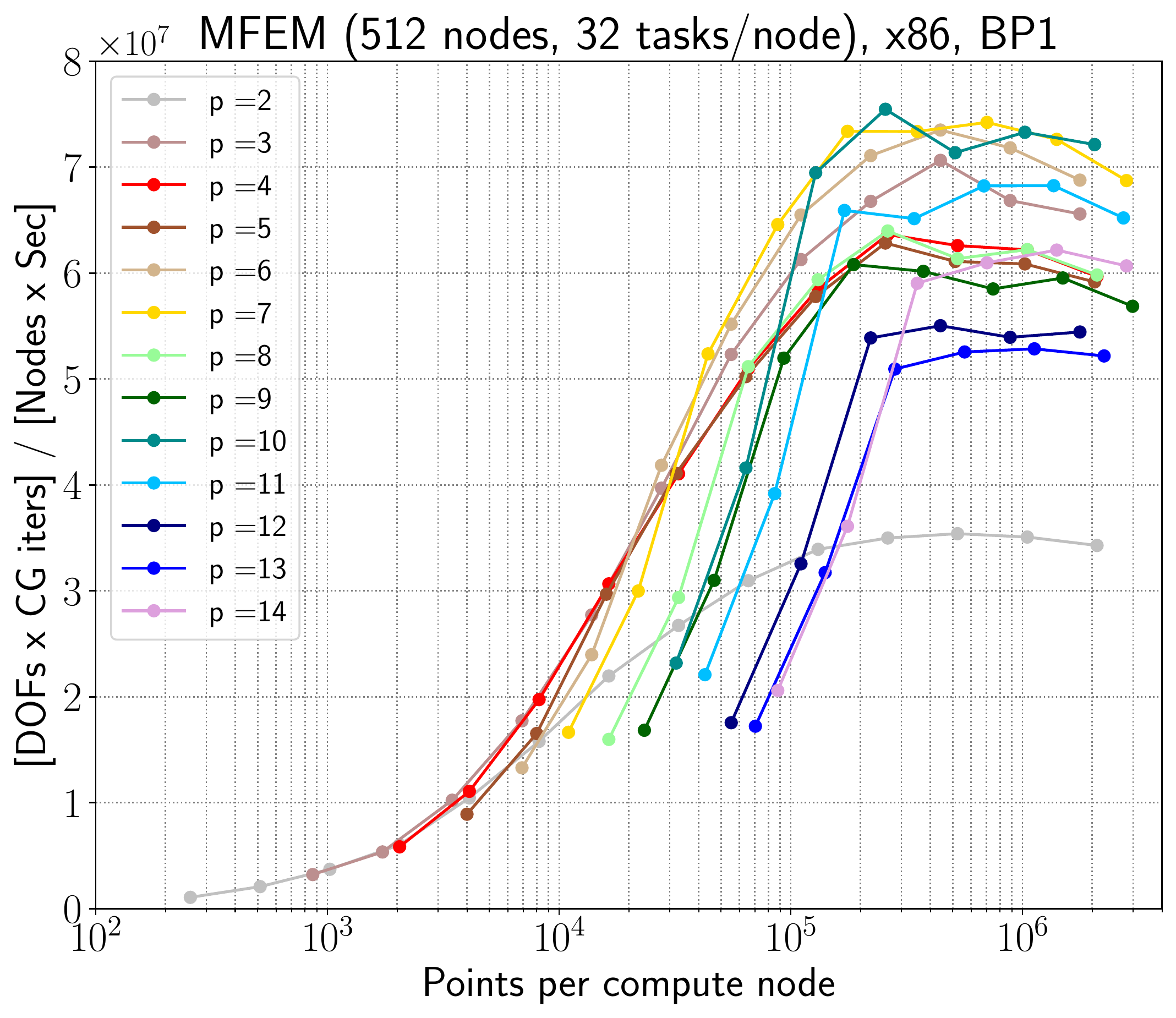}}}
 \caption{\label{fig:bp1_bgq_xlc}BP1 results with xlc compiler using 16,384 MPI ranks on 512 nodes of BG/Q; 
          varying polynomial order ($p=1,...,16$) and quadrature points ($q=p+2$).}
\end{figure*}

Figures \ref{fig:bp1_bgq_gcc}--\ref{fig:bp21_43_65_bgq_xlc} present the BP
results using the gcc and xlc compilers for Nek5000, MFEM, and deal.II.  In
each figure, each line represents a different polynomial order.  In all cases,
performance is strongly tied to the number of gridpoints per node, $n/P$.  In
the case of the gcc compilers, Nek5000 and MFEM generally exhibit a performance
plateau as $n/P$ increases, whereas deal.II shows a distinct peak.

In the discussion that follows, we focus primarily on the saturated (i.e., peak
observable) performance toward the right side of the graphs.  On the left
side, performance levels drop off to uninteresting values that users would
generally never experience.  This low-performance regime corresponds to
relatively few points per node and is easily avoided on distributed-memory
platforms by using fewer processors.  While the definition is not precise, the
point of rapid performance roll-off represents the {\em strong-scale limit} to
which most users will gravitate in order to reduce time per iteration.
Operating to the right of this point would incur longer run times.  This
transition point is thus the most significant part of the graph, and its
identification is an important part of the BP exercise.  A convenient
demarcation is $\n12$, which indicates the number of points per node where the
performance is 80\% of the realizable peak for the given polynomial order
$p$. (We reiterate that the peak is taken to be the peak across all codes in
the test suite, for each BP.)

\subsection{BP1}
Figures \ref{fig:bp1_bgq_gcc}--\ref{fig:bp1_bgq_xlc} present the BP results
for the mass matrix problem, BP1. Figure \ref{fig:bp1_bgq_gcc} uses the gcc
compilers, while Figure~\ref{fig:bp1_bgq_xlc} is based on xlc for Nek5000 and MFEM.
Nek5000+gcc sustains 27--33 million degrees of freedom per second (MDOFs or
{\em mega}-DOFs) for polynomial orders $p>4$, save for $p=14$ and 15, which saturate
around 25 MDOFs.  For MFEM+gcc, a peak performance of 42 MDOFs is realized for
$p=3$, which corresponds to $4 \times 4 \times 4$ bricks for each element.
MFEM realizes $>$ 32 MDOFs for $p=2$--4 and $\approx 20$ MDOFs for the
majority of the higher-order cases. With gcc, deal.II delivers an
impressive 54--64 MDOFs for $p>4$.  The highest values are attained for $n/P >450,000$.
The $\n12$ for deal.II is also high, however.
For example, for $n/P=100,000$, performance is below 30 MDOFs for all $p > 9$.
The rapid fall-off is related to the way deal.II distributes elements to MPI ranks.
The partitioner insists on having at least 8 elements per rank, rather than
insisting on a balanced load.  Having 8 elements per rank guarantees that 8-wide
vector instructions can be issued for any polynomial order but inhibits
strong scaling.

Figure~\ref{fig:bp1_bgq_xlc} again shows the BP1 results but now using
the xlc compiler for Nek5000 and MFEM.  In addition to the xlc compiler,
the Nek5000 results are using intrinsic-directed matrix-matrix product
routines for tensor contractions when the inner-product loop lengths
are multiples of 4. Separate timings (not shown) indicate
that most of the performance gains derive from the xlc compiler, with an
additional 5 to 30\% coming from the intrinsics, depending on $p$.
We see the advantage of xlc and the intrinsics, which boost the peak for
Nek5000 to 59 MDOFs for $p=7$ at $n/P=180,000$ and for MFEM to 54 MDOFs at
$n=190,000$ for $p=9$.  An interesting observation is that with xlc, $p=3$ is
the lowest performer (30 MDOFs) for MFEM (ignoring $p=1$), whereas it was the
highest (43 MDOFs) with gcc.

Inspired by the deal.II results on x86, the MFEM team also investigated the
use of xlc/x86 intrinsics, which allow vectoriziation over blocks of elements.
Panel (c) of Figure~\ref{fig:bp1_bgq_xlc}, denoted as MFEM xlc/x86,
shows the dramatic improvement resulting from this change.
For $p=6$, 7, and 10, MFEM xlc/x86 peaks at around 75 MDOFs,
and all polynomial orders except $p=2$ peak at over 50 MDOFs.

\subsection{BP3}
\begin{figure*} 
 \centering
 \subfloat[Nek5000 gcc]{{\includegraphics[width=0.32\textwidth]{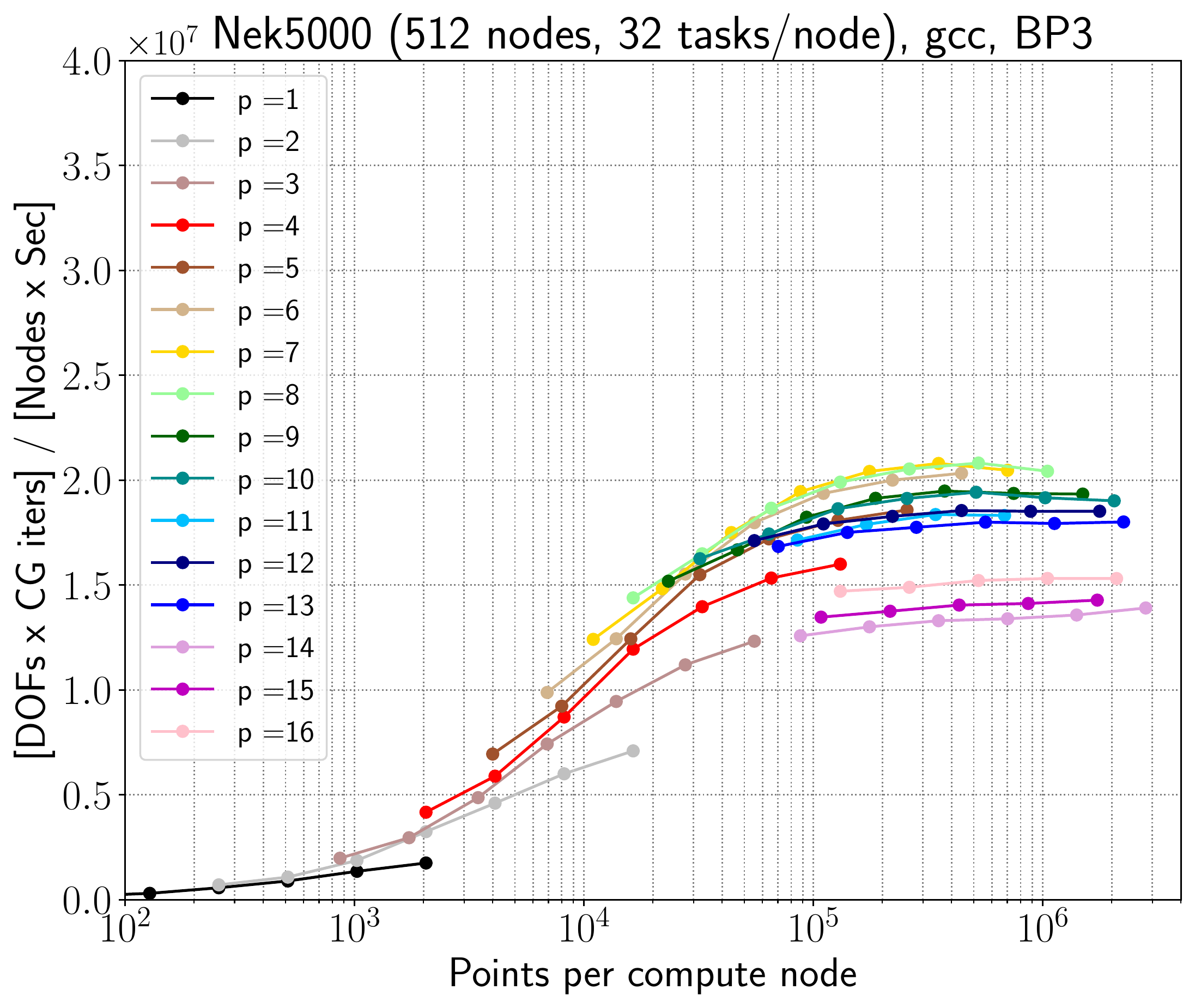}}}
 \hskip.1in
 \subfloat[MFEM    gcc]{{\includegraphics[width=0.32\textwidth]{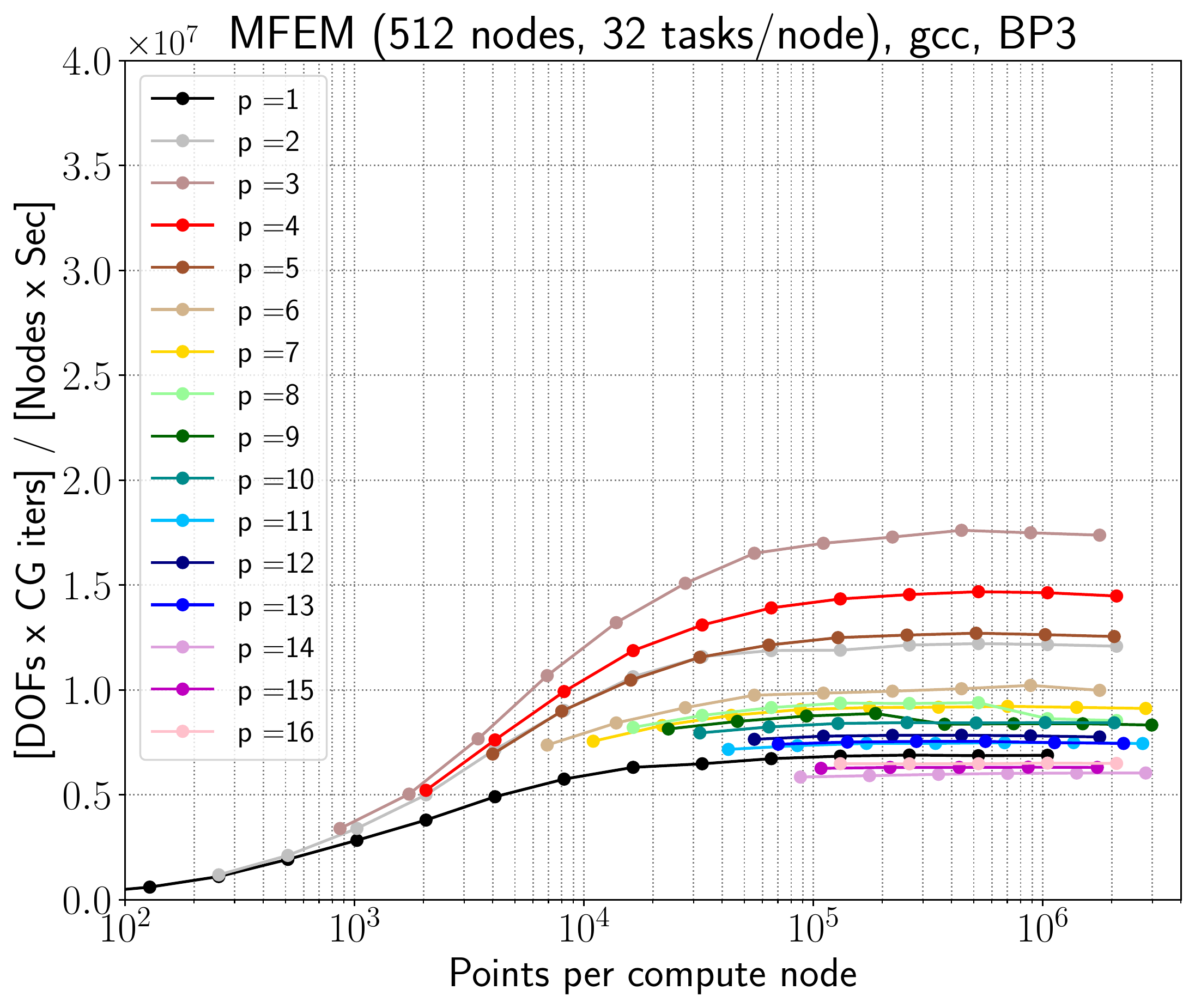}}}
 \hskip.1in
 \subfloat[deal.II gcc]{{\includegraphics[width=0.32\textwidth]{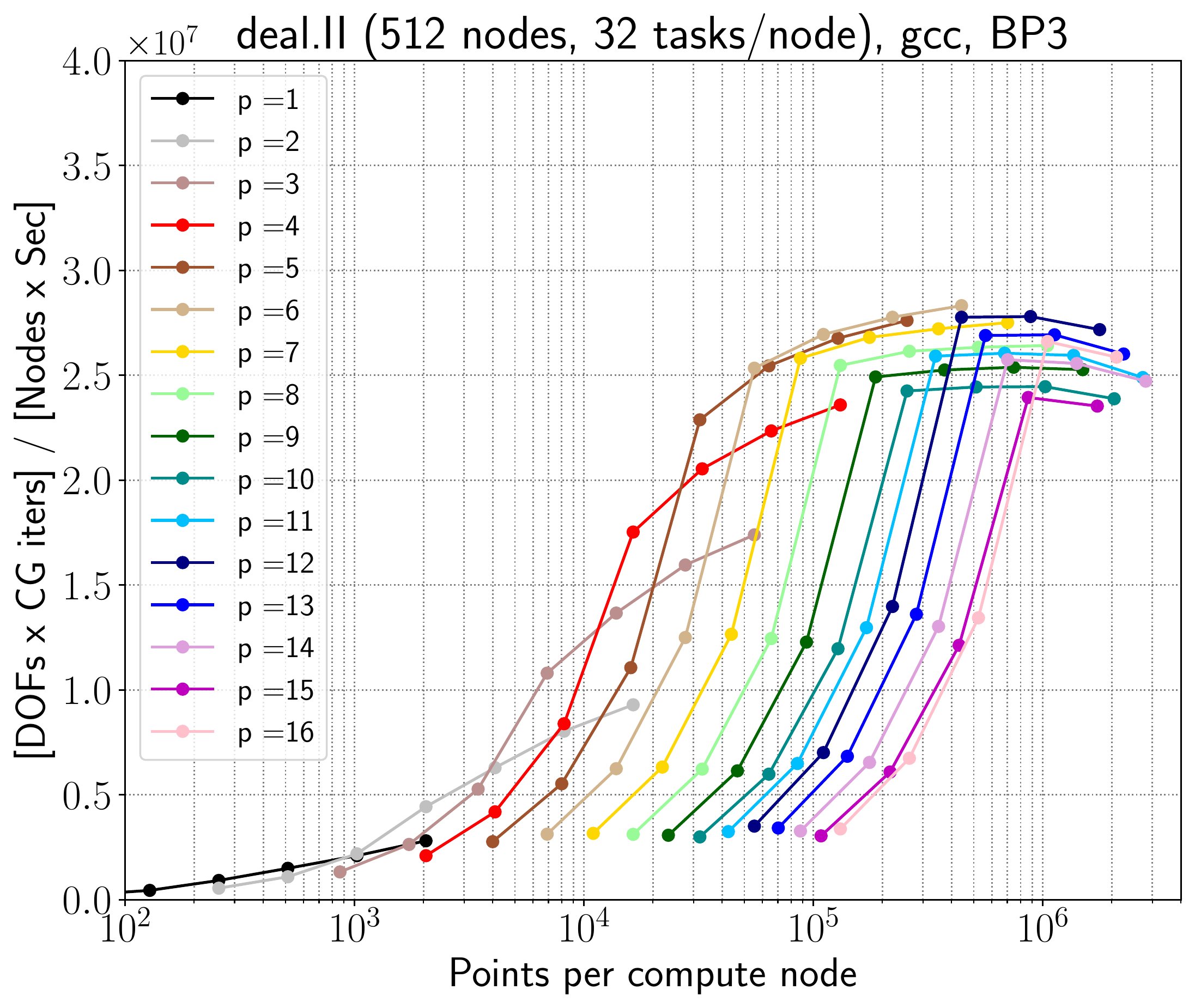}}}
 \caption{\label{fig:bp3_bgq_gcc}BP3 results with gcc compiler on 16,384 MPI ranks on 512 nodes of BG/Q;
           varying polynomial order ($p=1,...,16$) and quadrature points ($q=p+2$).}
\end{figure*}

\begin{figure*}
 \centering
 \subfloat[Nek5000 xlc]{{\includegraphics[width=0.32\textwidth]{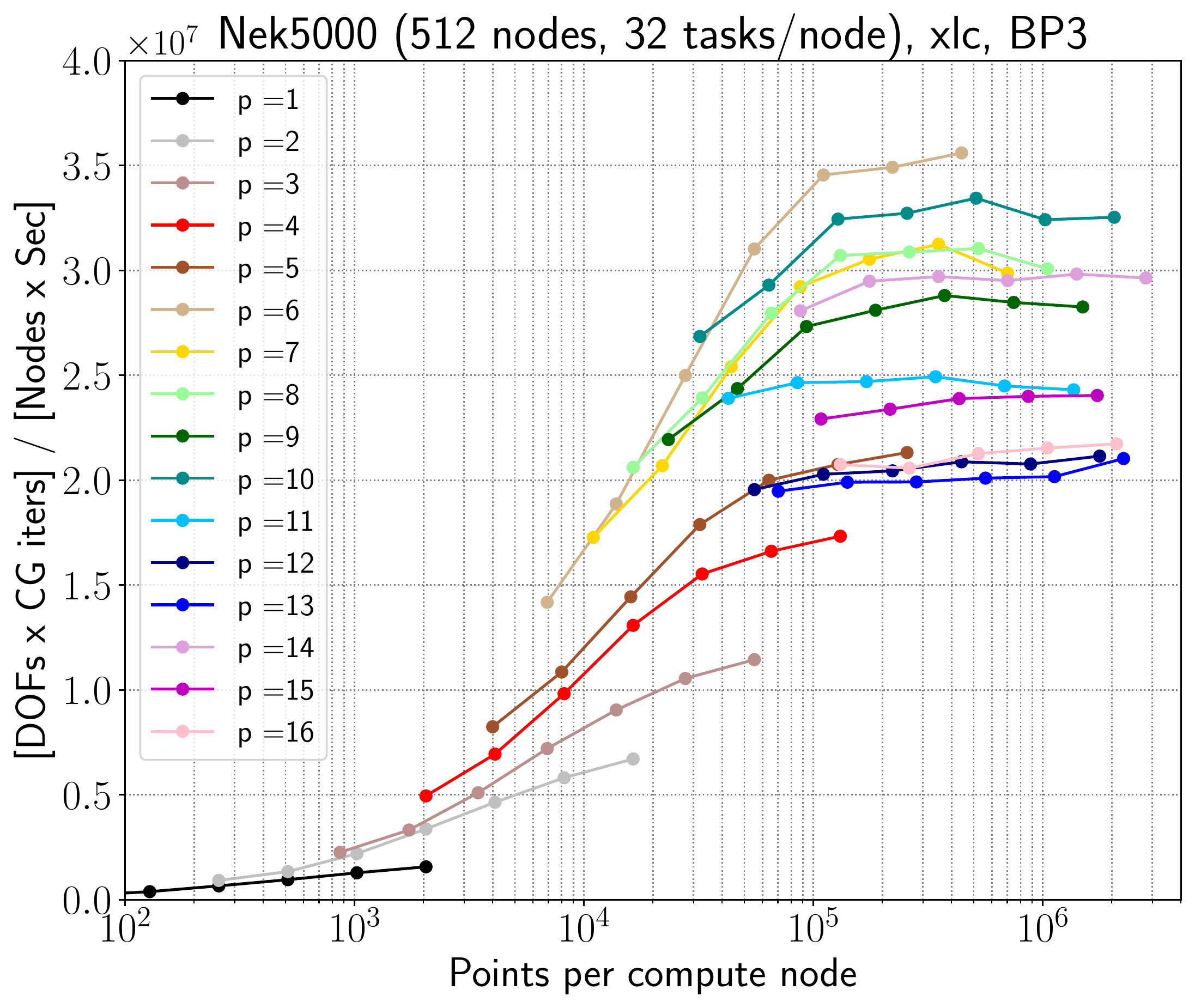}}}
 \hskip.1in
 \subfloat[MFEM    xlc]{{\includegraphics[width=0.32\textwidth]{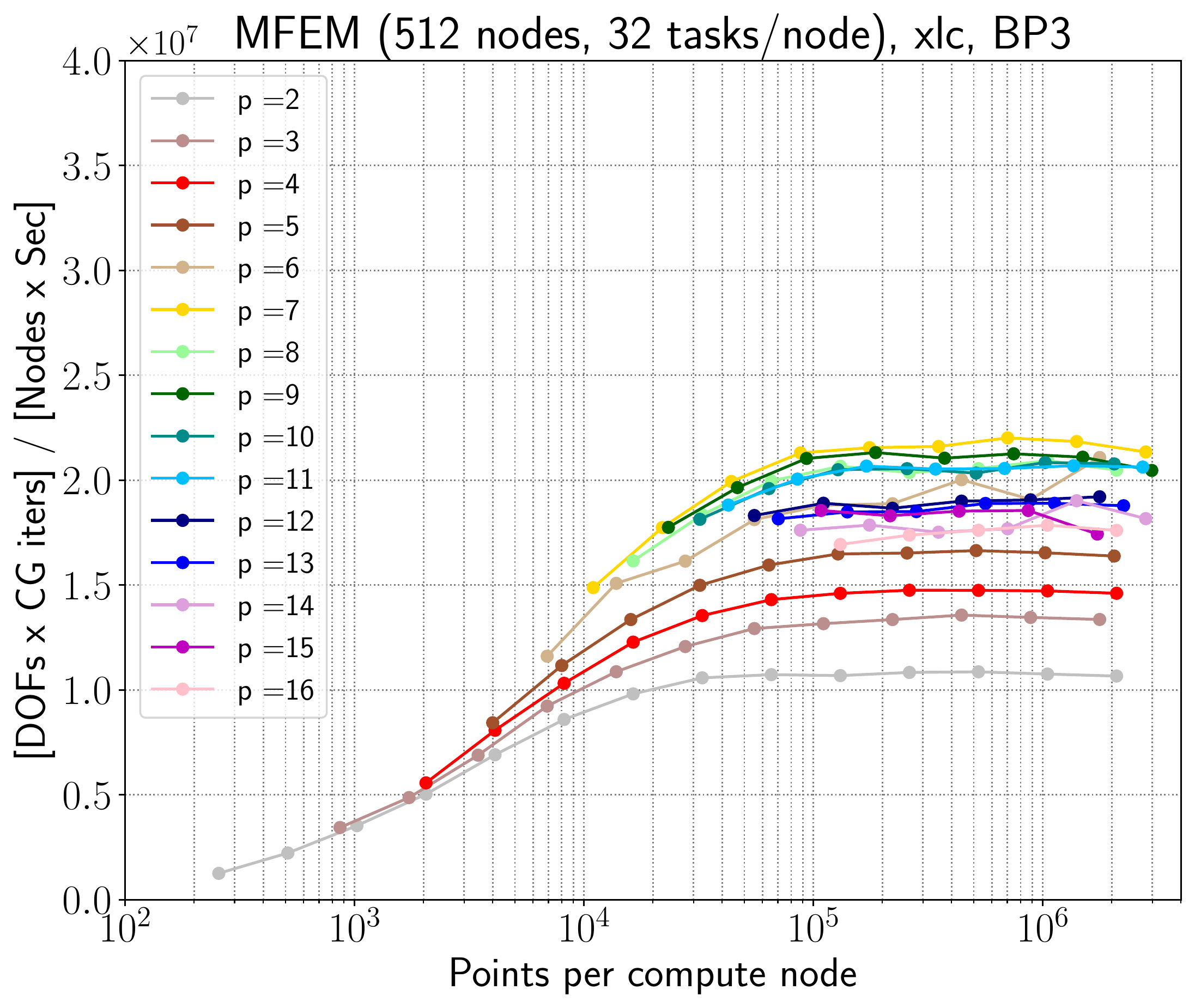}}}
 \hskip.1in
 \subfloat[MFEM    xlc/x86]{{\includegraphics[width=0.32\textwidth]{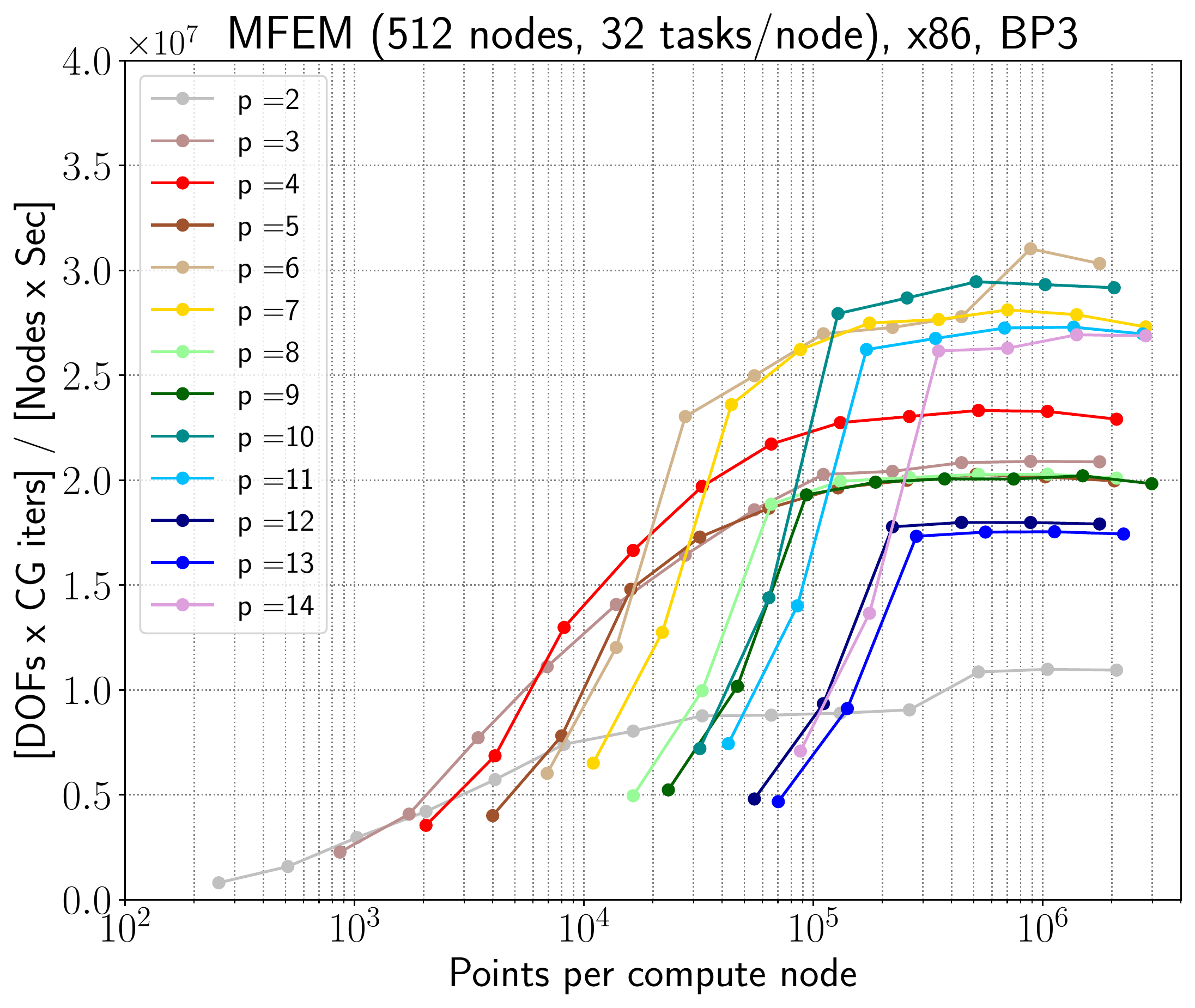}}}
 \caption{\label{fig:bp3_bgq_xlc}BP3 results with xlc compiler on 16,384 MPI ranks on 512 nodes of BG/Q;
           varying polynomial order ($p=1,...,16$) and quadrature points ($q=p+2$).}
\end{figure*}

Figures \ref{fig:bp3_bgq_gcc}--\ref{fig:bp3_bgq_xlc} present the BP results for the
stiffness-matrix problem, $A\uu=\ur$, with integration based on $q=p+2$ GL points
in each direction for each element.  The results are similar to those with BP1.  With
gcc, Nek5000 realizes 20 MDOFs for $p > 5$; MFEM achieves a peak of 18 MDOFs
for $p=3$, and deal.II reaches a peak of 28 MDOFs.  With xlc,
Nek5000 reaches a peak of 30--35 MDOFs for $p=6$--8 and 10, and MFEM reaches 20--22 MDOFs for
$p=7$--10.  The Nek5000 peak for $p=6$ corresponds to $q=8$ quadrature points,
for which the intrinsic, BLAS3-based tensor contractions are highly optimized.
For MFEM, the largest gains once again derive from the switch to using intrinsics,
which lift the MFEM xlc/x86 peak to around 30 MDOFs for $p=6$, 7, and 10.

\subsection{BP5}
\begin{figure*} 
 \centering
 \subfloat[Nek5000 gcc]{{\includegraphics[width=0.32\textwidth]{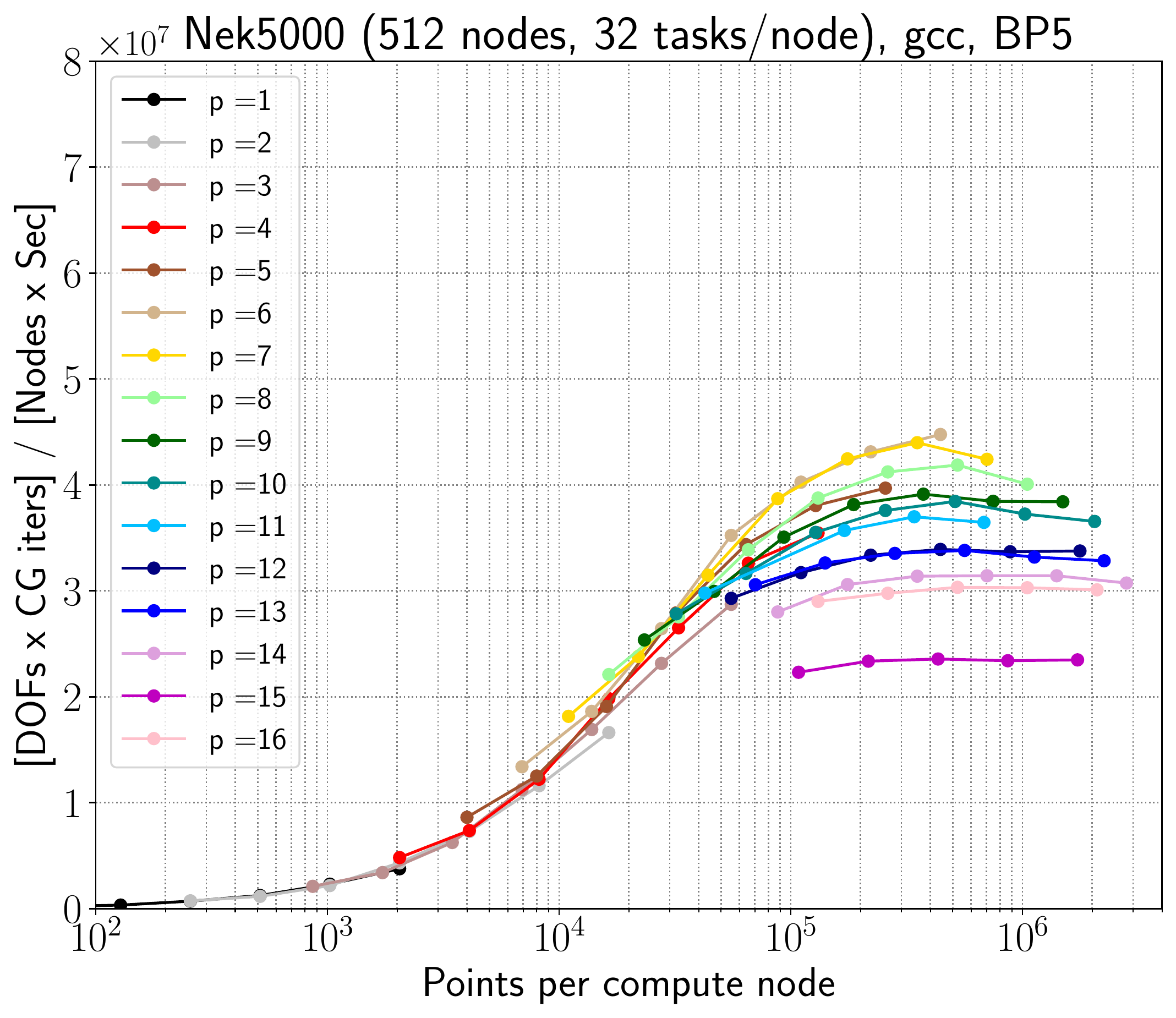}}}
 \hskip.1in
 \subfloat[MFEM    gcc]{{\includegraphics[width=0.32\textwidth]{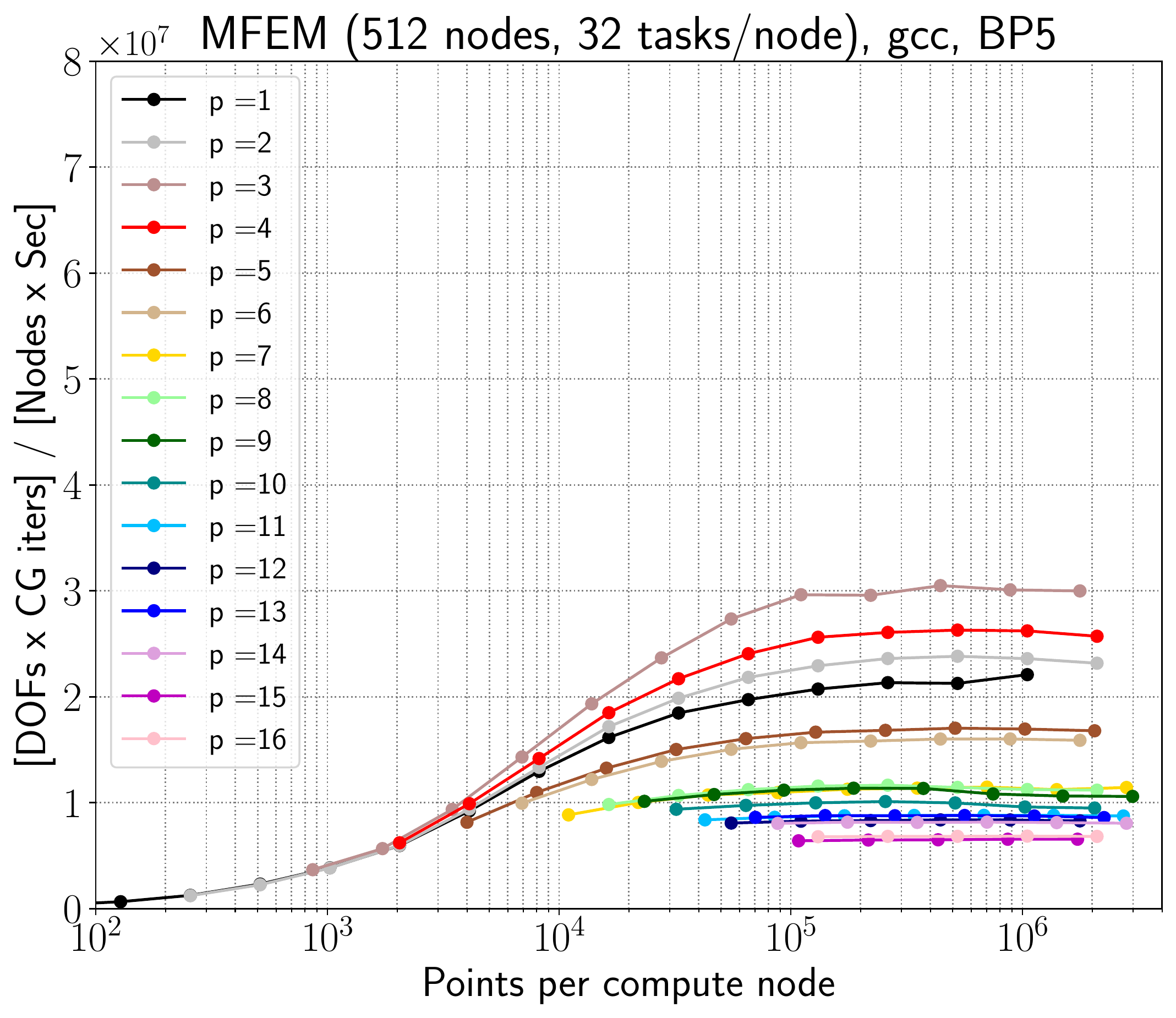}}}
 \hskip.1in
 \subfloat[deal.II gcc]{{\includegraphics[width=0.32\textwidth]{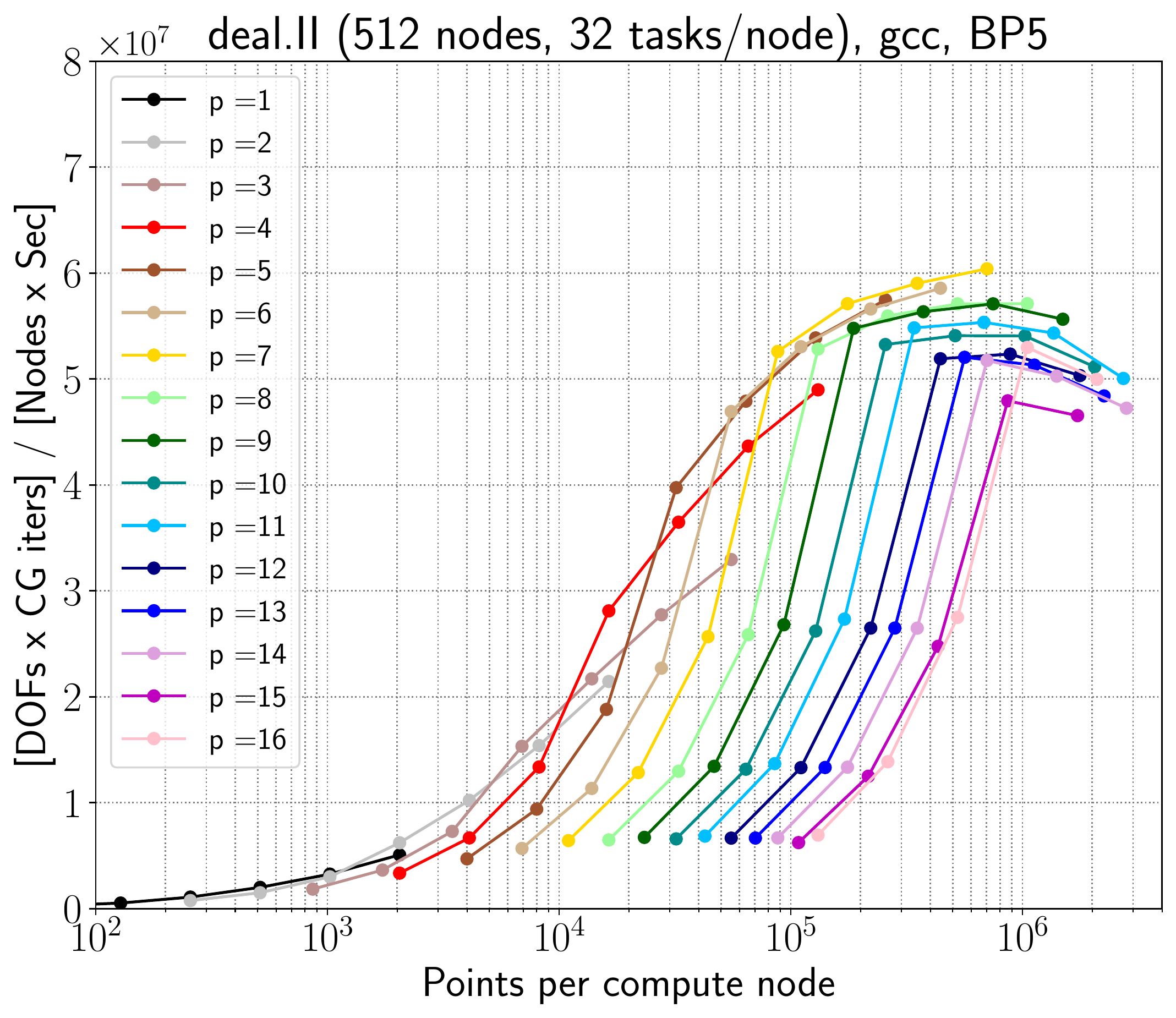}}}
 \caption{\label{fig:bp5_bgq_gcc}
           BP5 results with gcc compiler on 16,384 MPI ranks on 512 nodes of BG/Q;
           varying polynomial order ($p=1,...,16$) and quadrature points ($q=p+1$).}
\end{figure*}

\begin{figure*}
 \centering
 \subfloat[Nek5000 xlc]{{\includegraphics[width=0.32\textwidth]{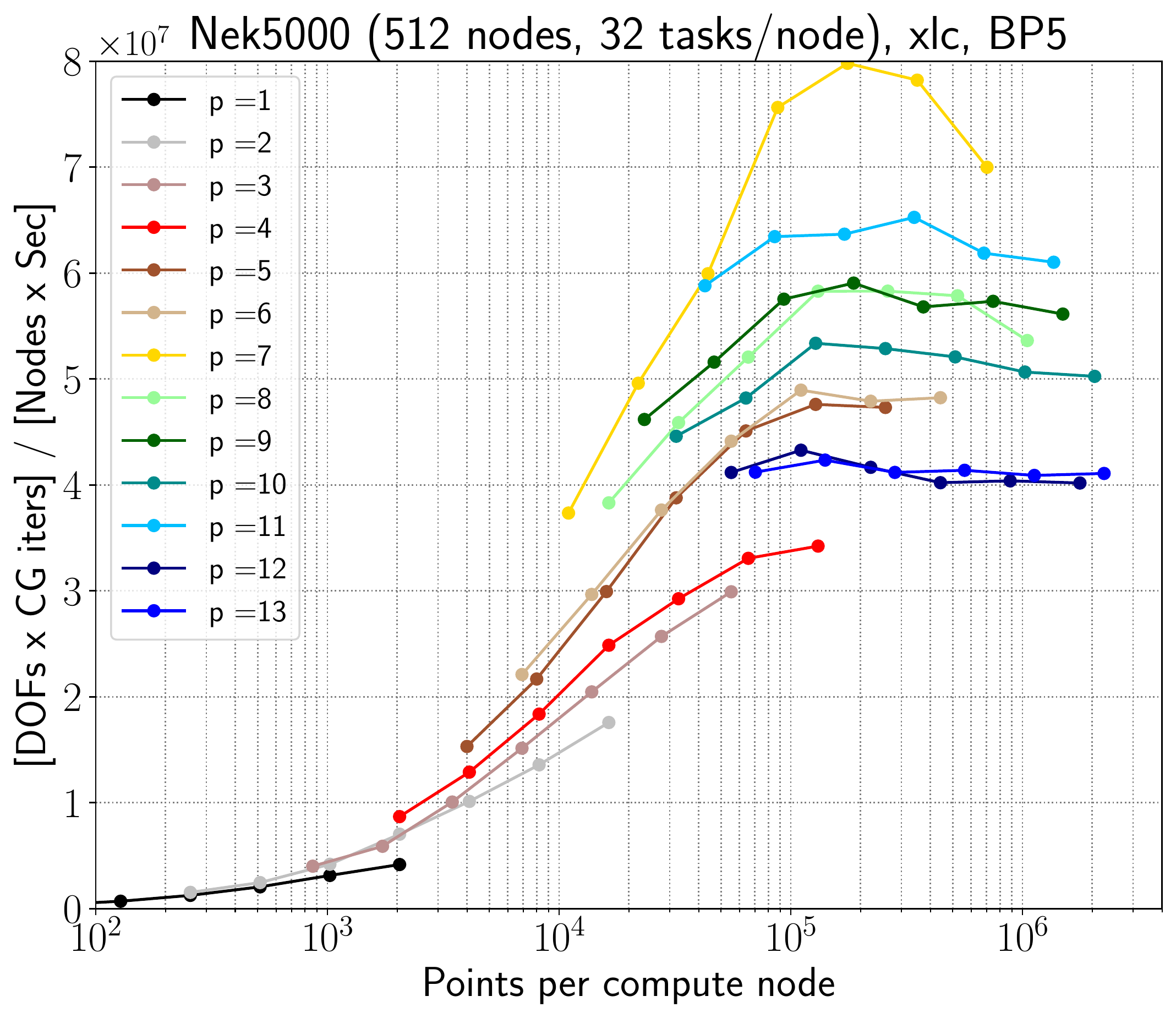}}}
 \hskip.1in
 \subfloat[MFEM    xlc]{{\includegraphics[width=0.32\textwidth]{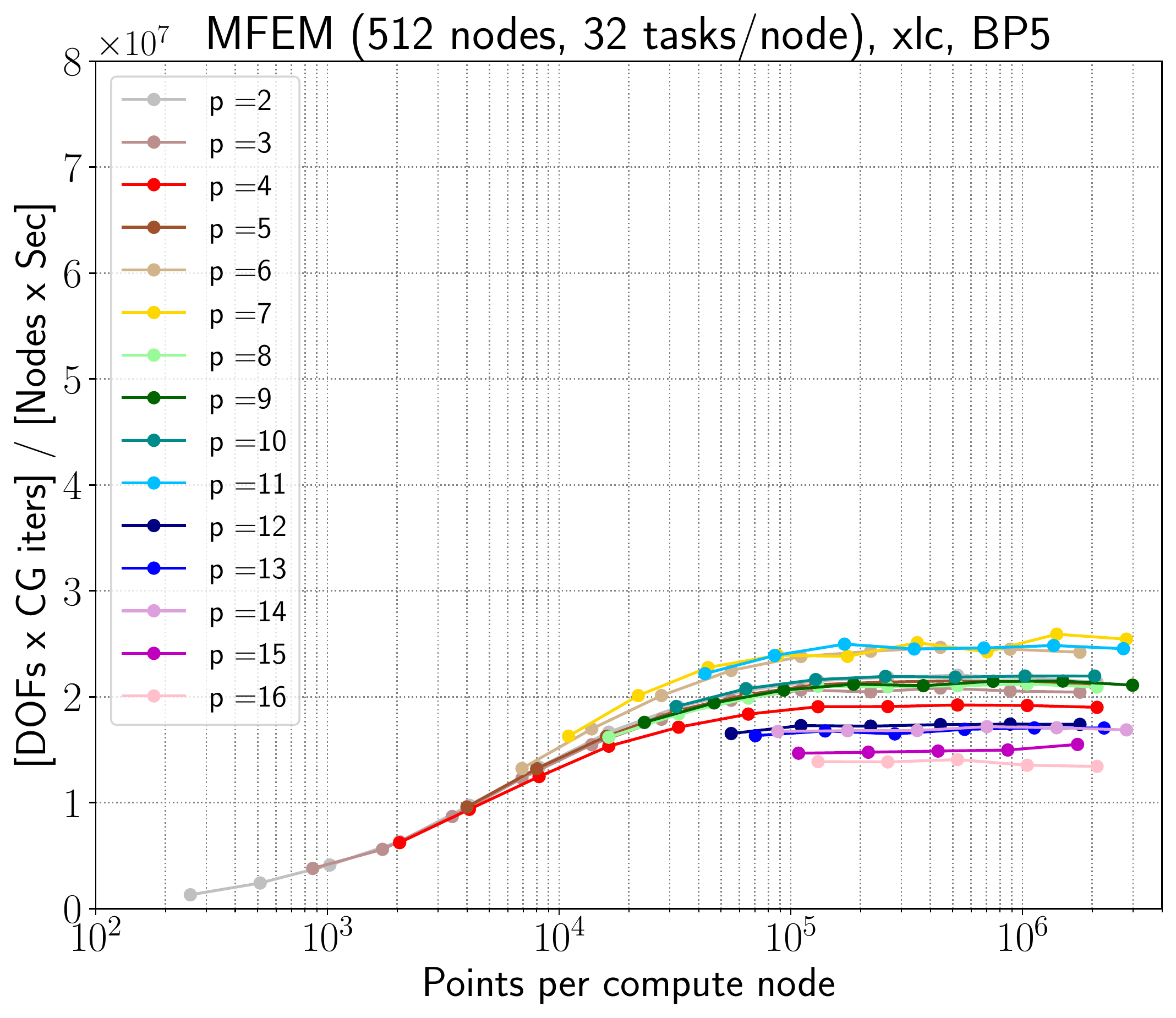}}}
 \hskip.1in
 \subfloat[MFEM    xlc/x86]{{\includegraphics[width=0.32\textwidth]{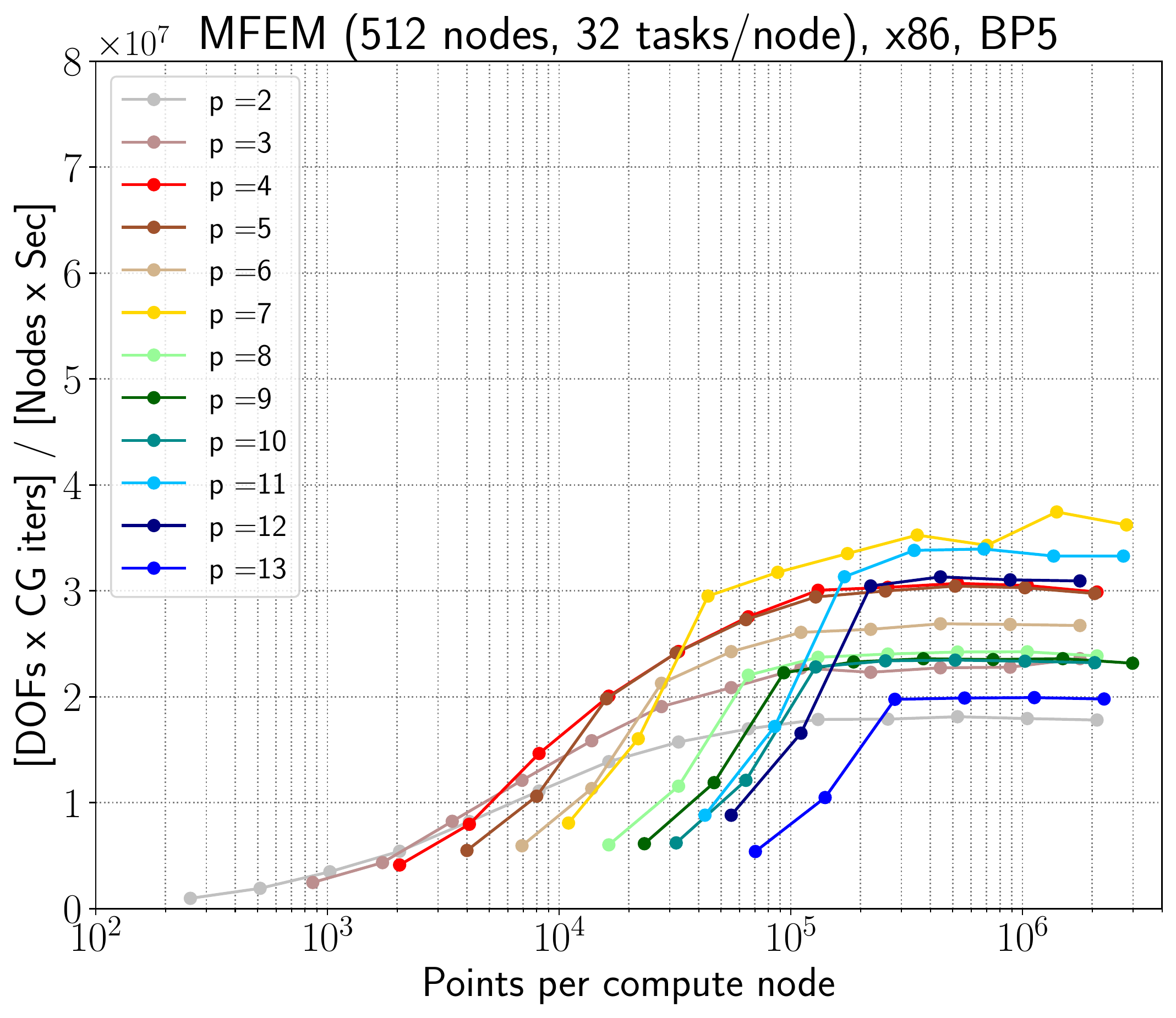}}}
 \caption{\label{fig:bp5_bgq_xlc}
            BP5 results with xlc compiler on 16,384 MPI ranks on 512 nodes of BG/Q;
           varying polynomial order ($p=1,...,16$) and quadrature points ($q=p+1$).}
\end{figure*}

BP5 (Figures \ref{fig:bp5_bgq_gcc}--\ref{fig:bp5_bgq_xlc}) solves a Poisson
problem using the standard spectral element stiffness matrix in which quadrature
is based on the $q=p_1$ GLL points, which  obviates the need for interpolation,
resulting in a shift from (\ref{eq:gradr}) to the simpler form (\ref{eq:gradr2}).
The net result is a nearly twofold increase in MDOFs across all cases.
Respectively for Nek5000, MFEM, and deal.II, BP5-gcc realizes peaks of 32, 30, and
60 MDOFs, in contrast to 18, 18, and 28 MDOFs for BP3-gcc.  For BP5-xlc, the corresponding
peaks are 80 and 25 MDOFs for Nek5000 and MFEM.  We note that the MFEM xlc/x86
code is not optimized for BP5 because it does a redundant interpolation from mesh
nodes to quadrature nodes in this case.  The xlc/x86 performance, Figure
\ref{fig:bp5_bgq_xlc}(c), is nonetheless marginally improved over the results
of Figure \ref{fig:bp3_bgq_xlc}(c) because of the slight reduction in the
number of quadrature points.

\subsection{BP2, BP4, and BP6}

\begin{figure*}
 \centering
 \subfloat[BP2 xlc]{{\includegraphics[width=0.32\textwidth]{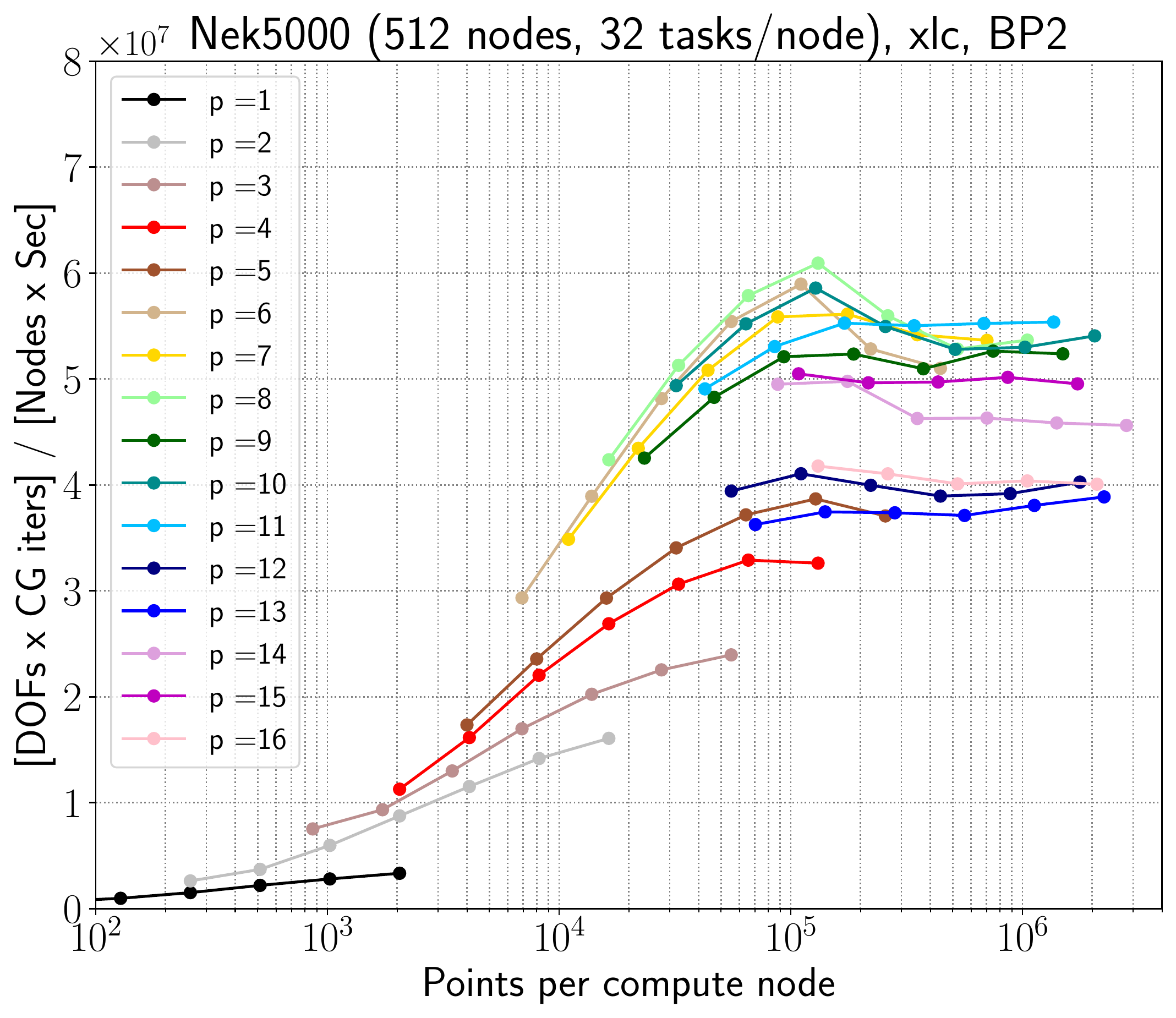}}}
 \hskip.1in
 \subfloat[BP4 xlc]{{\includegraphics[width=0.32\textwidth]{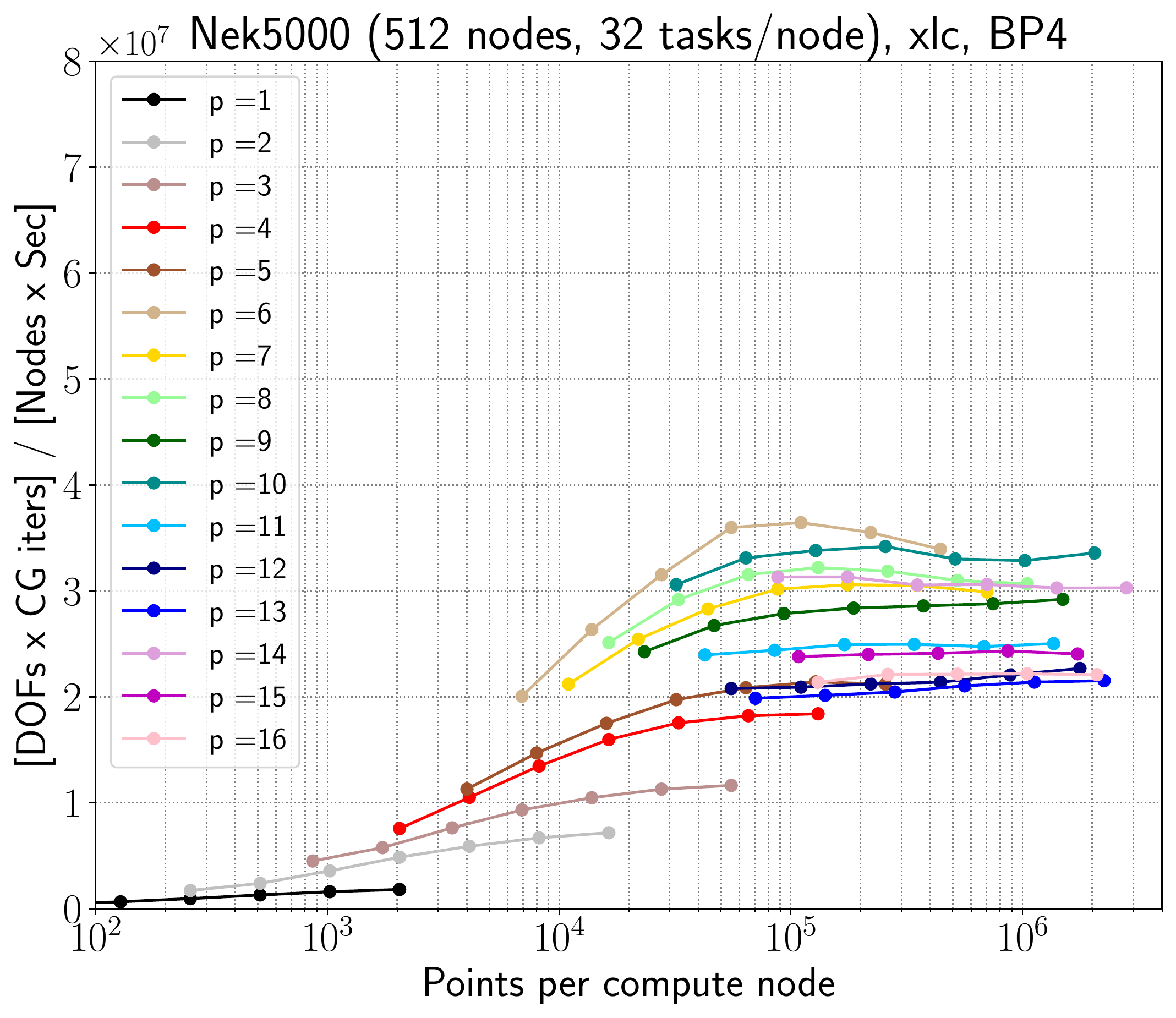}}}
 \hskip.1in
 \subfloat[BP6 xlc]{{\includegraphics[width=0.32\textwidth]{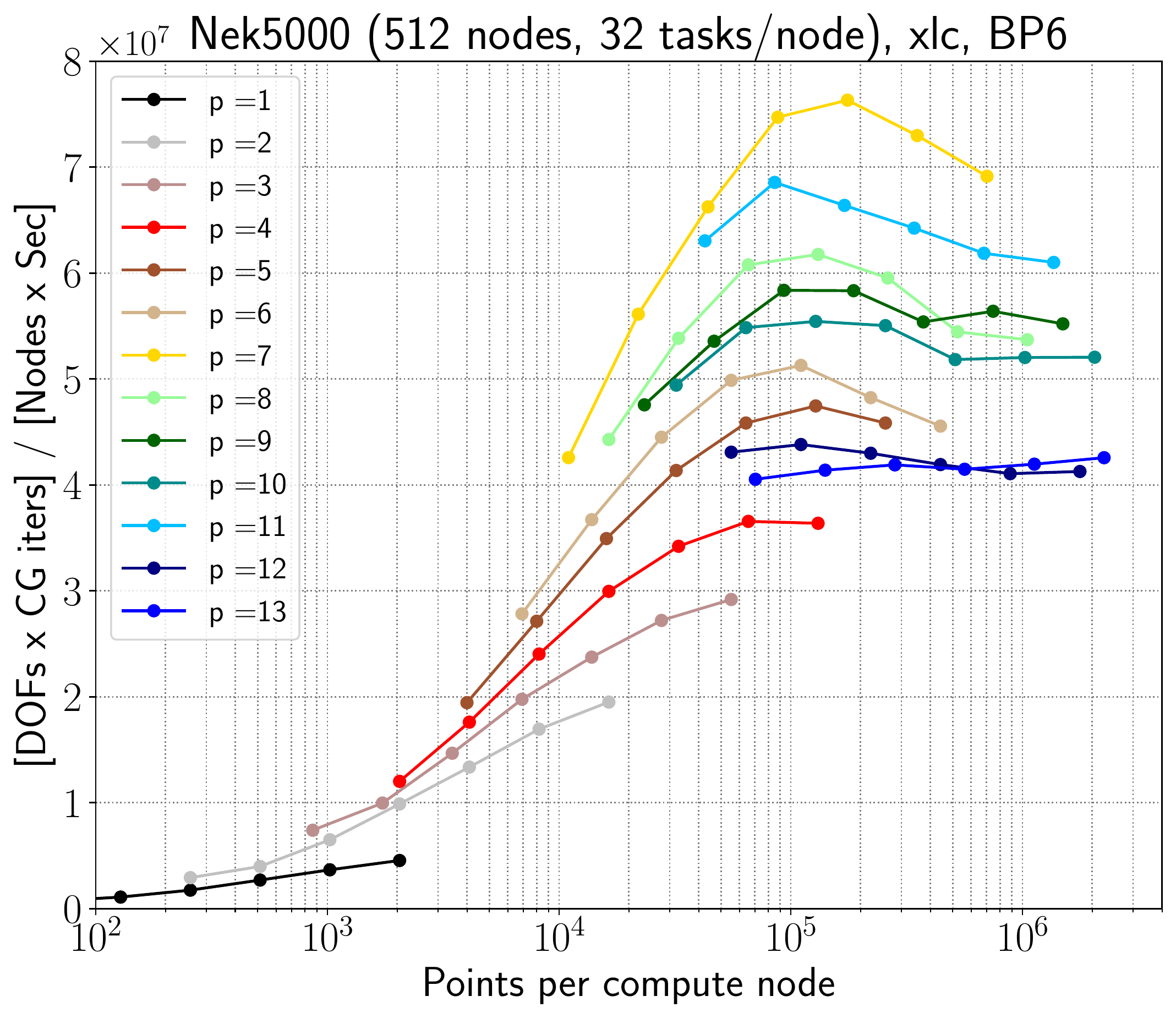}}}
  \caption{\label{fig:bp2_4_6_bgq_xlc}Nek5000 results with xlc compiler 
           on 16,384 MPI ranks on 512 nodes of BG/Q with varying polynomial order ($p=1,...,16$). }
\end{figure*}

Results for the vector-oriented BPs are shown in
Figure~(\ref{fig:bp2_4_6_bgq_xlc}) for Nek5000 only.
For each of these cases, we solve three systems
with three different right-hand sides.\footnote{In practice,
the systems may differ slightly because of the nature of the boundary
conditions for each equation, e.g., as is the case when solving
the Navier-Stokes equations with slip conditions for velocity.}
The performance results for BP2, 4, and 6 are similar to the corresponding
scalar results for BP1, 3, and 5, with the most evident change being that BP6
realizes only 76 MDOFs for $p=7$ instead of the 80 MDOFs attained for BP5.

\begin{figure*}
 \centering
 \subfloat[BP2/BP1]{{\includegraphics[width=0.32\textwidth]{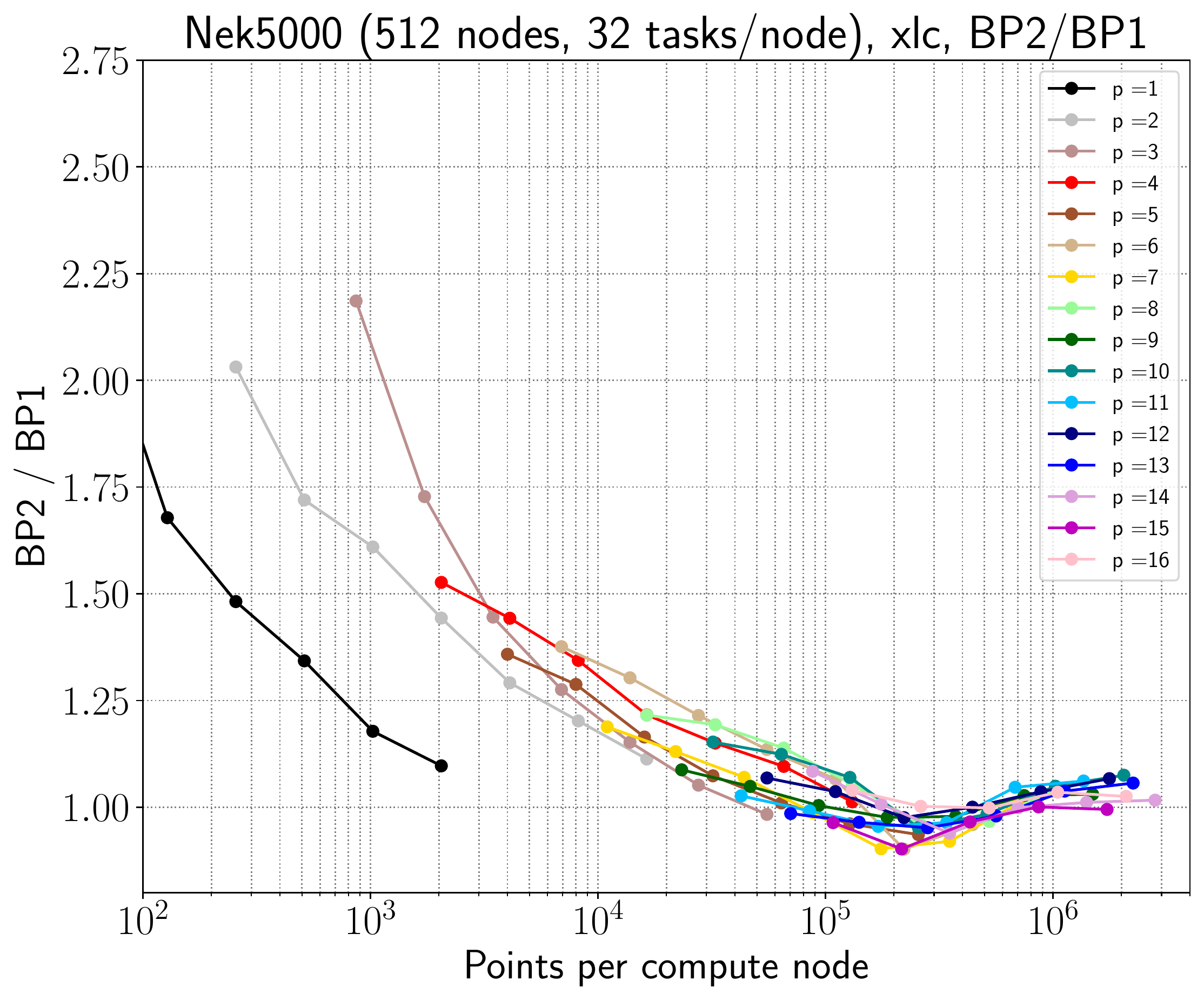}}}
 \hskip.1in
 \subfloat[BP4/BP3]{{\includegraphics[width=0.32\textwidth]{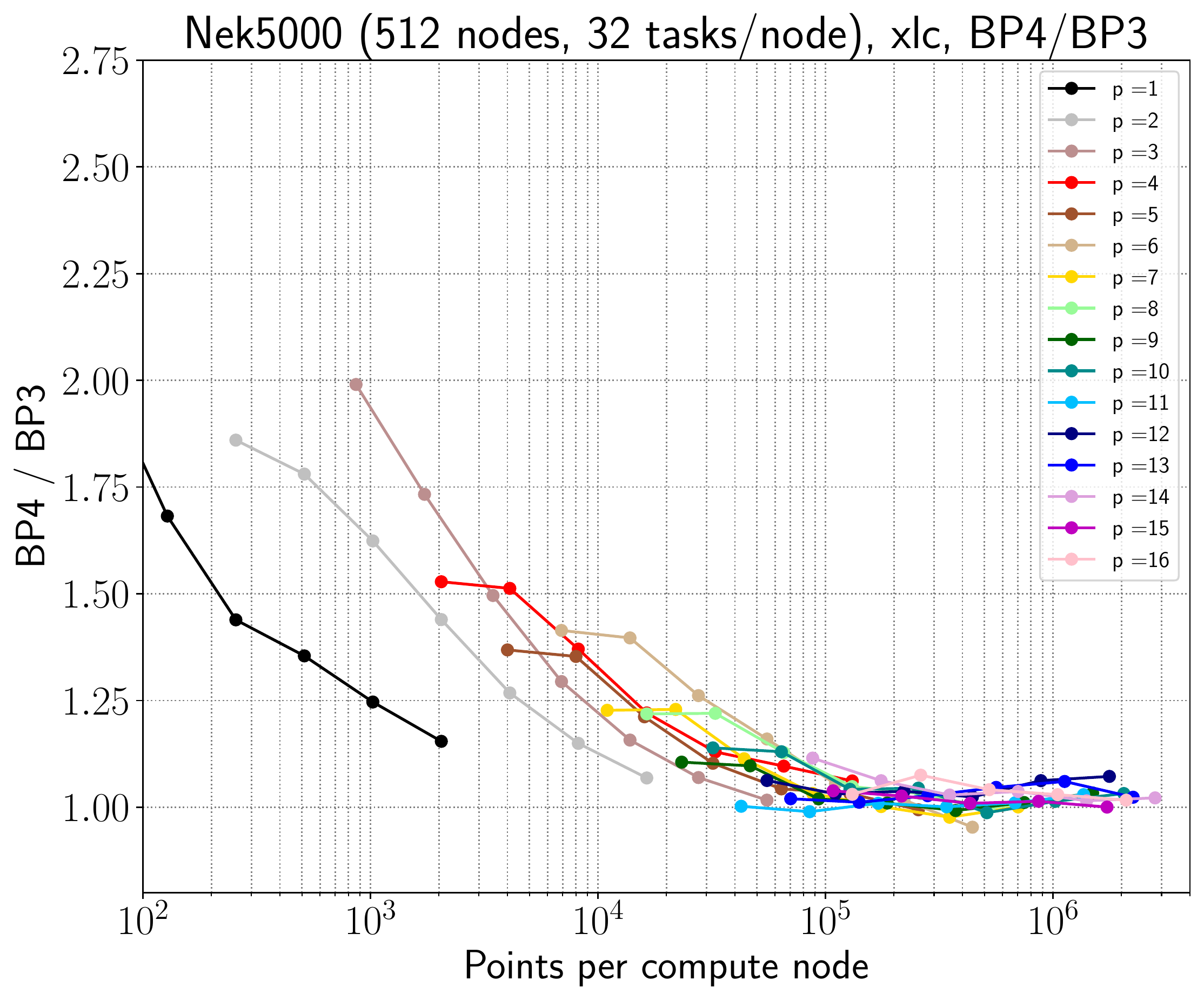}}}
 \hskip.1in
 \subfloat[BP6/BP5]{{\includegraphics[width=0.32\textwidth]{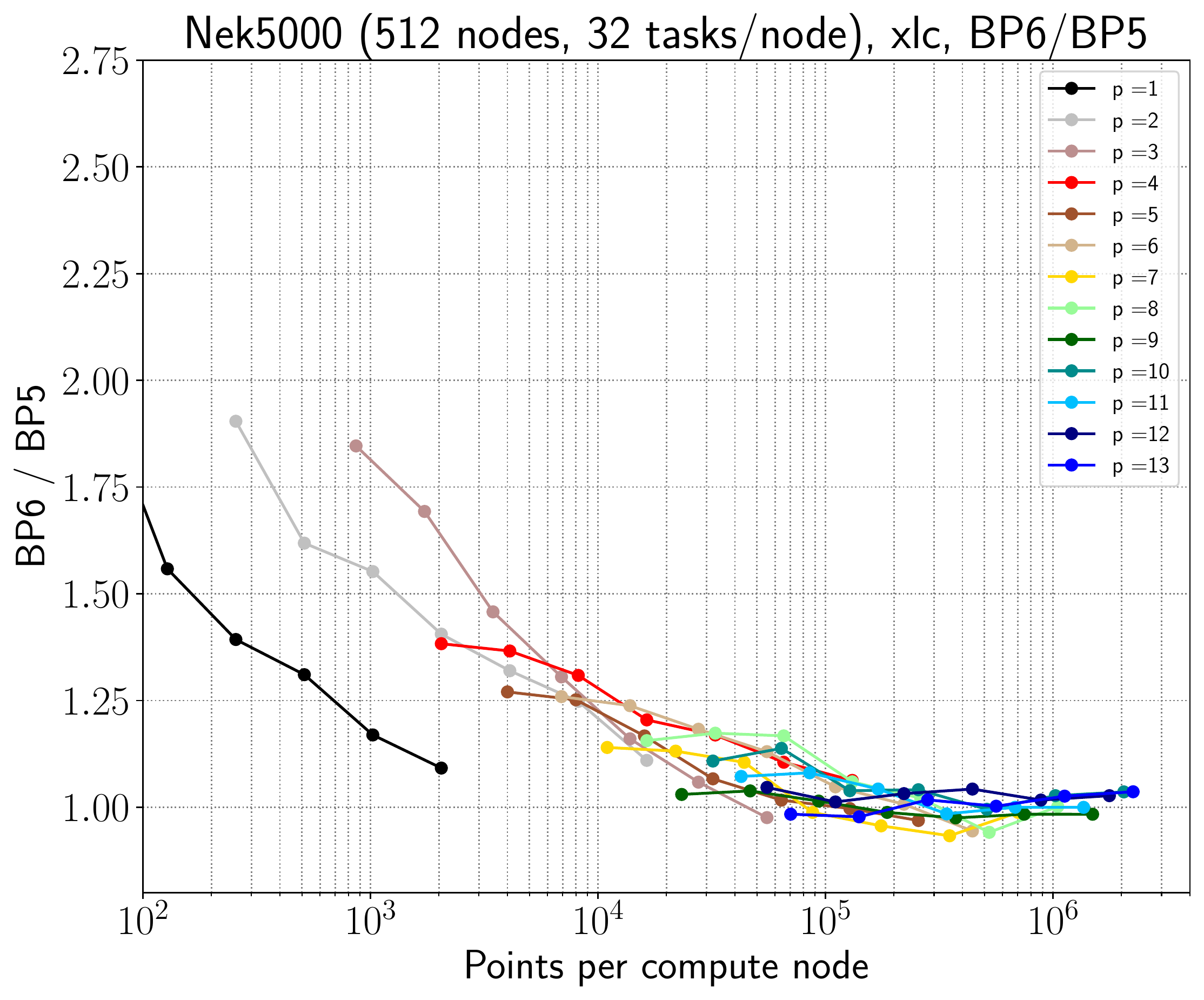}}}
  \caption{\label{fig:bp21_43_65_bgq_xlc}Nek5000 results with xlc compiler on
   the performance ratios of BP2/BP1, BP4/BP3, and BP6/BP5 with varying polynomial order ($p=1,...,16$). } 
\end{figure*}

To better illustrate the potential gain of the vector approach, we
plot in Figure~\ref{fig:bp21_43_65_bgq_xlc} the ratio of Nek5000 results in MDOFs 
for BP2/BP1, BP4/BP3, and BP6/BP5.  The gains for very low
$n/P$ must be discounted because users will generally
not run there, especially for $p=$1, where Nek5000 is not performant.
For large $n/P$, the ratios are generally slightly in excess of unity,
implying that the Nek5000 multicomponent solver is able to effectively
amortize the memory accesses for the geometric components in (\ref{eq:gij}).
These ratios trend upwards for smaller $n/P$ values. Near the strong-scale
limit, which is in the range $n/P \approx 50,000$--100,000 for Nek5000,
we see up to a 1.25-fold increase in MDOFs, which implies that the
vector-oriented problems could potentially run with half as many
processors with the same efficiency as their scalar counterparts.

\section{CPU Performance Analysis on BG/Q} \label{sec:cpu}

The results of the preceding sections clearly demonstrate the importance of
vector intrinsics in realizing high processing rates.  We further remark on
some of the cross-code performance variations.  One of the optimizations used
by deal.II is to exploit the bilateral symmetry of the GL and GLL points in order to cut
the number of operations in the tensor contractions by a factor of 2 by using an
even-odd decomposition (\cite{solomonoff92,Kopriva09}).  The other, as
previously mentioned, organizes the data into 8-element blocks in order to combine
favorable vector sizes that help in realizing improved peak performance.
Moreover, since the objective of CEED is to develop polyalgorithmic back-ends
that will deliver the best performance to the users, our objective is to
realize the optimal performance hull over the various implementations.  All
scaling analysis therefore is based on the best-realized values, not on
the values realized by a particular implementation.

\subsection{Strong-Scaling Analysis}
The broad trends evident in Figures
\ref{fig:bp1_bgq_gcc}--\ref{fig:bp2_4_6_bgq_xlc} also illustrate the importance
of the local workload ($n/P$) in realizing peak performance.  As noted earlier,
for a given problem size $n$, users will generally run on as many processors
as possible until the efficiency drops to a tolerable level.  This $P$-fold
performance multiplier is the most significant factor in realizing large (i.e.,
thousand-fold or million-fold) speedups in HPC, so understanding inhibitors to
increasing $P$ is of paramount importance.  For BG/Q, the strong-scale
inhibitor for linear systems solves of the type considered here is generally
internode latency and not internode bandwidth, as discussed in
\cite{fischer15} and \cite{sc17}.   We return to this observation shortly.

We see in Figure \ref{fig:bp5_bgq_xlc}(a) that, for $p=7$,
Nek5000 realizes 80 MDOFs for $n/P  \approx 175,000$, whereas deal.II achieves
a peak 60 MDOFs for $n/P \approx 700,000$. The net result is that for
approximately the same work rate (DOFs), whether measured in total energy
consumed or node-hours spent, the ability to strong-scale implies that the
Nek5000 implementation will run
fully 4$\times$ faster because, for a fixed $n$, one can use four times the
number of processors.  That is, acceptable performance is realized with
$n/P=175,000$ rather than the much larger $n/P=700,000$.  The point here is not
that one code is superior to another.  In fact, we do not yet have the xlc data
for deal.II and expect that it could be improved substantially.
Furthermore, the results show that deal.II's current mesh partitioning
into 8-element blocks clearly inhibits the strong scaling.
The point is to stress the importance of achieving reasonable performance at a
relatively low value of $n/P$; that is, to be able to {\em strong-scale.}

The importance of strong scaling is that it has a direct impact on time
to solution when solving problems on large HPC systems.  For problems
such as considered here, where the work scales linearly with $n$, the
time per iteration at 80\% efficiency will take the form
\begin{eqnarray} \label{eq:speed}
   t_{0.8} &=& C \frac{\n12}{0.8 S},
\end{eqnarray}
where $C$ is a problem-dependent constant, $0.8 S$ is 80\% of the peak
realizable speed per node (e.g., the peak MDOFs shown in
Figures~(\ref{fig:bp1_bgq_gcc}--\ref{fig:bp2_4_6_bgq_xlc})), and $\n12$ is
value of $n/P$ where performance matches 80\%.  The smaller the value of
$\n12$, the more processors that can be used and the faster the calculation
will run.    
We stress that reduction of time to solution (at the fast, strong-scale limit) 
depends critically on minimization of $n_{0.8}/S$ and not solely on increasing
the speed of the node, $S$.

\subsection{ 80\% Peak vs. Minimum Time}
\begin{figure*}[!htb]
  \subfloat[BP1 $t_{0.8}$]{{\includegraphics[width=0.32\textwidth]{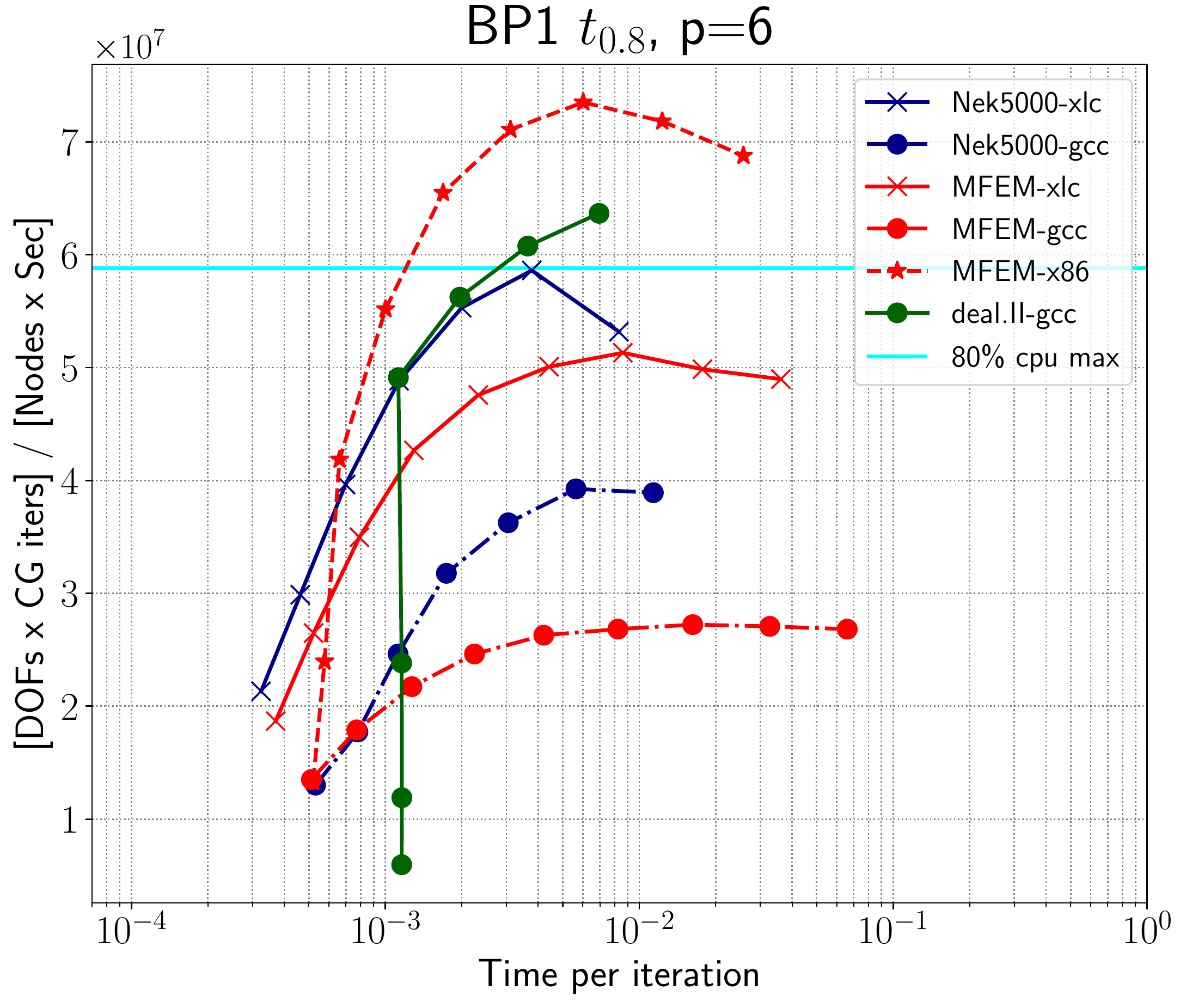}}}
  \hskip.1in
  \subfloat[BP3 $t_{0.8}$]{{\includegraphics[width=0.32\textwidth]{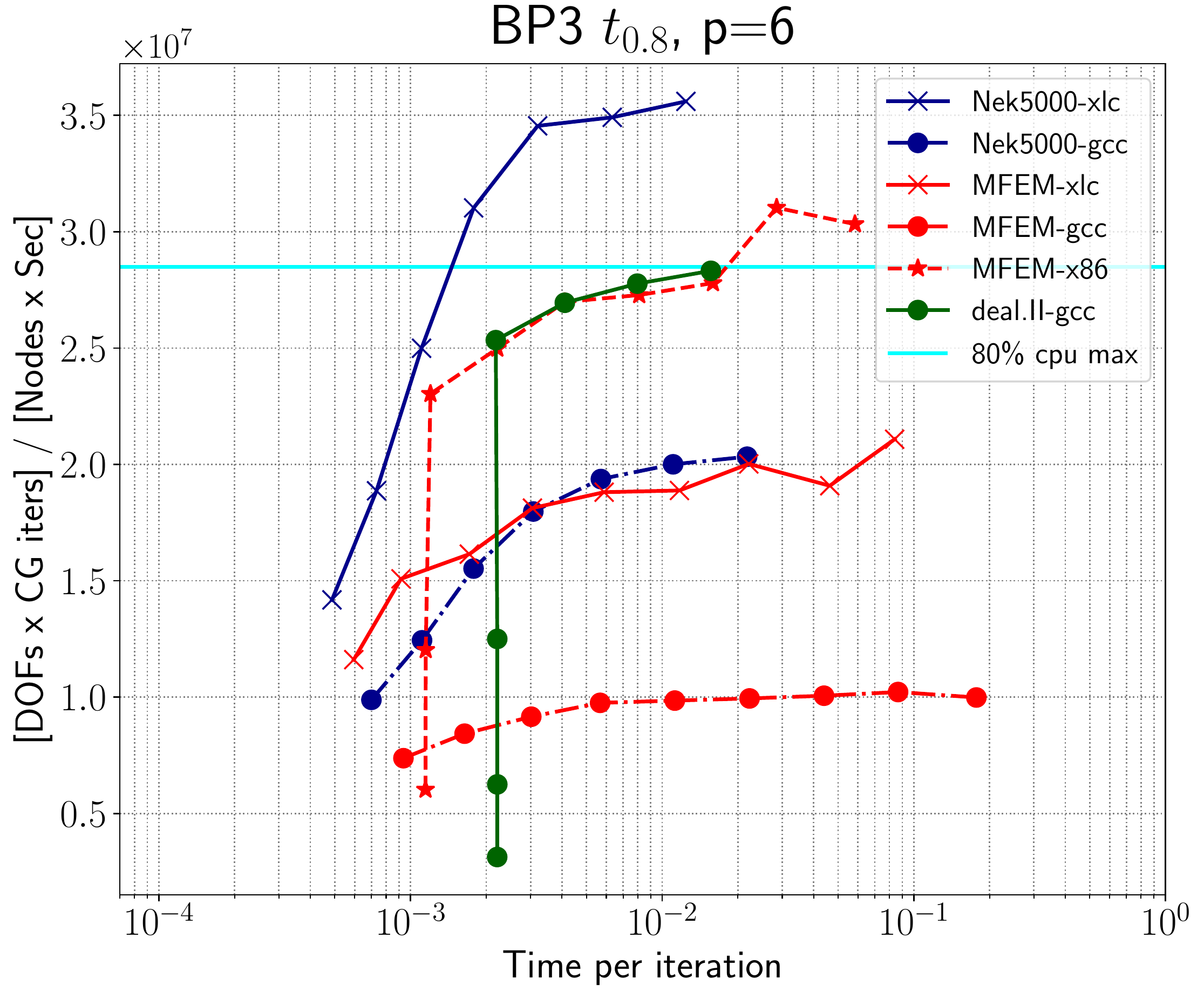}}}
  \hskip.1in
  \subfloat[BP5 $t_{0.8}$]{{\includegraphics[width=0.32\textwidth]{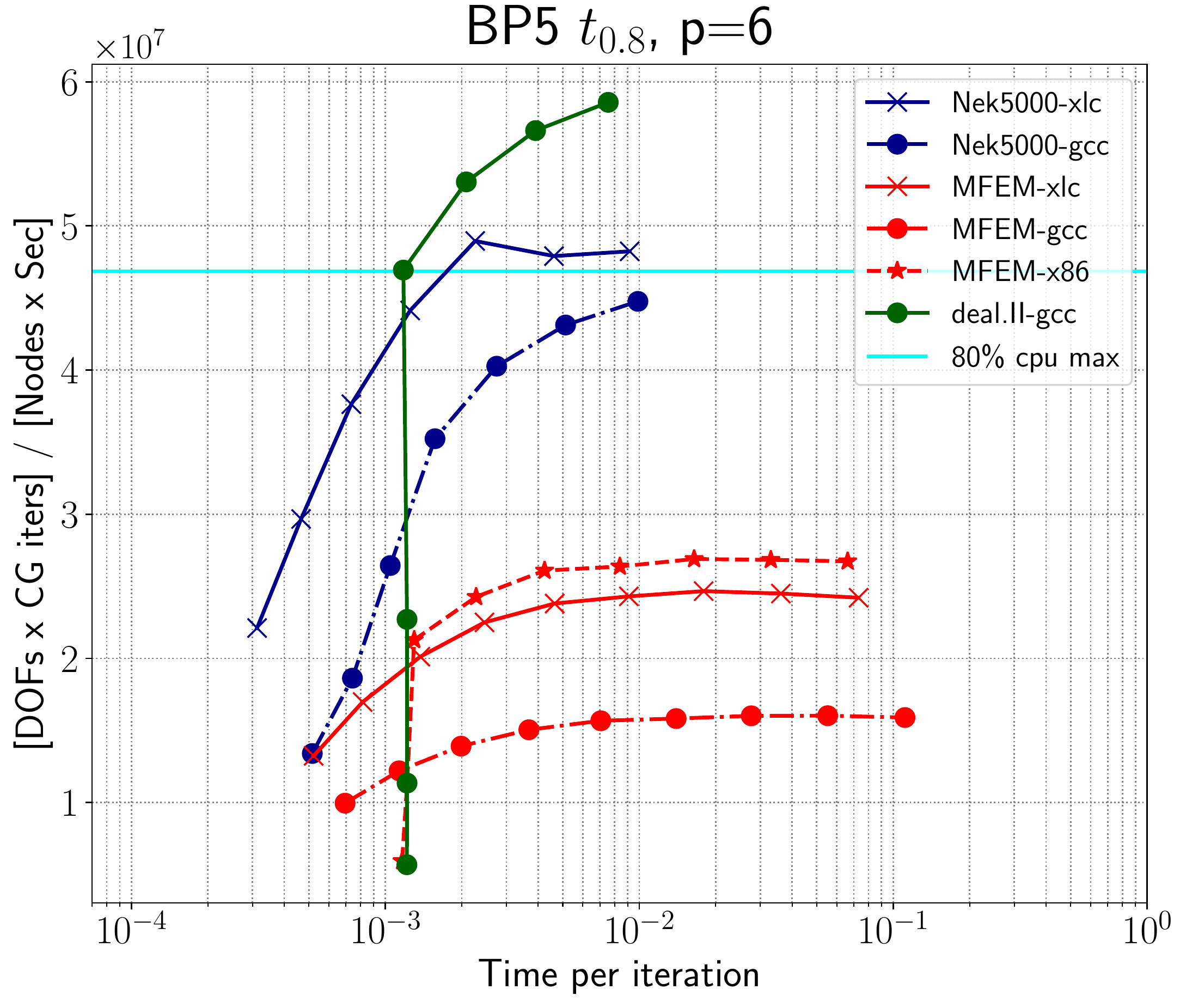}}}
  \caption{\label{fig:t80_p06}DOFs vs time per iteration for BP1, BP3, and BP5 on BG/Q
            with $p=6$ for Nek5000, MFEM, deall.II.}
\end{figure*}

\begin{figure*}[!htb]
 \subfloat[BP1 $t_{0.8}$]{{\includegraphics[width=0.32\textwidth]{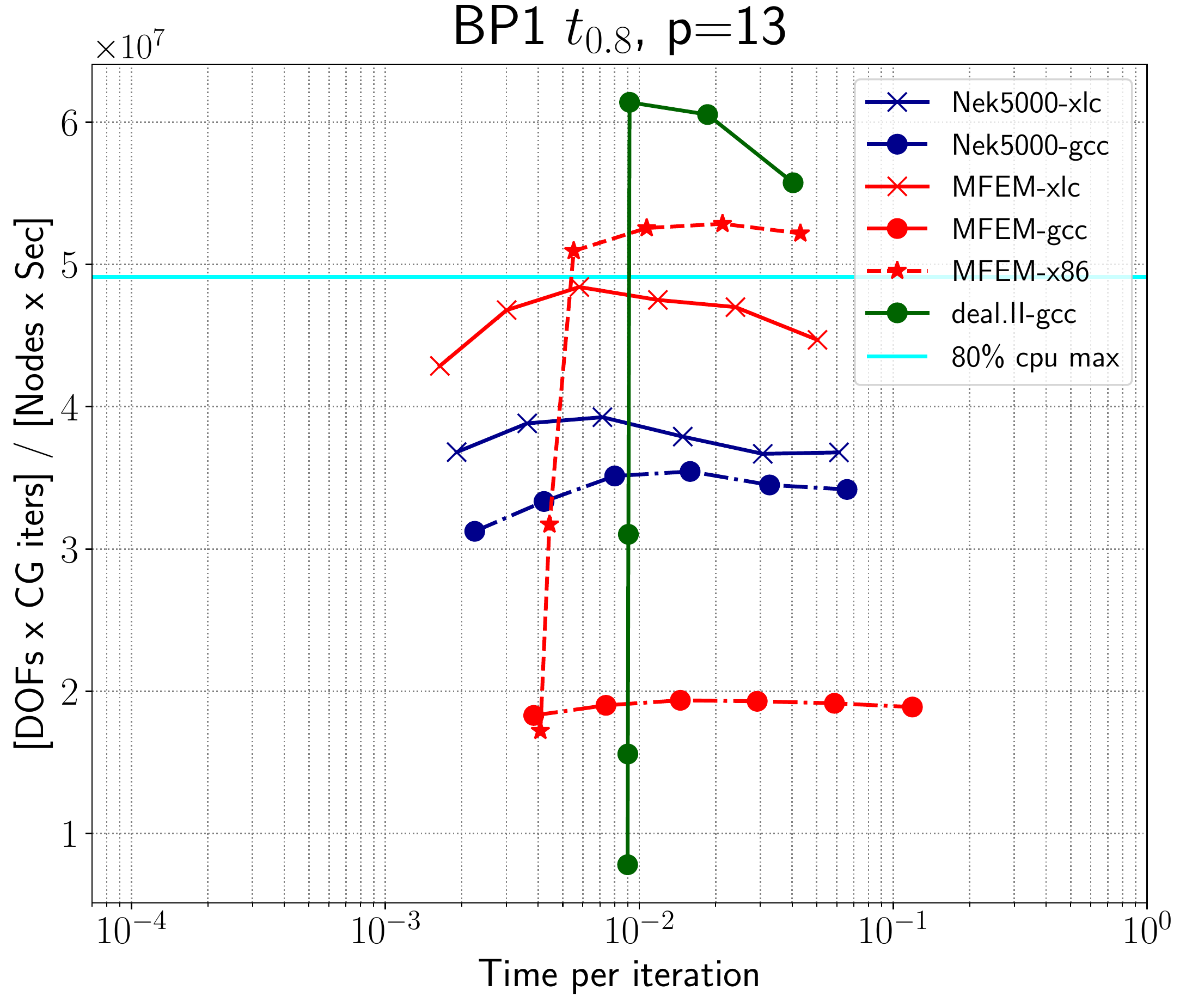}}}
  \hskip.1in
 \subfloat[BP3 $t_{0.8}$]{{\includegraphics[width=0.32\textwidth]{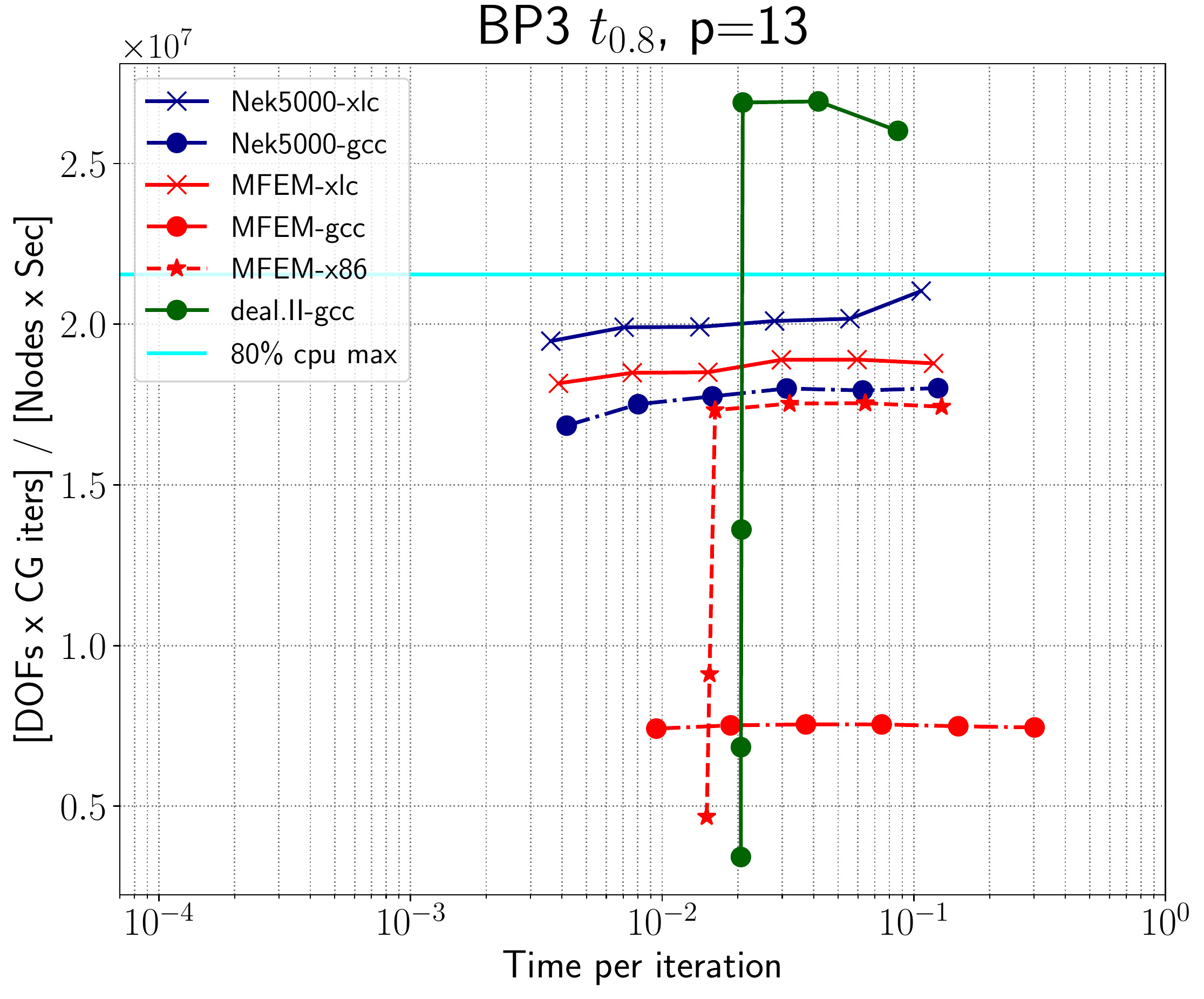}}}
  \hskip.1in
 \subfloat[BP5 $t_{0.8}$]{{\includegraphics[width=0.32\textwidth]{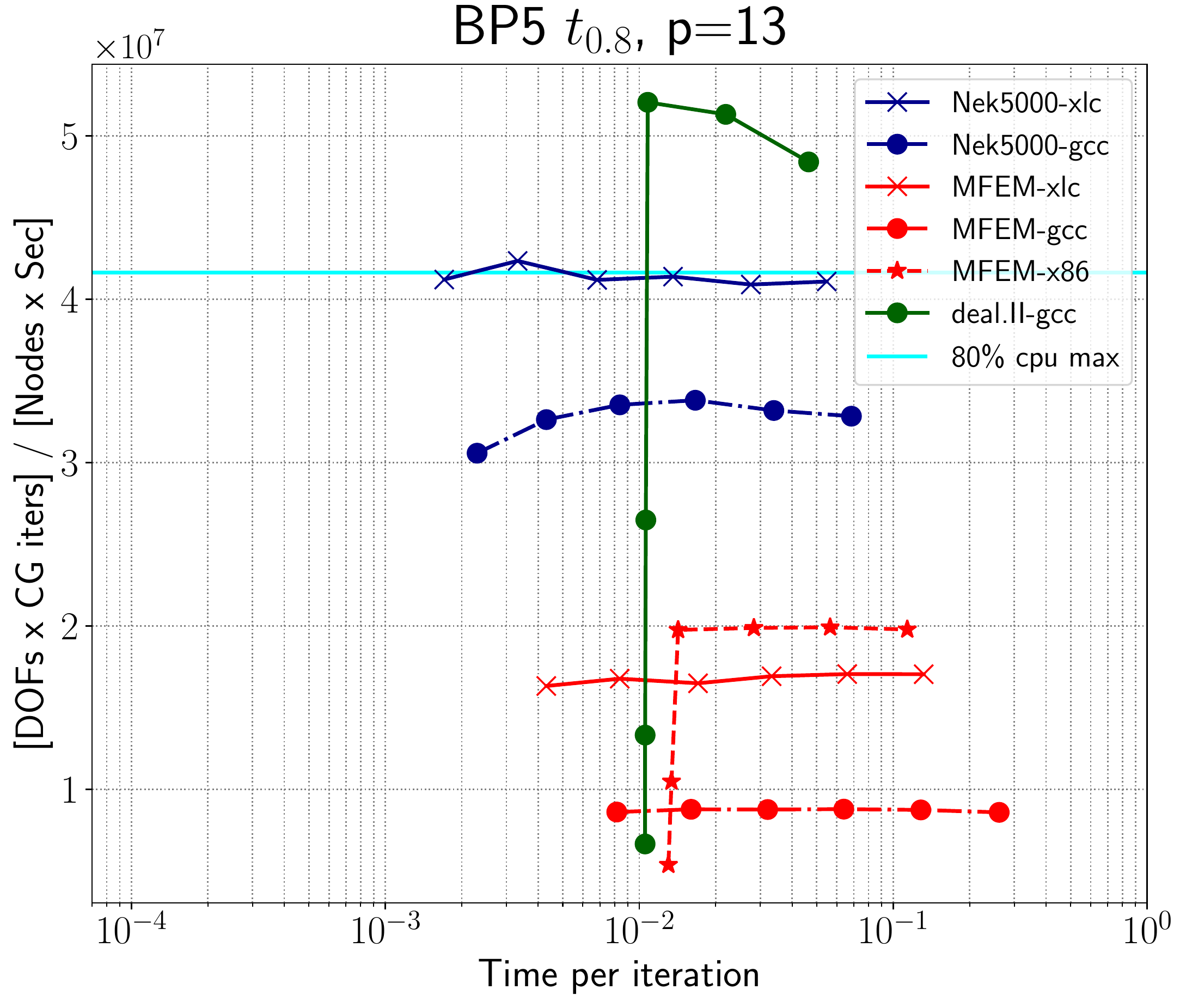}}}
 \caption{\label{fig:t80_p13}DOFs vs time per iteration for BP1, BP3, and BP5 on BG/Q 
            with $p=13$ for Nek5000, MFEM, deall.II.} 
\end{figure*}

Another way of quantifying minimum time per iteration is to plot parametric
curves of DOFs vs. time per iteration,  where the independent parameter is $n/P$.
Figures \ref{fig:t80_p06}--\ref{fig:t80_p13} show a series of such plots for
BP1, BP3, and BP5 at order $p=6$ and 13.  In addition, we plot the lines
corresponding to 80\% of peak, which allow identification of the minimum time
at which 80\% or greater performance is realized.

In the case of $p=6$, the peak for each of the BPs is attained by a different
code:  MFEM for BP1, Nek5000 for BP3, and deal.II for BP5. The minimum time at
$\eta=0.8$ is realized for the same BP/code pairings: 1.2 ms for MFEM--BP1,
1.5 ms for Nek500--BP3, and 1.2 ms for deal.II--BP5.   Each time is
determined as the intercept of the DOFs time curve with the $0.8 \times
$DOFs$_{\footnotesize peak}$ horizontal line in Figure \ref{fig:t80_p06}.  
By contrast, for $p=13$, the {\em peak} performance is realized by deal.II for all 
three BPs, whereas the minimum times at $\eta=0.8$ are realized by different codes.
For BP1, MFEM-xlc/x86 realizes the minimum time of 5.5 ms; this is the
left-most point in Figure \ref{fig:t80_p13} that is above the 80\% line.  For
BP3, deal.II has the minimum time-per iteration at 20 ms; none of the other
codes are above the 80\% mark.  For BP5, Nek5000 realizes a minimum time of 2.2
ms at $\eta=0.8$.

For both the $p=6$ and 13 cases, we see that the minimum-time graphs for the
MFEM-xlc/x86 and deal.II cases exhibit vertical asymptotes as the times are
reduced.  That is, they cannot deliver smaller times once $n/P$ reaches a
critical value.  This behavior results from their approach to performance,
namely, vectorizing over multiple elements within a single MPI rank.  For
Nek5000, the minimum times are governed either by communication overhead,
as in Figure \ref{fig:t80_p06} for $p=6$, or by granularity, as in Figure
\ref{fig:t80_p13} for $p=13$, where the DOFs vs. time curve is essentially
flat.  For larger values of $p$ (e.g., $p>13$), the Nek5000 work rates are nearly
$p$-independent because there is sufficient work to dominate communication
overhead, even at the minimum $n/P$ values where there is only one element
per MPI rank.

\begin{figure*}[!htb]
 \centering
 \subfloat[BP1 \& BP2 $t_{0.8}$]{{\includegraphics[width=0.32\textwidth]{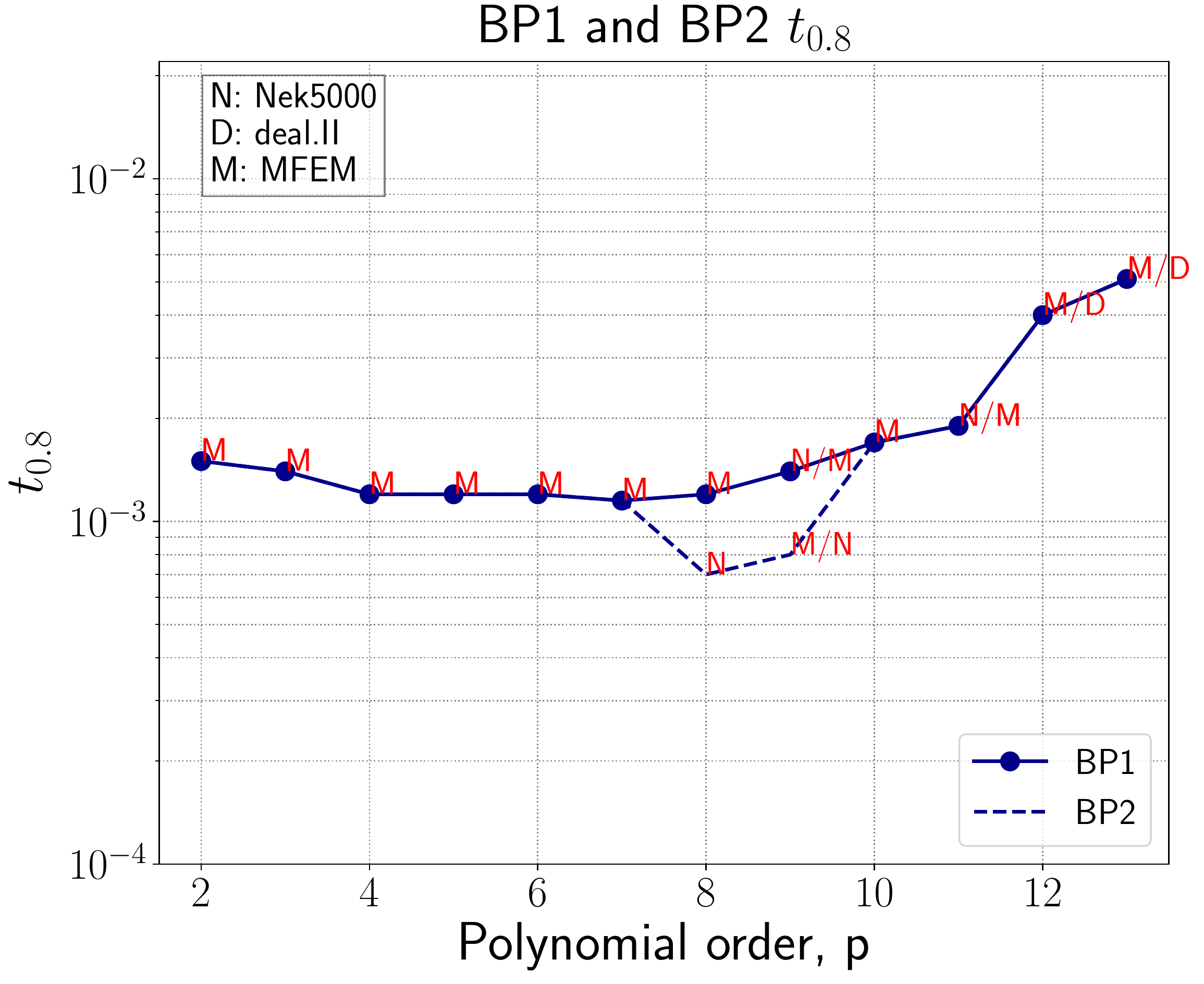}}}
 \hskip.1in
 \subfloat[BP3 \& BP5 $t_{0.8}$]{{\includegraphics[width=0.32\textwidth]{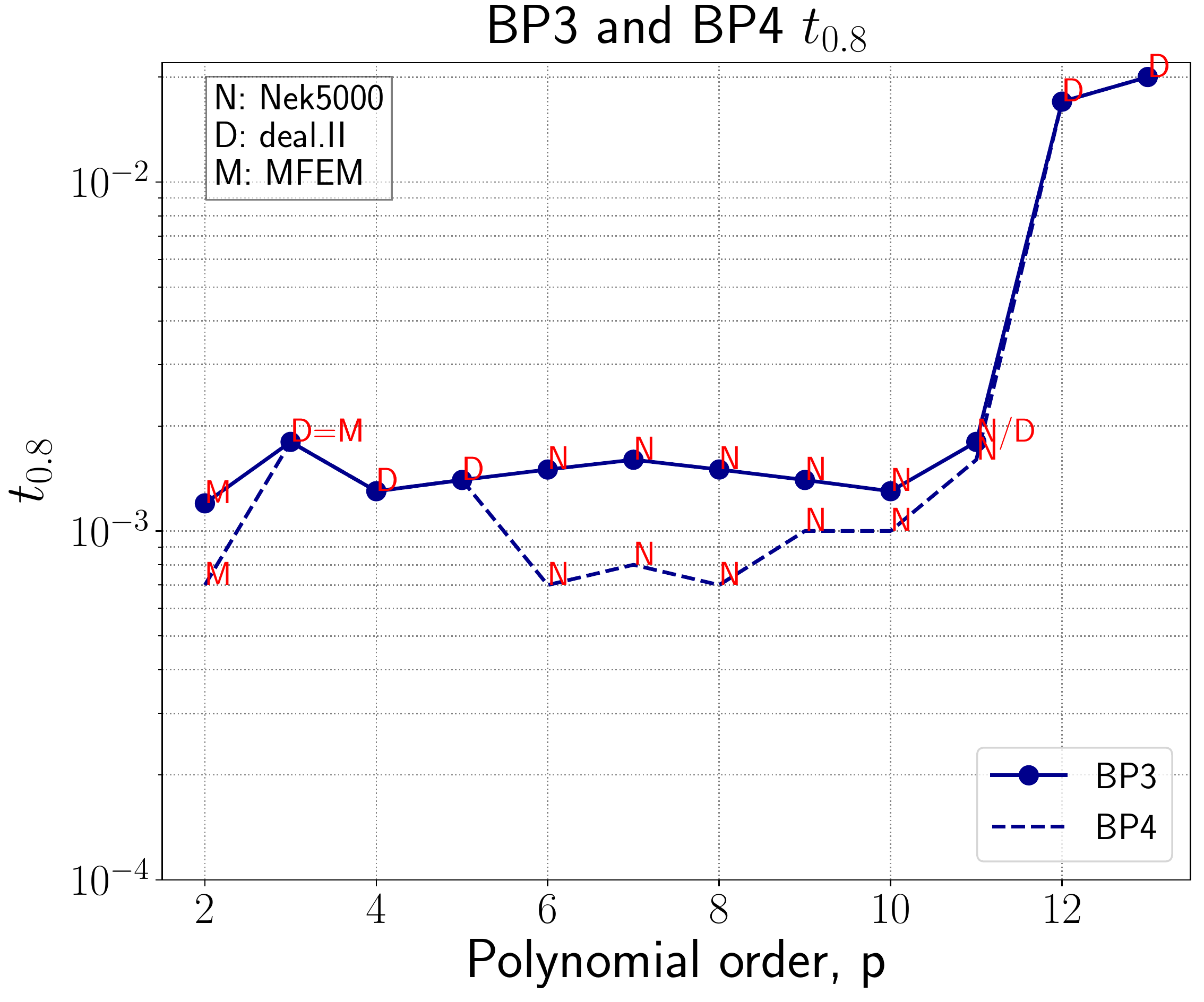}}}
 \hskip.1in
 \subfloat[BP5 \& BP6 $t_{0.8}$]{{\includegraphics[width=0.32\textwidth]{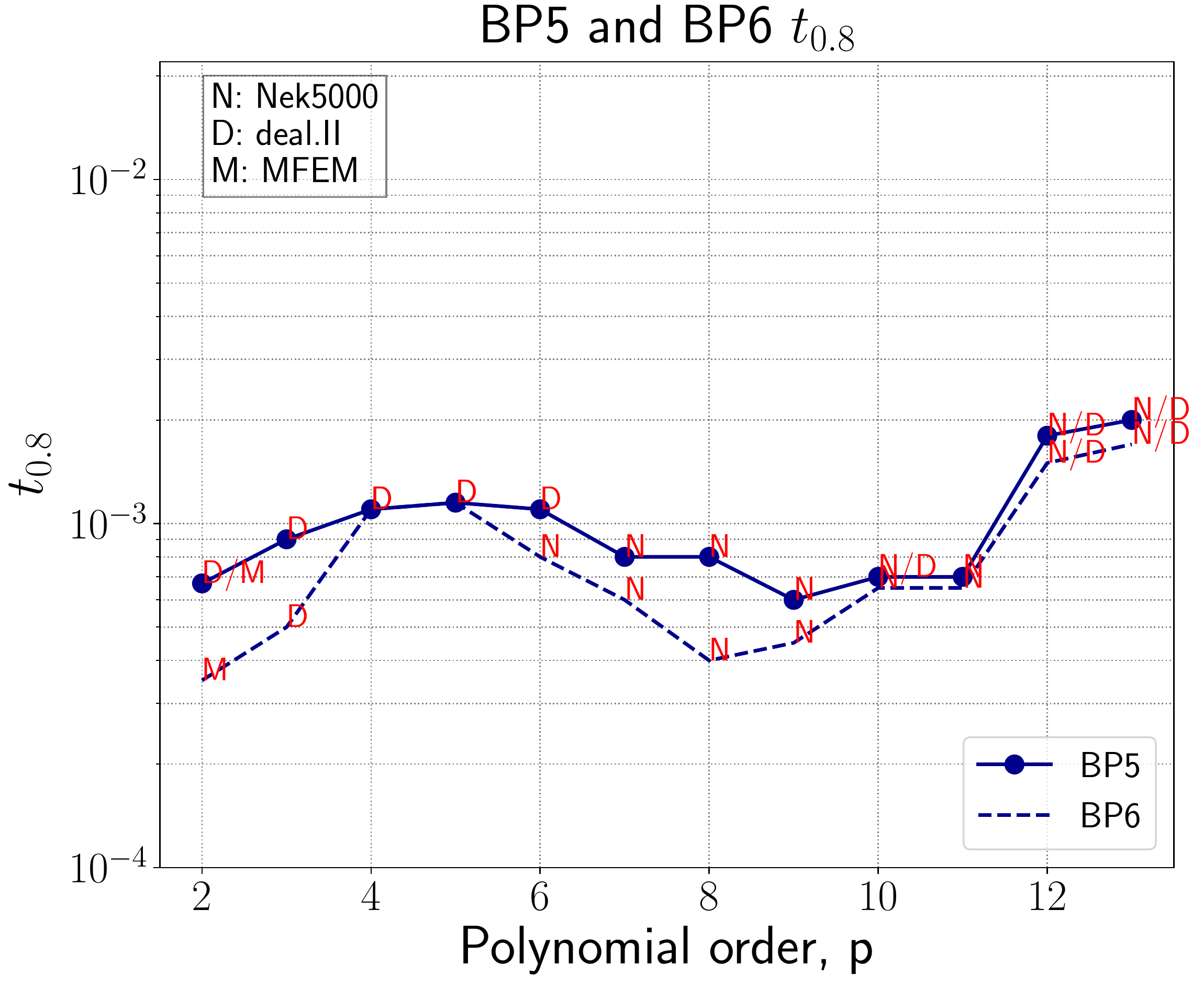}}}
  \caption{\label{fig:t80} Minimum time per iteration at 80\% of peak, $t_{0.8}$, 
   for BP1, BP3, and BP5 (solid) and BP2, BP4, and BP6 (dashed) vs polynomial order $p$. 
   A/B indicates the associated codes
   where code A realizes the minimum time and code B realizes the peak performance.}
\end{figure*}

\begin{figure*}[!htb]
 \centering
  \subfloat[BP1 \& BP2 $n_{0.8}$]{{\includegraphics[width=0.32\textwidth]{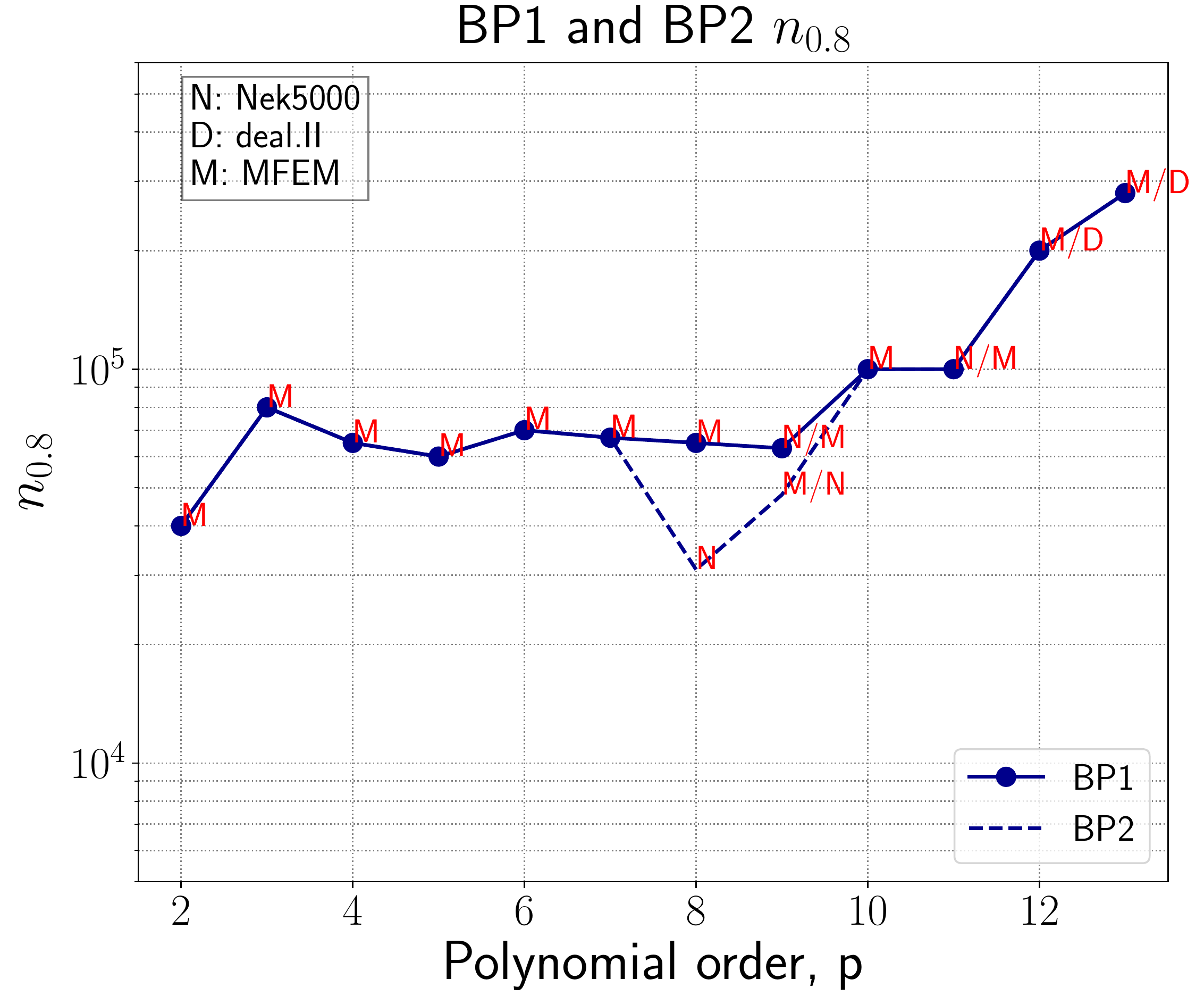}}}
  \hskip.1in
  \subfloat[BP3 \& BP4 $n_{0.8}$]{{\includegraphics[width=0.32\textwidth]{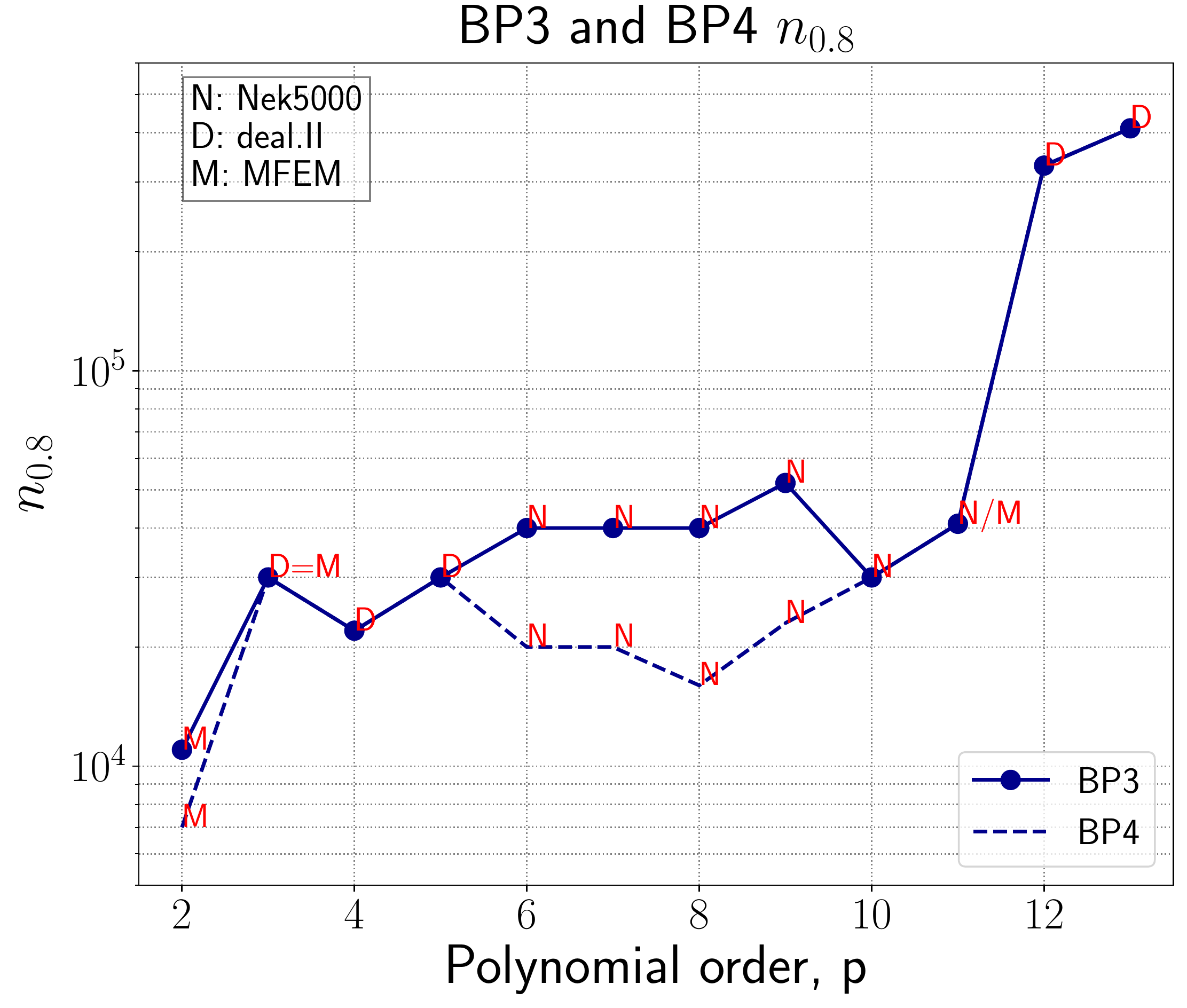}}}
  \hskip.1in
  \subfloat[BP5 \& BP6 $n_{0.8}$]{{\includegraphics[width=0.32\textwidth]{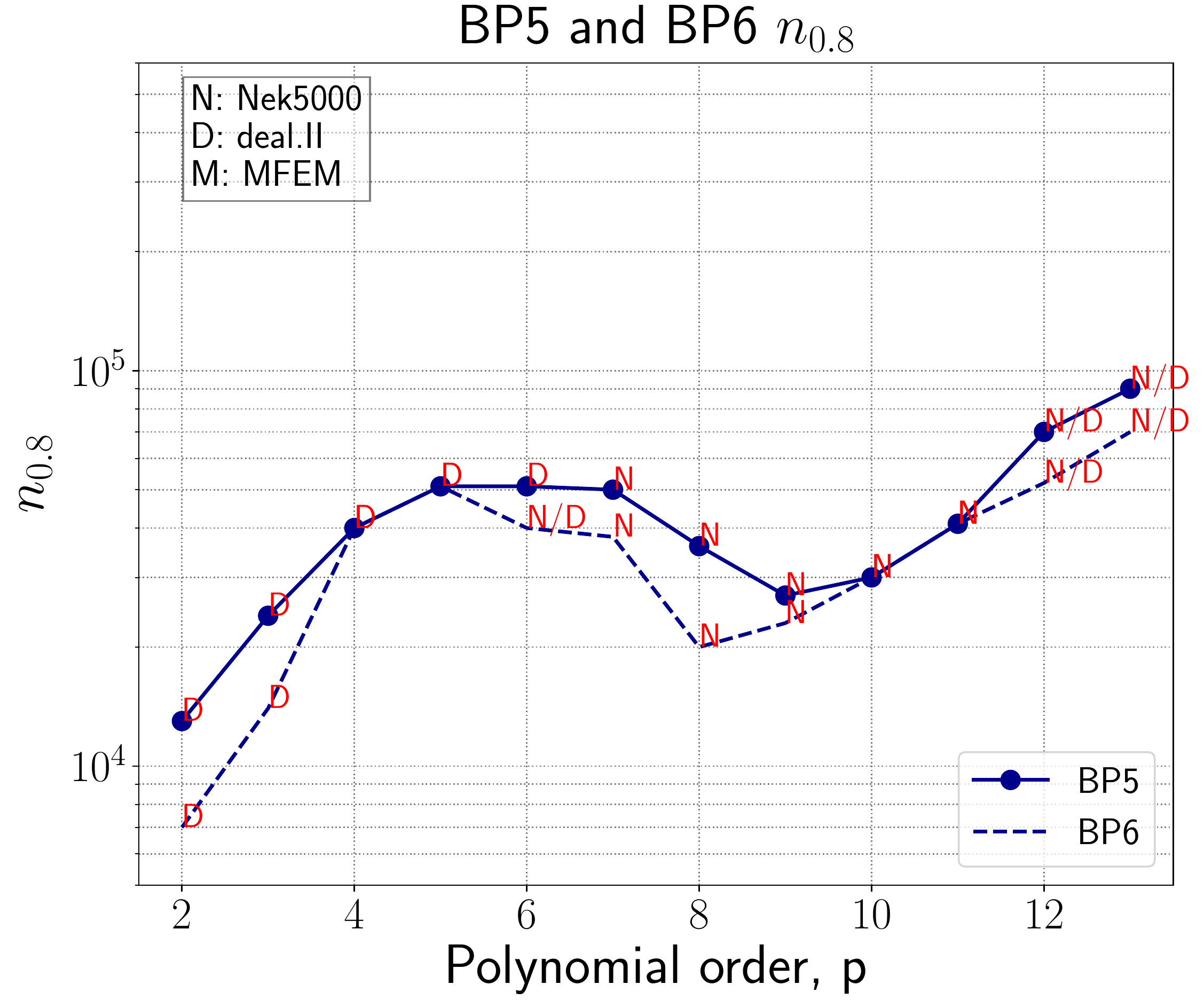}}}
  \caption{\label{fig:n80}Points per node, $n_{0.8}$,
          corresponding to the $t_{0.8}$ values of Figure~\ref{fig:t80} with the associated 
          the codes.}
\end{figure*}

\begin{figure*}[!htb]
 \centering
  \subfloat[BP1 $r_{0.8}$]{{\includegraphics[width=0.32\textwidth]{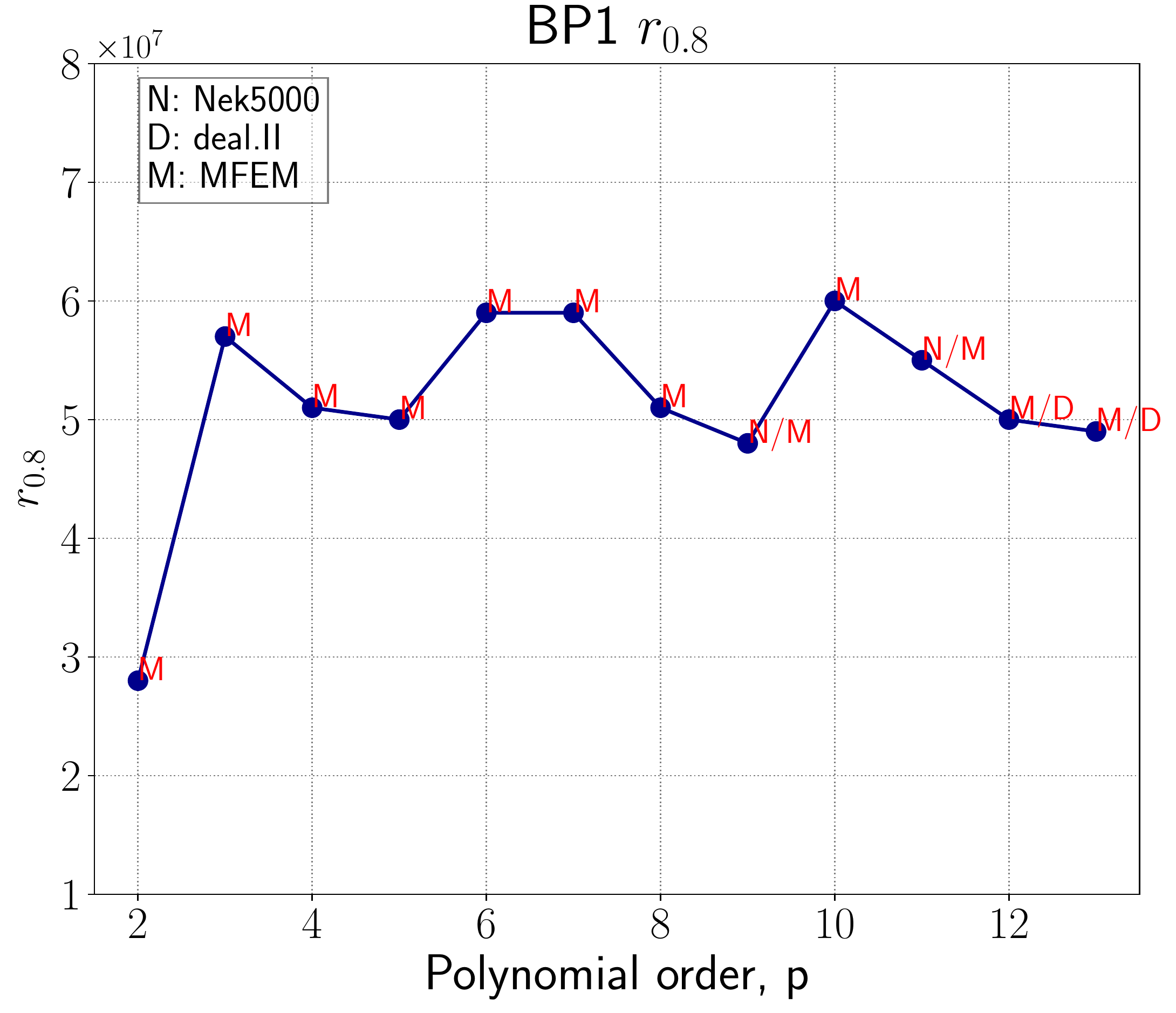}}}
  \hskip.1in
  \subfloat[BP3 $r_{0.8}$]{{\includegraphics[width=0.32\textwidth]{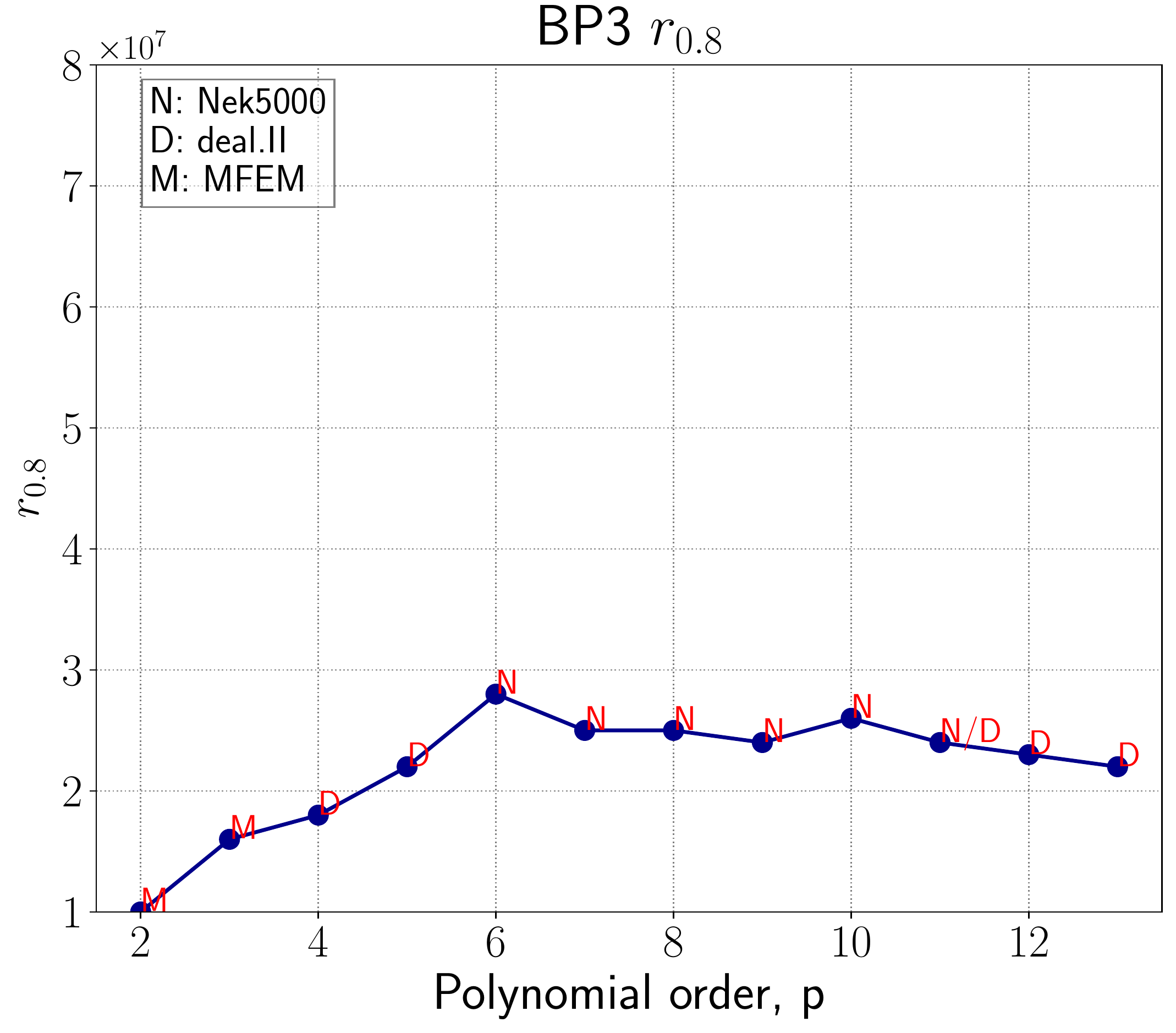}}}
  \hskip.1in
  \subfloat[BP5 $r_{0.8}$]{{\includegraphics[width=0.32\textwidth]{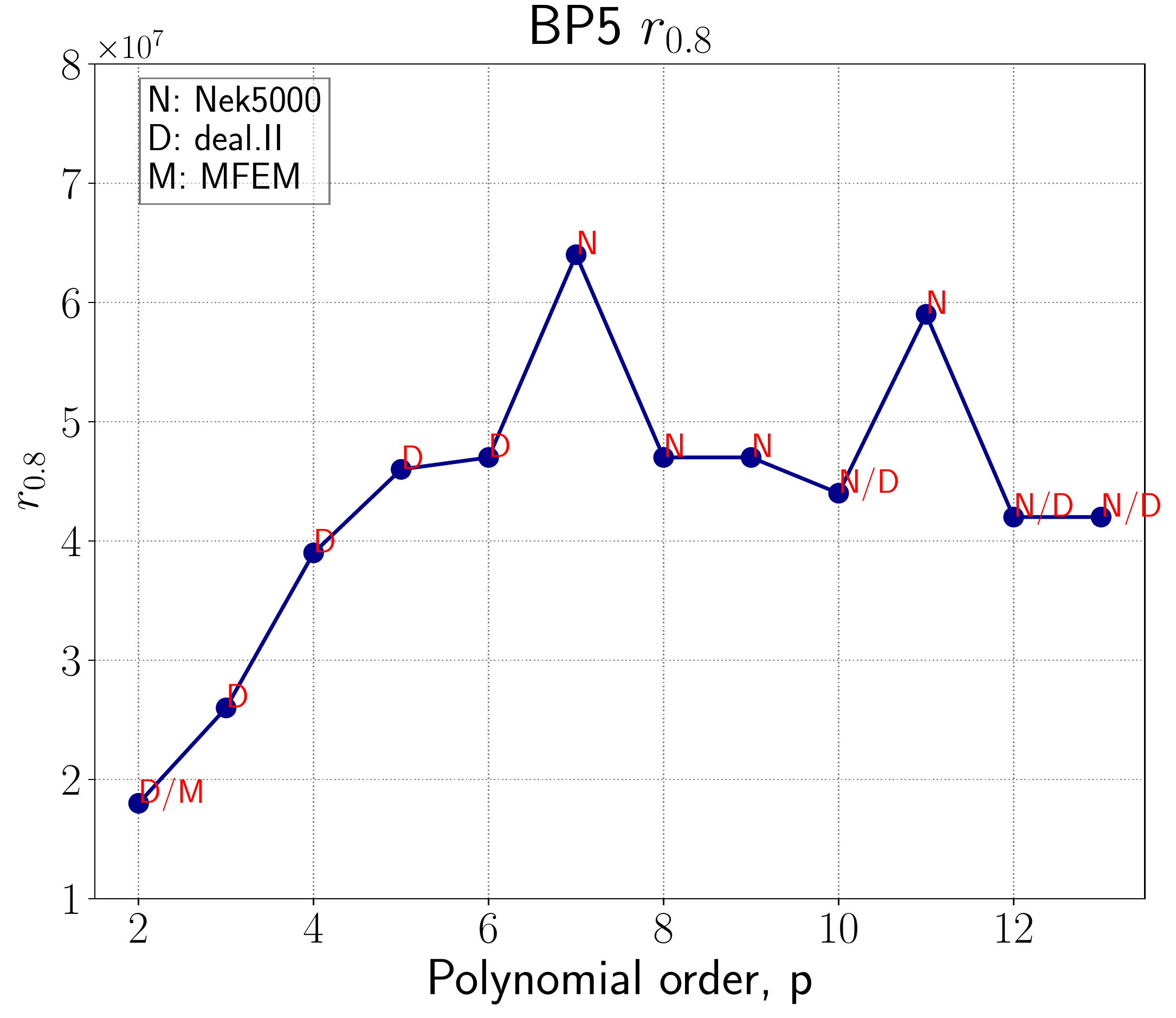}}}
  \caption{\label{fig:r80} Work rate $r_{0.8}=$ 
      [DOFs x CG Iters.] / [Nodes x Sec.]
      corresponding to the values in Figures~\ref{fig:t80}--\ref{fig:n80} with the
      associated codes.  (The even BP results are the same as the odd.)}
\end{figure*}

We have extended the preceding analysis to all polynomial orders $p$ and summarize
the results in Figures \ref{fig:t80}--\ref{fig:r80}.  Figure \ref{fig:t80}
shows the minimum time per iteration that sustains (at least) 80\% of peak.
The symbol indicates which code realizes this minimum time for the given $p$
(D=deal.II, M=MFEM, N=Nek5000).  Where two symbols appear at a given data point,
the first indicates the code realizing the minimum time while the second
indicates the code that established the peak rate at the given $p$.  Figure
\ref{fig:n80} shows the number of points per node, $n_{0.8}$, associated with
each minimum-time point from Figure \ref{fig:t80}.  Figure \ref{fig:r80}
indicates the corresponding work rate (DOFs) for these points. This value will
generally be $r_{0.8}=0.8\times$DOFs$_{\footnotesize peak}$ unless the performance at
the minimum time point exceeds that value.  The graphs show results for the
odd (scalar) cases as solid lines and for the even (vector) cases as dashed
lines.

We begin with a discussion of the scalar cases.
For polynomial orders $p=1$--11, the minimum time per iteration is in the range
0.6 to 2 ms.  The implication of Figure \ref{fig:t80} is that, for BG/Q, {\em
no simulation will be able to execute PCG iterations at rates faster than 0.5
ms per iteration for BP5 or 1 ms for BP1 and BP3 unless that simulation
runs at $\eta < 0.8$.}  Adding more processors does not help reduce the
time per iteration.  These are the strong-scale bounds, and they are essentially
$P$ independent, following the logic that initiates with Figures
\ref{fig:strong1}--\ref{fig:strong2}.

The range and near-uniformity of these minimal iteration times 
are not surprising.  In the strong-scale limit, communication is on par
with local work.  Moreover, the size of the local messages is relatively small,
$m \leq (p+1)^2 \ll m_2$. Here,  $m_2$ is the size of
a message (in 64-bit words) that costs 2$\alpha$, where $\alpha$ is the
latency, or 1/2-round-trip message-passing time for a single 64-bit word.
For BG/Q, $\alpha = 3.8$ $\mu$s and $m_2 =845$, as noted by \cite{fischer15}.
At the finest granularity of a single element per processor, there are
26 messages to be exchanged at all polynomial orders, each having a cost
between $\alpha$ and 2$\alpha$.  On BG/Q, vector reductions are performed in
hardware with all-reduce times of approximately $4 \alpha$ for all $P$
(\cite{fischer15}).  Given the two vector reductions required for PCG, we
conclude that message-passing costs for a single element per rank on BG/Q lie
between
$(26+8)\alpha \approx 0.13$ ms and
$(52+8)\alpha \approx 0.23$ ms, which is within the range of
20\% of of 0.6 ms (the minimum time observed) to 2.0 ms.   Larger overheads
for the larger times (e.g., 20\% of 2.0 ms = 0.4 ms) are likely due to the
fact that there are typically more than 26 messages when there is more than one
element per rank.\footnote{One of the attractions of high-order
{\em discontinuous} Galerkin methods is that they need only 6 (possibly 12)
data exchanges per element, resulting in considerably lower message-latency
overhead in the strong-scale limit.}

We see in Figure \ref{fig:t80} that beyond $p=11$, there is a rise in
minimum iteration times with increasing $p$, with the the most rapid rise
occurring for BP3, which is the most work-intensive.  According to
(\ref{eq:work2})--(\ref{eq:work3}), the leading-order complexity for BK3 is
$\approx$ 24--40$p^4E$, versus $\approx 12p^4E$ for BK1 and BK5.  Most of the
rise in time per iteration, however, is observed for cases where deal.II
performs extremely well and the other codes do not deliver 80\% of peak.
Consequently, the times rise significantly (by a factor between 4 and 8) because
the peak for deal.II is realized only for a minimum of 8 elements per MPI rank.

One of the more significant observations from Figure \ref{fig:t80} is that
elliptic problems can be solved with polynomial orders up to $p \approx 11$ in
the same (or less) time per iteration as their low-order counterparts.
(Effective preconditioning is a different topic, studied elsewhere, e.g.,
\cite{lottes05,pedro19}, and is slated for a future bake-off study.)   In fact,
the minimum times for BP5 are realized for $p=9$--11.  To understand why, we
plot in Figure~\ref{fig:n80} the number of points per node that corresponds to
the time points of Figure~\ref{fig:t80}.  Here we can see that the decrease in
minimum time for BP5 for $p=7$--9 corresponds to a decrease in the number of
points where the minimum time was realized.  Fewer points are required to meet
the 80\% criterion with $p=9$ than with $p=7$, for example.  To further test
this hypothesis, we plot the corresponding DOFs rates in Figure \ref{fig:r80},
which is effectively the ratio of the data in Figures \ref{fig:n80} and
\ref{fig:t80}.  Across the BPs, there is a general upward trend in DOFs as $p$
increases from 2 to 6. In addition, there are distinct peaks where the mod 4
intrinsics apply for Nek5000 (e.g., $p=6$ and 10 for BP3, which correspond to
$q=8$ and 12, and $p=7$ and 11 for BP5, for which $q$=8 and 12).  The minima in
Figure \ref{fig:t80} for BP3 and BP5 correspond primarily to locations where
smaller problems are running faster, rather than where the processing rates are
higher.  Overall, we note that the minimum times in the range $p=2$ to 11
correspond to $n_{0.8}\approx$ 30,000--70,000, or 1000--2000 points per rank, in
accordance with the 80\% efficiency-point observed in Figure \ref{fig:strong2}(a).

The vector results in Figure \ref{fig:t80} demonstrate that up to a twofold
improvement in minimum time per iteration is realized for several cases.
As seen in Figure \ref{fig:n80}, this reduction comes through a reduction
in the minimum problem size that is capable of sustaining 80\% of peak
through amortized messaging overhead.  While some performance gain would
be expected through reuse of the geometric factors $G_{mn}$
in (\ref{eq:bk5x}), that is not the leading source of performance improvement.
If it were, it would be manifest in the performance-saturated limit, which is
not the case.   Note that we elected in Figures \ref{fig:t80}--\ref{fig:n80} to
base the 80\% performance threshold on the {\em scalar} peak.  The rationale
for doing so is that we view the switch to vector solvers as an improvement on
a baseline scalar solver and consequently measure performance gains against
this scalar baseline.

 \section{Bake-Off Performance on Summit}
\label{sec:gpu}

\begin{figure*}[!htb]
 \subfloat[BK5 Tuning]{{\includegraphics[width=0.42\textwidth]{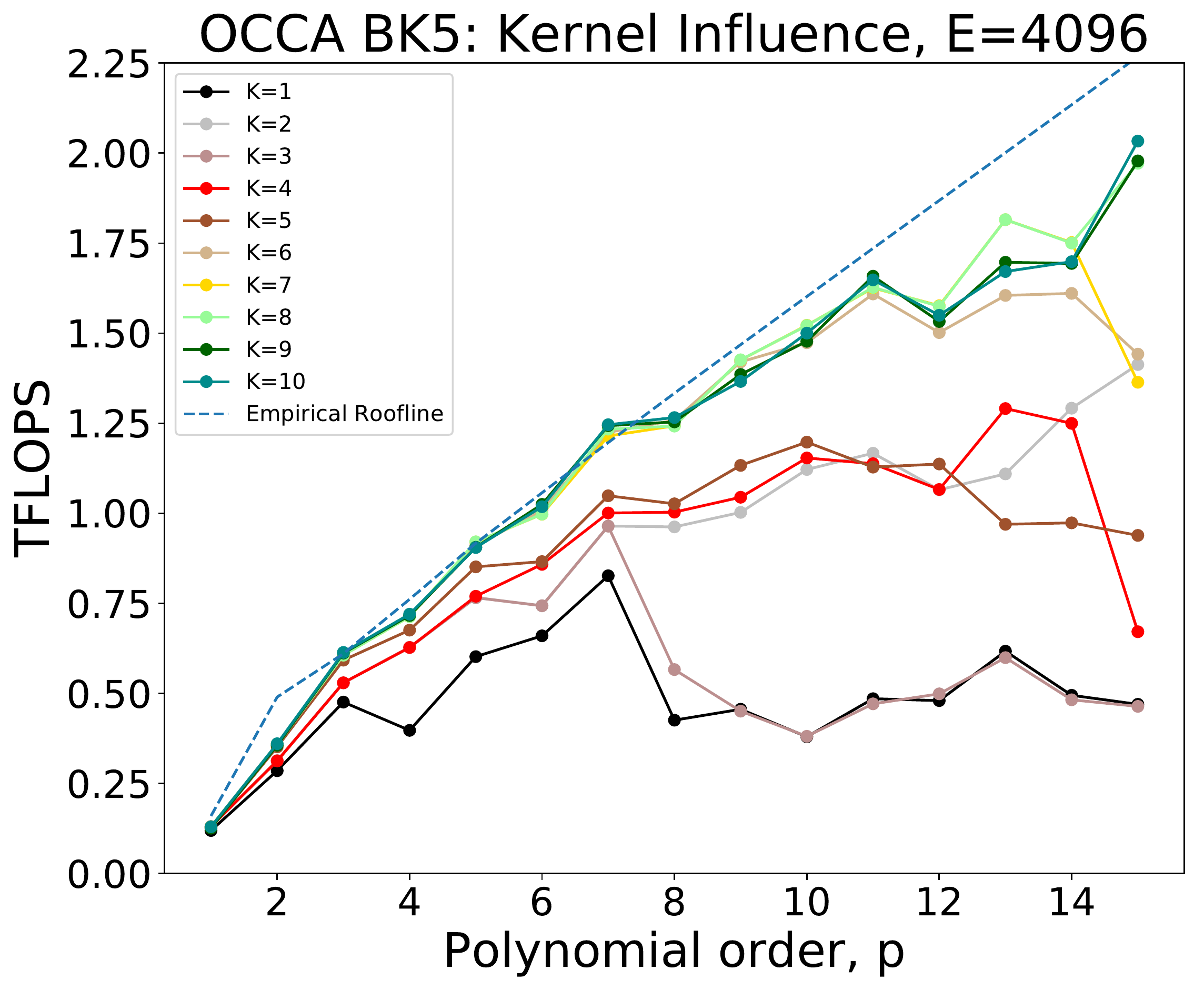}}}
 \hskip.2in
 \subfloat[BK5 Tflops]{{\includegraphics[width=0.42\textwidth]{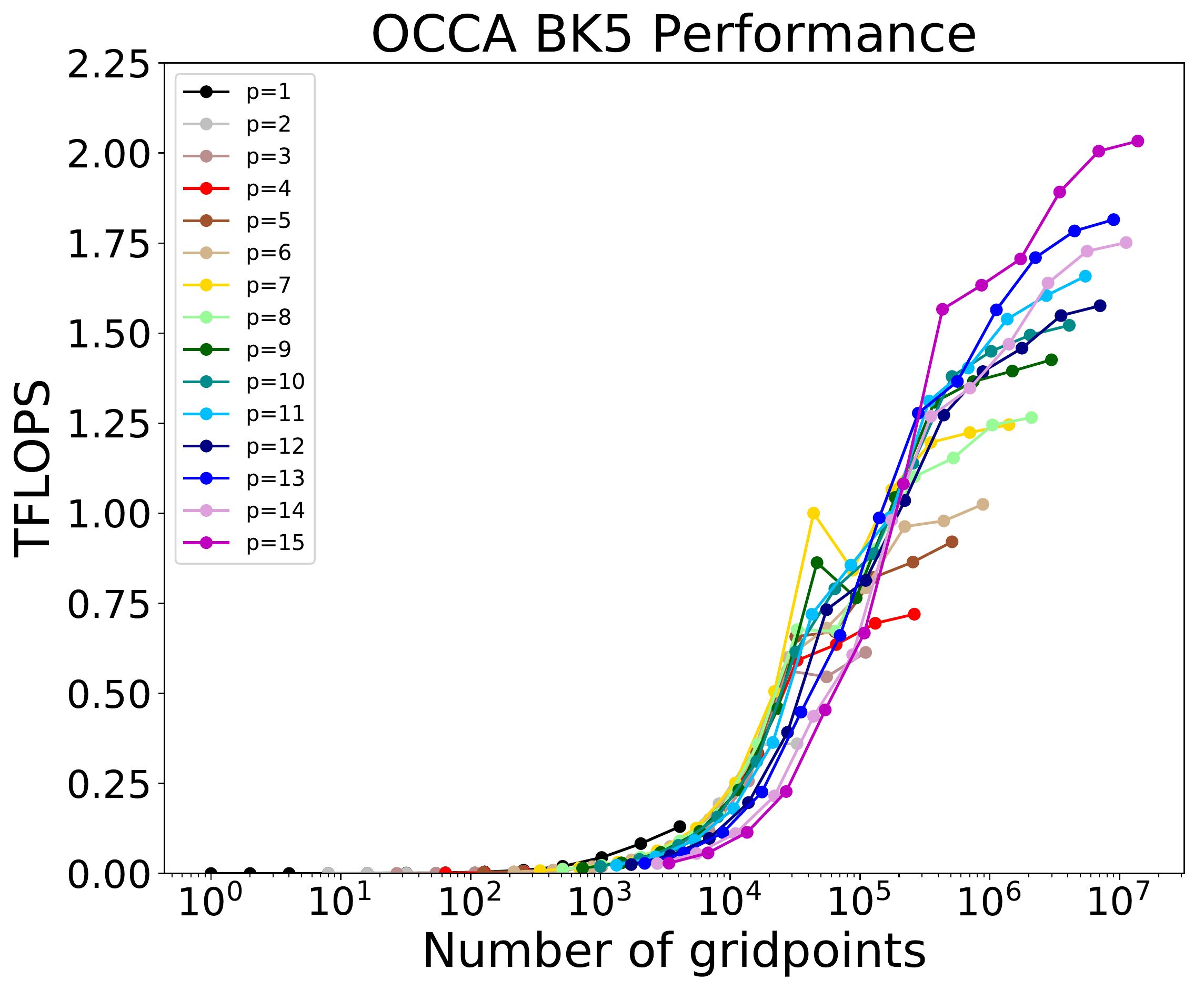}}}
 \\
 \subfloat[BK5 Timings]{{\includegraphics[width=0.42\textwidth]{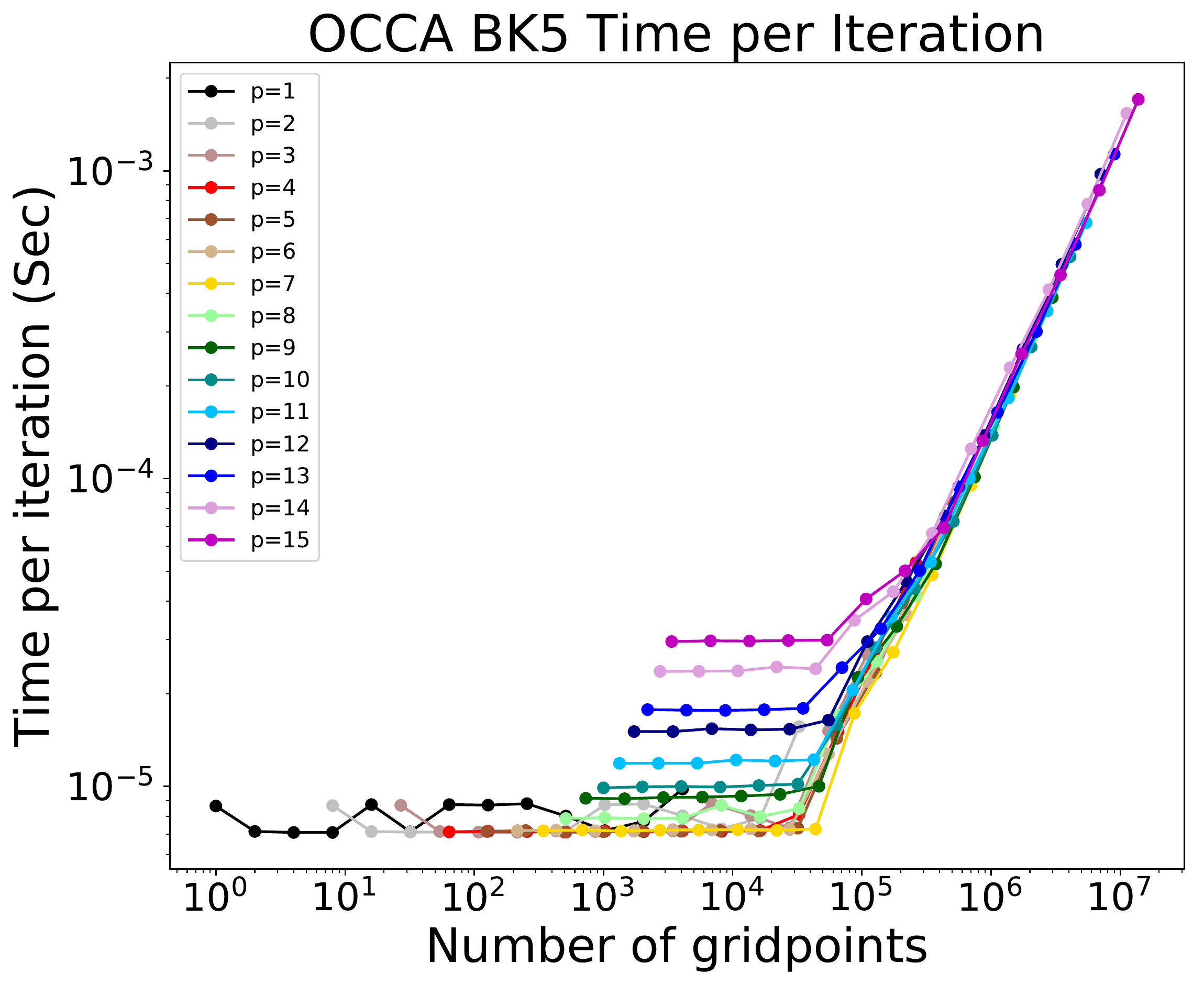}}}
 \hskip.2in
 \subfloat[BK5 Timings per point]{{\includegraphics[width=0.42\textwidth]{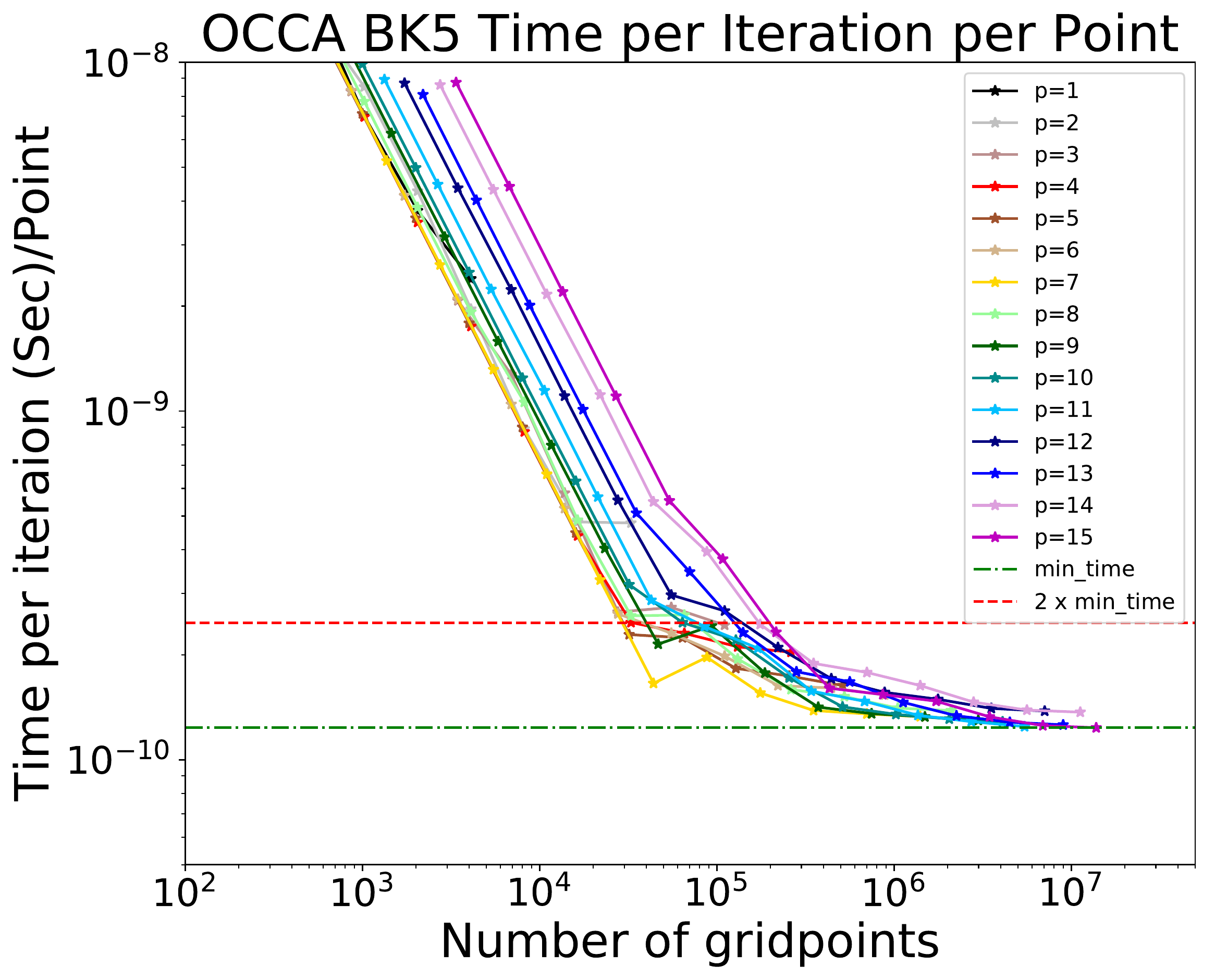}}}
 \caption{\label{fig:bk5_perform} Single GPU performance:
         (a) TFLOPS for different kernel tunings.
         (b) TFLOPS versus problem size $n$ for different polynomial orders, $p$.
         (c) Execution time versus $n$ for varying $p$.
         (d) Execution time per point versus number of points, $n$. }
\end{figure*}

With multinode scaling issues addressed through the BP studies, we turn now to
node scaling on next-generation accelerator-based architectures.  Specifically,
we consider the performance of BK5 and BP5 implemented on the Nvidia V100 core on
the OLCF Summit using libParanumal, which is an SEM-based high-performance high-order library
written in the OCCA kernel language (\cite{occa}).
The Summit system configuration can be found in Table~\ref{tab:sys}.


\subsection{BK5 on Summit}

BK5 amounts to evaluation of the matrix-vector product
$\uw_L = A_L \uu_L$, where $A_L$=block-diag($A^e$), $e=1,\dots,E$.  This kernel
is fully parallel because $A_L$ represents the {\em unassembled} stiffness
matrix using the formulation in (\ref{eq:bk5x}).
For performance tuning on the V100, we initially consider only the kernel
(BK5), not the full miniapp (BP5), which means we are ignoring communication
and device-to-host transfer costs.  Our intent is to explore the potential
and the limits of GPU-based implementations of the SEM.  We seek to determine
what is required to get a significant fraction of the V100 peak performance in
the context of distributed-memory parallel computing architectures such as
OLCF's Summit.

Figure~\ref{fig:bk5_perform}(a) shows the performance for (\ref{eq:bk5x}) on a
single GPU core of Summit through a sequence of OCCA tuning steps, referred to
as kernels $K=1,...,10$, for BK5.  The number of elements is $E=4096$, and the
polynomial order $p$ varies from 1 to 15 ($q=2$ to 16). Each tuned version $K$
is run multiple times, and the time for the kernel is taken by dividing total
time by the total number of iterations.  This procedure is done to smooth out
the noise and to be sure that we are not misguided by the clock resolution
limitations.

  In the OCCA implementation, each spectral element is assigned to a separate
thread block on the GPU with a 2D thread structure.  Threads within a block are
assigned in a 2D layout with an $i$-$j$ column of the $i$-$j$-$k$ spectral
element data assigned to each thread.  The individual kernel formulations,
$K=1$ to 10, constitute successive improvements in memory management
that are described briefly in the Appendix and in detail in \cite{warburton2019}.
Kernel 1 sustains 500 GFLOPS for $p=8$--15, while Kernels 9 and 10 reach a
peak of 2 TFLOPS for the highest polynomial orders.  While the V100 is formally
capable of 7 TFLOPS in 64-bit arithmetic, it is well known that memory bandwidth
demands constrain the realizable peak to a modest fraction of this value. 
Figure~\ref{fig:bk5_perform}(a) also shows a graph of the empirical
roofline as a function of $p$, which is the theoretical maximum performance
that can be realized for BK5 based on the number of operations and bytes
transferred.  We note that the most highly tuned kernels, $K=9$ and 10, attain
a significant fraction of the roofline model.  The results here and below
thus establish upper bounds on performance to be expected in production
simulations.

The tuning curves of Figure~\ref{fig:bk5_perform}(a) were for $E=4096$, which
is the largest case considered.  Because GPUs are themselves parallel
computers, they have an intrinsic strong-scale limit when the number of points
per GPU, $\nl=n/P$, is insufficient to sustain high work rates.  (In this
section, we define $P$ to be the number of GPUs, i.e., the number of V100s,
used on Summit.)  To determine this intrinsic limit, we performed a sequence of
BK5 timings using $E=1,\, 2,\, 4,,\,\dots,\,4096$ for a single GPU, $P=1$.
Figure~\ref{fig:bk5_perform}(b) shows the V100 TFLOPS for BK5 as a function of
$\nl=p^3E$ for $p=1$ to 15.  The tight data collapse demonstrates that $\nl$ is
a leading indicator of performance, but the polynomial order is also seen to be
important, with $p=15$ realizing a peak of 2 TFLOPS.

Figure~\ref{fig:bk5_perform}(c) shows the execution time for the cases of
Figure~\ref{fig:bk5_perform}(b).  Linear dependence is evident at
the right side of these graphs, while at the left the execution time approaches
a constant for $\nl < 10^5$.  In the linear regime, the time scales as $O(\nl)$,
independent of $p$, despite the $O(p\nl)$ scaling of the FLOPs count.  This
behavior is expected from the roofline analysis---the performance of BK5 is
bandwidth limited even for the largest values of $p$ considered.  The
lower-bound plateaus in Figure~\ref{fig:bk5_perform}(c) are readily understood.  At
the smaller values of $\nl$, the number of elements is less than 80, which is
the number of streaming multiprocessors (SMs, individual compute units) on the
V100.  Element counts below this value constitute situations in which SMs are
idled and the time per iteration is consequently not reduced.

Further insight into the $(p,E,P)$ performance trade-offs can be gained by
looking at the execution time per point, shown in Figure
\ref{fig:bk5_perform}(d), which also shows the minimal time and the 2$\times$
line, which is twice the minimal execution time per point.
For a fixed total problem size, $n$, moving horizontally on this plot
corresponds to increasing $P$ and reducing $\nl$ such that $n = \nl P$.  In the
absence of communication overhead, one gains a full $P$-fold reduction in the
execution if the time per point does not increase when moving to the left.  We
see in this plot that $p=7$ appears to offer the best potential for high
performance, where even at $\nl=$30,000 the execution time per point is within a
small multiple of the minimum realized over all cases.  This low value of $\nl$
is in sharp contrast with the $p=14$ and 15 cases, which cross the $2 \times$
line at $\nl=$200,000.  Thus, through additional internode parallelism, the $p=7$
case affords a potential 200/30 $\approx$ 7-fold performance gain over the
larger $p$ cases.
Of course, this analysis must be tempered by consideration of a full solver
that includes communication, particularly for Poisson problems, which
require communication-intensive multilevel solvers for algorithmic
efficiency.  In the next section, we take a step in that direction by analyzing
the BP5 performance on Summit.

\subsection{BP5 on Summit}

\begin{figure*}[t]
 \subfloat[BP5 $p=1,...,10$]{{\includegraphics[width=0.32\textwidth]{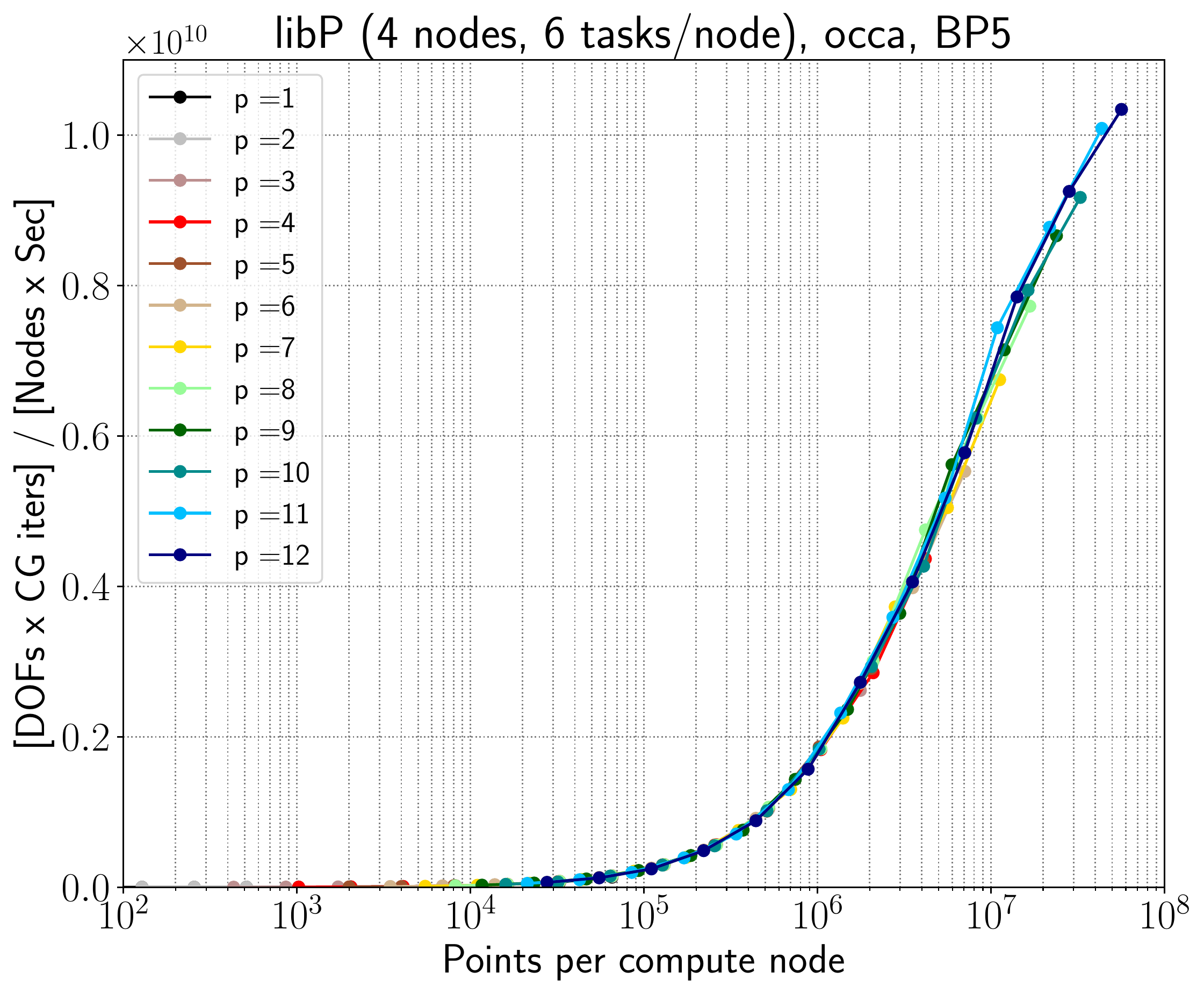}}}
 \hskip.1in
 \subfloat[BP5 $p=8$]{{\includegraphics[width=0.32\textwidth]{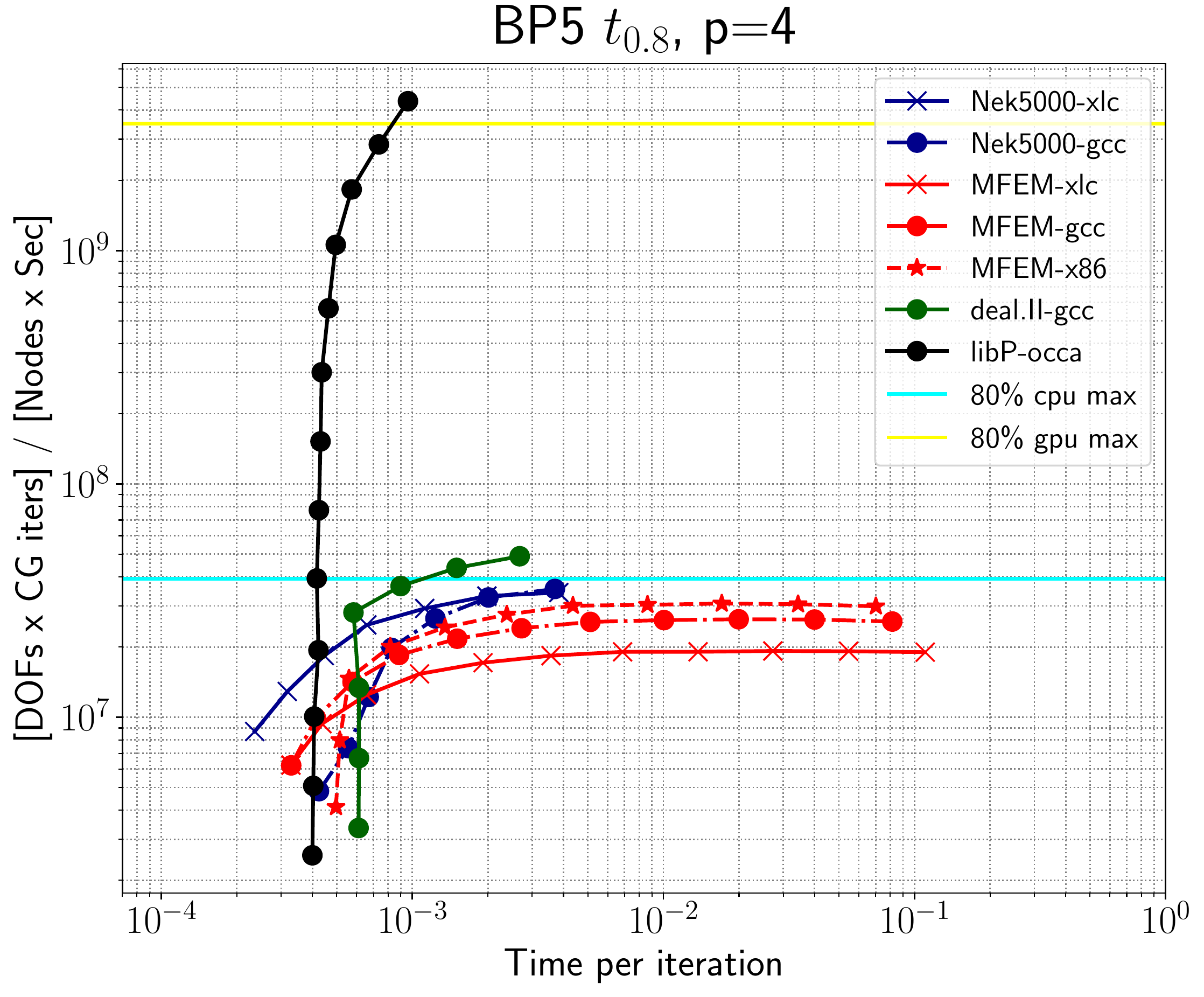}}}
 \hskip.15in
 \subfloat[BP5 $p=10$]{{\includegraphics[width=0.32\textwidth]{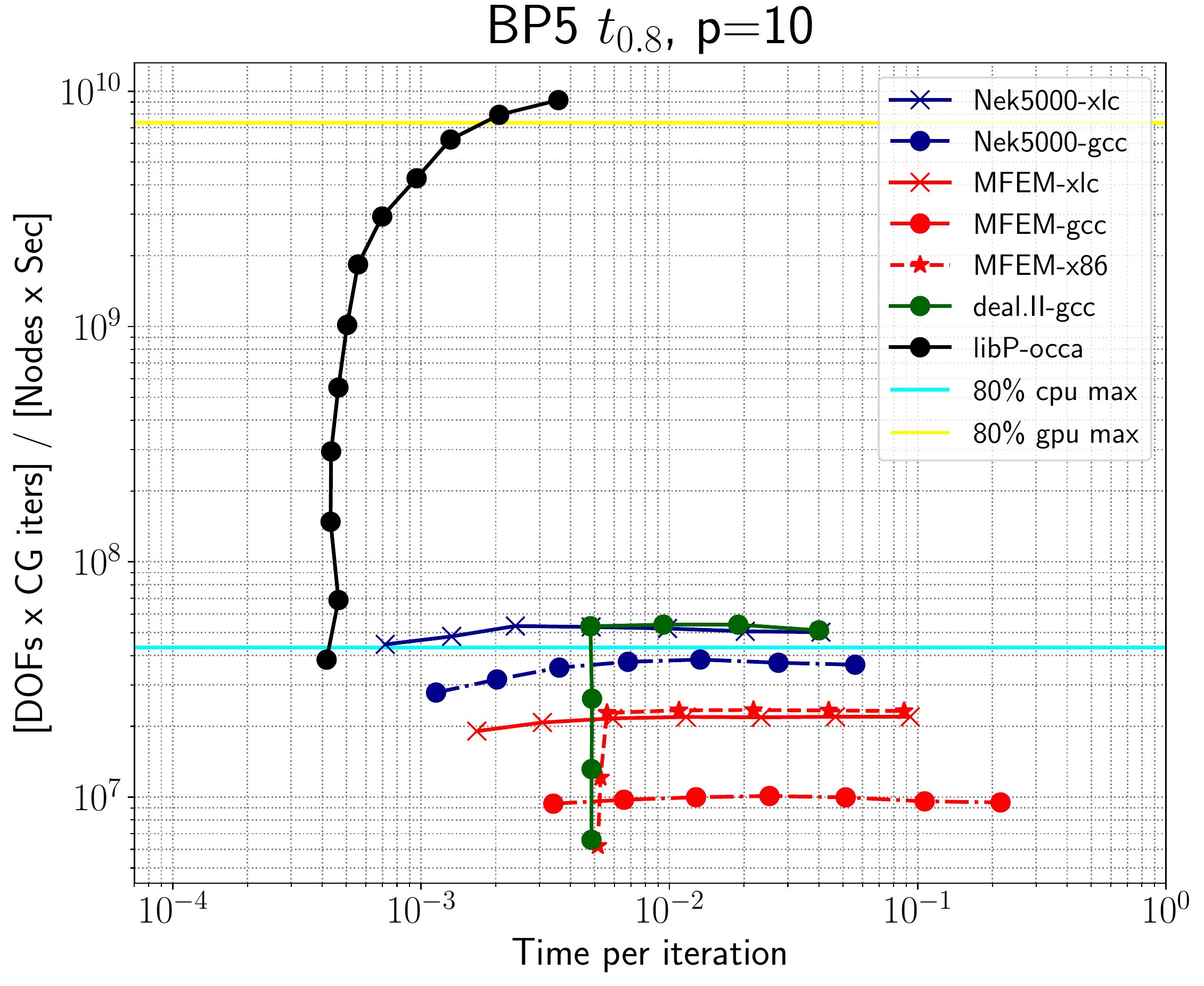}}}
 \caption{\label{fig:libP_BP5} Multi GPU performance using 24 V100s:
   (a) BP5 DOFS vs. grid points per node plot for libParanumal (libP) on Summit
      Summit with $p=1$--10.
   (b) DOFS vs time per iteration comparison on Summit and BG/Q for $t_{0.8}$ with $p=8$.  
   (c) DOFS vs time per iteration comparison on Summit and BG/Q for $t_{0.8}$ with $p=10$.}  
\end{figure*}


\begin{figure*}
 \subfloat[BP5 $t_{0.8}$]{{\includegraphics[width=0.32\textwidth]{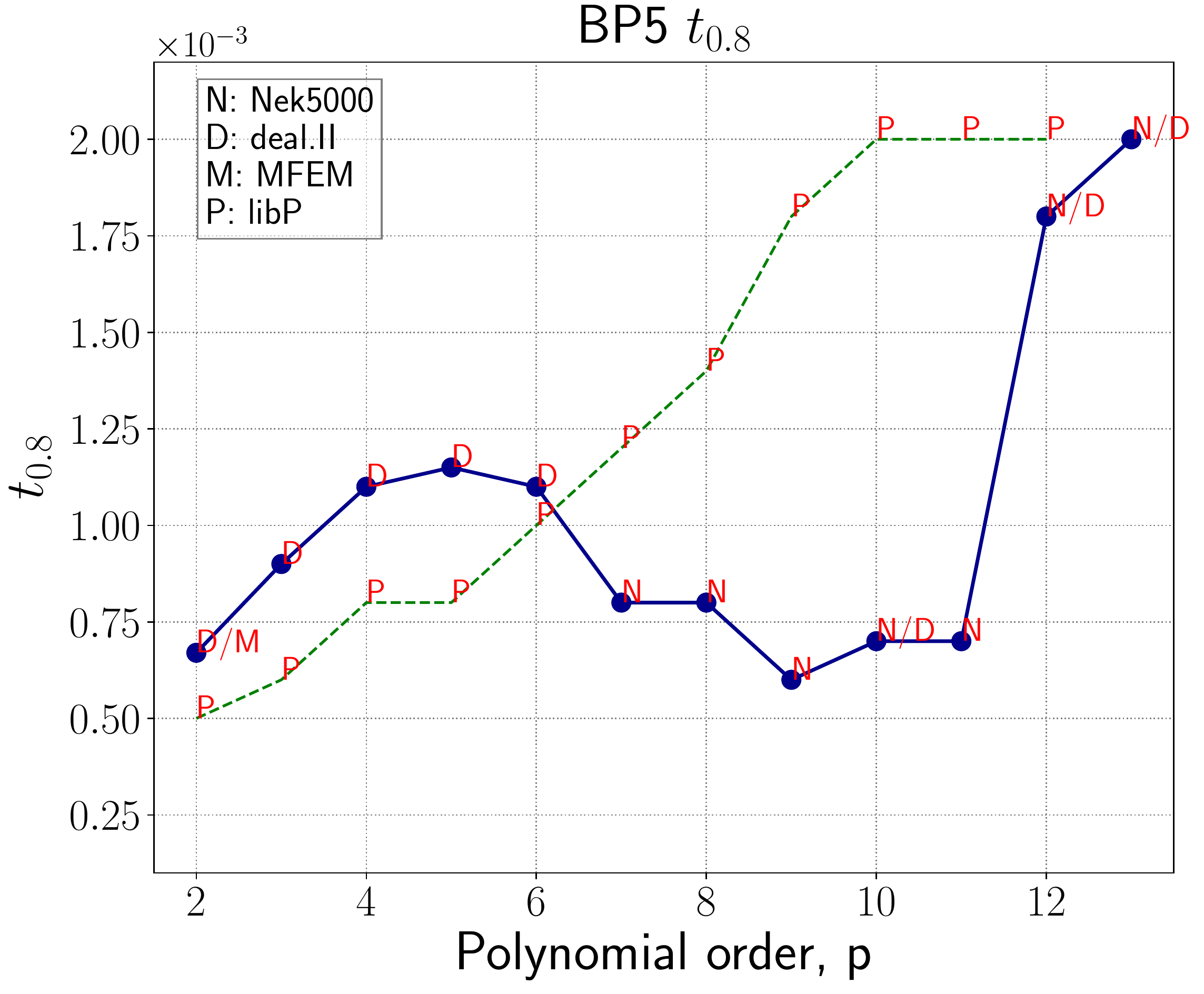}}}
 \hskip.1in
 \subfloat[BP5 $n_{0.8}$]{{\includegraphics[width=0.32\textwidth]{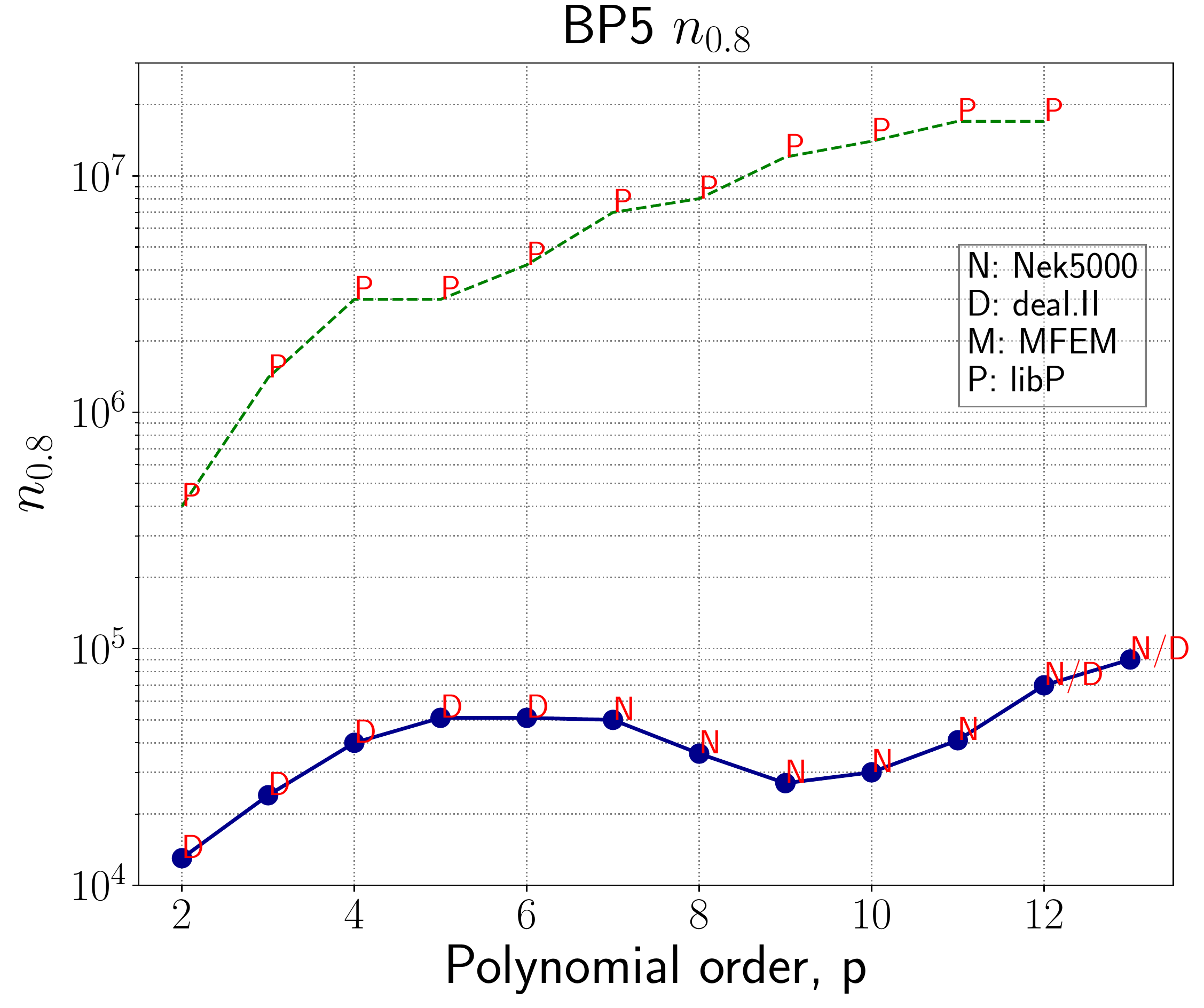}}}
 \hskip.1in
 \subfloat[BP5 $r_{0.8}$]{{\includegraphics[width=0.32\textwidth]{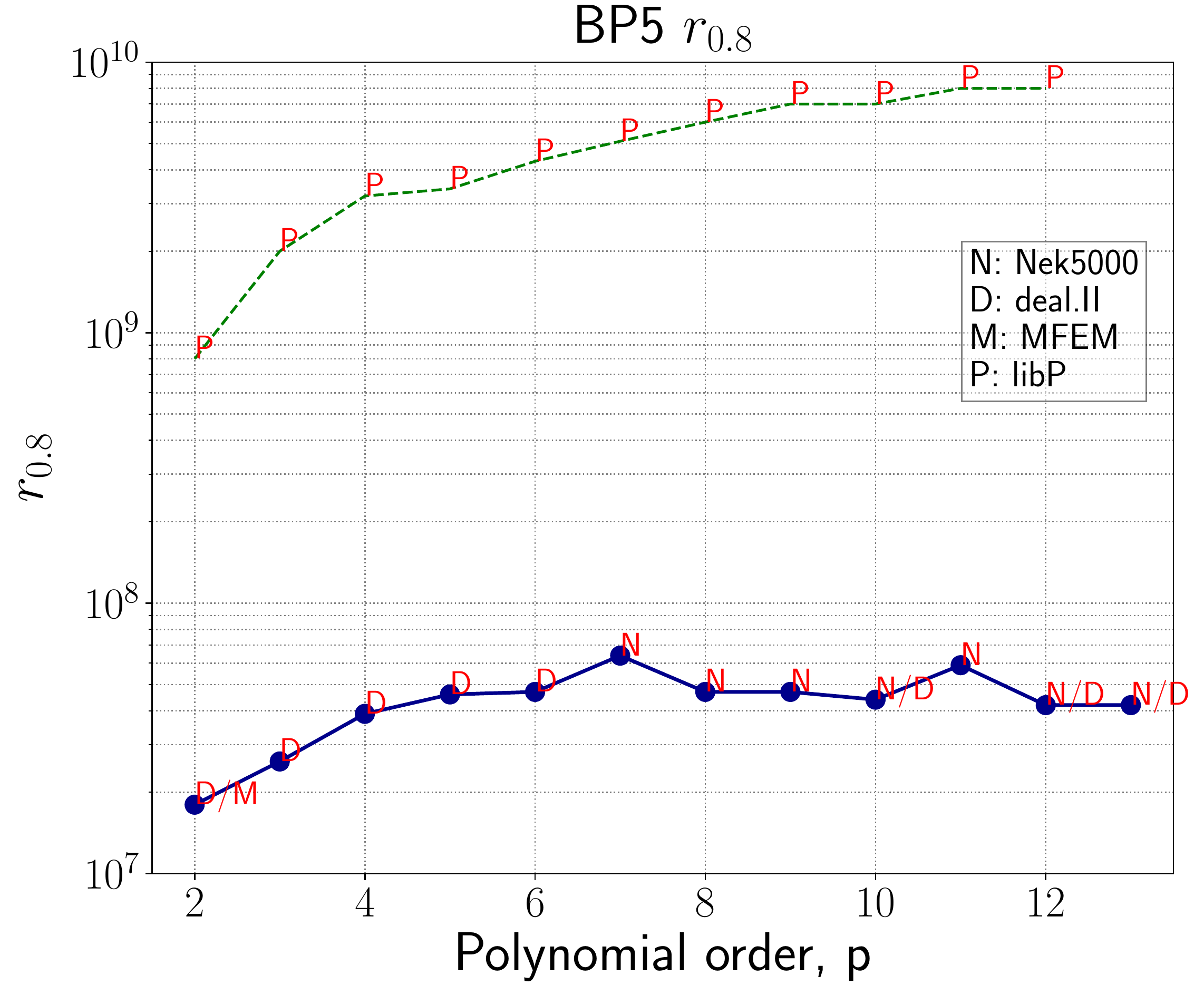}}}
 \caption{\label{fig:BP5_all} BP5 time ($t_{0.8}$), grid points ($n_{0.8}$), and rate ($r_{0.8}$),
  running from libParanumal (libP) on Summit, and Nek5000, deal.II, and  MFEM on BG/Q.}
\end{figure*}

The BP5 implementation on the V100s employs the optimally performing BK5 kernel
of the preceding section.  All vectors are stored in their {\em local form}, following
the Nek5000 storage approach described in Appendix A1.  The vector operations for
PCG, including the diagonal preconditioner, are straightforward streaming operations
with a FLOPs count of $n$ or $2n$ and data reads involving only one or two 64-bit
words per operation.   Assembly for the matrix-vector products (the $QQ^T$ operation
in (\ref{eq:qqt})) is invoked in two stages.   Values corresponding to vertices shared
within a GPU are condensed on the GPU. Values for vertices shared between GPUs
are sent through the host and then condensed through a pairwise exchange and sum
across the corresponding MPI ranks.   For all of the studies, we use a total of
$P=24$ GPUs: six GPUs on each of four nodes of Summit.

Figure \ref{fig:libP_BP5}(a) shows DOFS vs. $n/P$ curves analogous to Figures
\ref{fig:bp5_bgq_gcc}--\ref{fig:bp5_bgq_xlc}.  At 10,000 MDOFs, the peak V100
performance is substantially higher than the peak of 80 MDOFs realized for a
single node of BG/Q.  The $n_{0.8}$ is also substantially higher on the
V100--around 15 million for one node of Summit vs. a maximum of 80,000 for
BG/Q. In Figures \ref{fig:libP_BP5}(b) and (c), we plot DOFS vs. time for $p=4$
and 10 for the V100 and for the bake-off codes on BG/Q.  
We see that, for $p=4$, the minimum time for Summit at 80\% of its 
{\em realized peak}\footnote{For technical reasons, these tests do not take
the problem size for the lower $p$ values to saturation, which slightly
skews the discussion.  In practice, however, it's not clear that any user
will be able or willing to run the V100s in saturation mode.  These results
are thus indicative rather than definitive.}
is smaller than the $t_{0.8}$ value realized on BG/Q.
For $P=10$, Summit can also realize a smaller time per iteration
than BG/Q, but not at the realized 80\% efficiency value.  At that value,
the min-time is $0.002$ seconds, which is about three times slower than
$t_{0.8}=0.0007$ seconds attained for $p=10$ on BG/Q.

Using plots similar to Figures \ref{fig:libP_BP5}(b) and (c), we generate time
($t_{0.8}$), size ($n_{0.8}$), and DOFs ($r_{0.8}$) curves for libParanumal on
Summit and compare these with their BG/Q counterparts in Figure \ref{fig:BP5_all}.
For $p<6$, Summit is able to deliver relatively small values of $t_{0.8}$, subject
to the caveat that the $r_{0.8}$ values for Summit in Fig. \ref{fig:BP5_all}(b)
are not fully saturated.  For midrange polynomial orders, $p=7$--10, the
minimum times on the V100 are roughly a factor of two to three higher than on BG/Q.
While the V100 execution rates are
higher (Figure \ref{fig:BP5_all}(c)), they are not sufficiently high to make up
for the increased problem size required to sustain 80\% of the observed peak.
This imbalance results in increased time per iteration,
as indicated in (\ref{eq:speed}).

We close this section by noting that the strong-scale time per iteration is not
the only performance metric of interest in evaluation of exascale platforms.
Cost and energy consumption, which are not assessed in the present studies, are also
important concerns in the drive to exascale, and accelerators have been shown to
offer advantages with respect to these important metrics.

 \section{Conclusion}

We have presented performance results for highly optimized matrix-free
implementations of high-order finite element problems on large-scale parallel
platforms and state-of-the-art accelerator nodes.
The data are the results of an initial suite of benchmark problems that
will be extended in the future to include preconditioning comparisons,
nonsymmetric operators, and other geometric configurations.
Four teams participated in the study---three running on the BG/Q, Cetus,
at ALCF, and one running on the OLCF V100-based platform, Summit.

For BG/Q, no single strategy was best across all the BPs.  Vectorization across
elements (i.e., in cases with more than one element per MPI rank)
could in some cases yield greater than 1.25$\times$ speedup over
single-element performance, implying that 80\% efficiency could not be realized
at the granularity limit of a single element per processor. Reduction in
floating-point operation counts can be realized by exploiting the bilateral
symmetry of the nodal and quadrature point distributions.  Effective time
reductions for this approach requires careful local memory management.

For $p<12$, the minimum time per iteration at 80\% efficiency is fairly
uniform with  $t_{0.8} \approx$ 0.6--1.2 ms at $n/P$ values of 30,000--70,000
points per node for the midrange polynomial orders.   For BP3 and BP5,
the midrange processing rates were substantially higher than the lower
order polynomial cases, implying that, in the same time, the higher-order
solutions could process more points (i.e., higher resolution) and
achieve higher accuracy (higher-order polynomial approximation).
In several cases, a twofold reduction in $t_{0.8}$ could be realized by solving
block diagonal systems simultaneously in order to amortize communication
latency at the strong-scale limit, as demonstrated in the comparison of
BP2, BP4, and BP6 with their scalar counterparts.

On Summit, we found that highly tuned OCCA kernels could deliver
2 TFLOPS, which is a significant fraction of the peak 7 TFLOPS cited
for a single V100 GPU.   Moreover, the OCCA performance is near the bounds
established by bandwidth-limited roofline performance model for BK5 (and, for
other kernels, as shown by \cite{warburton2019}).  For BP5, Summit realizes
in excess of 10,000 MDOFs per node, a factor of 125 larger than a single
node of Cetus.  The $n_{0.8}$ values, however, are also much larger--
approximately $10^7$ vs. $10^5$ on Cetus.  Consequently,
$t_{0.8}$ values are superior to the values realized on Cetus
for $p < 6$, but are larger for $p=7$--11, with Summit times 2 to 3
times larger than $t_{0.8}$ found on Cetus.  For $p > 11$ the times
are again comparable, but are significantly larger (on both platforms)
than those for $p \leq 11$.

 \section*{Acknowledgments}
This research is supported by the Exascale Computing Project
(17-SC-20-SC), a collaborative effort of two U.S. Department of Energy
organizations (Office of Science and the National Nuclear Security
Administration) responsible for the planning and preparation of a capable
exascale ecosystem, including software, applications, hardware, advanced system
engineering and early testbed platforms, in support of the nation’s exascale
computing imperative.
The research  used resources of the Argonne Leadership Computing Facility,
which is supported by the U.S. Department of Energy, Office of Science, under
Contract DE-AC02-06CH11357.
This research also used resources of the Oak Ridge Leadership Computing Facility
at Oak Ridge National Laboratory, which is supported by the Office of
Science of the U.S. Department of Energy under Contract No. DE-AC05-00OR22725.
Contract DE-AC02-06CH11357.

\section*{Funding}
This material is  based upon work supported by the U.S. Department of Energy,
Office of Science, under Contract DE-AC02-06CH11357.
This Work is also performed under the auspices of the U.S. Department of Energy under
Contract DE-AC52-07NA27344 (LLNL-JRNL-782135).
Kronbichler is supported by the German Research Foundation (DFG) under the project
``High-order discontinuous Galerkin for the EXA-scale'' (ExaDG) within the priority
program ``Software for Exascale Computing'' (SPPEXA), grant agreement no. KR4661/2-1,
and the Bayerisches Kompetenznetzwerk f\"ur Technisch-Wissenschaftliches Hoch-und H\"ochstleistungsrechnen
(KONWIHR).



\section*{Appendix: Algorithmic Description}
Algorithmic approaches and their implementations are discussed for Nek5000,
MFEM, deal.II, and libParanumal.
\subsection*{Appendix A1: Nek5000}

A key feature of Nek5000 is that all data are stored in {\em local form}.  That
is, for any function in $X^N \subset C^0$, there exists a local vector
$\uu_L=Q\uu$.  (In fact, the local form, $\uu_L$, also exists for discontinuous
functions, but the global form, $\uu$, does not.) If the function is in $C^0$,
then shared face and edge values are stored redundantly, which means that
computations of inner products, daxpys, and other vector operations are doing
roughly a factor of $((p+1)/p)^3$ extra work.  Moreover, inner products need to
be weighted to account for repetition of the redundantly stored values unless
they are computed prior to residual assembly.  The advantage of the local
storage format is that the near-neighbor data exchange can be effected in a
single communication phase as follows.  First, we note that the matrix-vector
product $\uw = B \uv = Q^T B_L Q \uv $ in local form becomes
\begin{eqnarray} \label{eq:qqt}
\uw_L = Q \uw
= Q (Q^T B_L Q) \uv
= QQ^T \, B_L \,\uv_L.
\end{eqnarray}
The $QQ^T$ operation, often referred to as direct-stiffness summation (see
\cite{pat84}), corresponds to an exchange and sum of values between shared
vertices and is implemented in the general-purpose communication library {\em
gslib}, which supports other associative/commutative operations (e.g., min,
max, multiplication) for scalar and vector fields (see \cite{fischer08a}).
{\em gslib} implements the communication and local operator evaluation for
several Boolean matrix types other than the symmetric $QQ^T$ form.  An
advantage of using the form in (\ref{eq:qqt}) is that data that is interior to
elements or on domain boundaries is never moved.  Unlike $Q$ or $Q^T$ in
isolation, the symmetric form $QQ^T$ avoids reshuffling of data between global
and local layouts.

For Nek5000, (\ref{eq:bk5x}) is applied element-by-element,
with each local matrix vector product $\uw^e = A^e \uu^e$ applied
in three phases: {\em gradient of $u$, application of ${\bf G}$, followed
by gradient-transpose.}  These operations are detailed in subroutine
{\tt axe} in Algorithm~\ref{fig:nek5code}.  The tensor-product contractions
for the gradient and gradient-transpose operations are expressed as
BLAS3 matrix-matrix products---though rarely as {\tt dgemm}, since that
routine is typically optimized for much larger values of $p_1$ than encountered
in FEM/SEM applications.  Fast contractions are typically best realized
by writing different code for each polynomial order, illustrated for the
case $p_1=8$ shown in {\tt mxm8} in Algorithm~\ref{fig:nek5code} and
the BG/Q intrinsic version {\tt mxm\_bgq\_8} in Algorithm~\ref{fig:qpx}.
For application of ${\bf G}$, all six elements of the tensor are
stored for unit-stride access, as seen in {\tt axe.}

\begin{algorithm}
\caption{Nek5000 implementation pseudocode for (\ref{eq:bk5x}).
{\tt mxm8} is called for tensor contraction lengths of 8.  On
BG/Q, {\tt mxm8} is replaced by {\tt mxm\_bgq\_8} of Algorithm~\ref{fig:qpx}.
\label{fig:nek5code}}
\footnotesize
\begin{verbatim}
 subroutine ax(rho,wL,uL,g,mask,ur,ur,ut,n,nel,gsh)

! Compute wL=A*uL and rho = wL'*uL

 real wL(n*n*n,nel),ul(n*n*n,nel,g(6*n*n*n,nel)
 real mask(n*n*n*nel),ur(n*n*n),us(n*n*n),ut(n*n*n)
 integer gsh

 rho = 0
 do e=1,nel
   call axe(rho,wL(1,e),uL(1,e),g(1,e),ur,us,ut,n)
 enddo
 rho = glsum(rho,1)         ! sum over all P
 call gs_op(gsh,wL,'+')     ! wL = QQ^T wL

 do i=1,n*n*n*nel           ! Restrict Dirichlet
    wL(i,1)=mask(i)*wL(i,1) ! boundary values to 0
 enddo

 end
c--------------------------------------------------
 subroutine axe(rho,w,u,g,ur,us,ut,n)

 common /derivatives/ D(lnn),Dt(lnn)

 call loc_grad3  (ur,us,ut,u,n,D,Dt)

 do i=1,n*n*n
   wr = g(1,i)*ur(i) + g(2,i)*us(i) + g(3,i)*ut(i)
   ws = g(2,i)*ur(i) + g(4,i)*us(i) + g(5,i)*ut(i)
   wt = g(3,i)*ur(i) + g(5,i)*us(i) + g(6,i)*ut(i)
   ur(i) = wr
   us(i) = ws
   ut(i) = wt
 enddo

 call loc_grad3t (w,ur,us,ut,n,D,Dt)

 do i=1,n*n*n
    rho = rho + w(i)*u(i)
 enddo

 end
c-------------------------------------------------
 subroutine loc_grad3(ur,us,ut,u,n,D,Dt)
 real ur(n,n,n),us(n,n,n),ut(n,n,n),u(n,n,n)
 real D(n,n),Dt(n,n)

 call mxm(D ,n,u,n,ur,n*n)
 do k=1,n
    call mxm(u(1,1,k),n,Dt,n,us(1,1,k),n)
 enddo
 call mxm(u,n*n,Dt,n,ut,n)

 end
c-------------------------------------------------
 subroutine mxm8(a,n1,b,n2,c,n3)
 real a(n1,8),b(8,n3),c(n1,n3)

 do j=1,n3
 do i=1,n1
    c(i,j) = a(i,1)*b(1,j) + a(i,2)*b(2,j)
           + a(i,3)*b(3,j) + a(i,4)*b(4,j)
           + a(i,5)*b(5,j) + a(i,6)*b(6,j)
           + a(i,7)*b(7,j) + a(i,8)*b(8,j)
 enddo
 enddo

 end
\end{verbatim}
\end{algorithm}

\begin{algorithm}
\caption{QPX Intrinsics routine for $C=AB$ with inner-product length 8.
\label{fig:qpx} }
\footnotesize
\begin{verbatim}
subroutine mxm_bgq_8(a,n1,b,n2,c,n3)
real(8)  a(n1,8),b(8,n3),c(n1,n3)

vector(real(8)) av1,av2,av3,av4,av5,av6,av7,av8
vector(real(8)) bv1,bsv1,bsv2,bsv3,bsv4
vector(real(8)) bv2,bsv5,bsv6,bsv7,bsv8,cv

call alignx(32, a(1,1))
call alignx(32, b(1,1))
call alignx(32, c(1,1))

do i = 1, n1, 4
  av1 = vec_ld(0,a(i,1))
  av2 = vec_ld(0,a(i,2))
  av3 = vec_ld(0,a(i,3))
  av4 = vec_ld(0,a(i,4))
  av5 = vec_ld(0,a(i,5))
  av6 = vec_ld(0,a(i,6))
  av7 = vec_ld(0,a(i,7))
  av8 = vec_ld(0,a(i,8))

  do j = 1, n3
    bv1  = vec_ld(0,b(1,j))
    bv2  = vec_ld(0,b(5,j))
    bsv1 = vec_splat(bv1, 0)
    bsv2 = vec_splat(bv1, 1)
    bsv3 = vec_splat(bv1, 2)
    bsv4 = vec_splat(bv1, 3)
    bsv5 = vec_splat(bv2, 0)
    bsv6 = vec_splat(bv2, 1)
    bsv7 = vec_splat(bv2, 2)
    bsv8 = vec_splat(bv2, 3)

    cv =  vec_mul (av1, bsv1)
    cv =  vec_madd(av2, bsv2, cv)
    cv =  vec_madd(av3, bsv3, cv)
    cv =  vec_madd(av4, bsv4, cv)
    cv =  vec_madd(av5, bsv5, cv)
    cv =  vec_madd(av6, bsv6, cv)
    cv =  vec_madd(av7, bsv7, cv)
    cv =  vec_madd(av8, bsv8, cv)

    call vec_st(cv, 0, c(i,j))
  enddo
enddo
end 

end
\end{verbatim}
\end{algorithm}

\subsection*{Appendix A2: MFEM}

%
MFEM supports a set of HPC extensions targeting CPU architectures.
This functionality is implemented as
template classes that require the specification of parameters such as the solution's
polynomial order, the type of the mesh element (quad, hex, etc.), and the order of
the quadrature rule, to be given at compile time. This allows the compiler to
optimize the compute-intensive inner loops by using loop unrolling, inlining, and
even autovectorization in some cases. In addition to the template approach, the
HPC extensions implement better algorithms, taking full advantage of the tensor
product structure of the finite element basis on tensor product mesh elements: quads and
hexes. Furthermore, the HPC extensions support various levels of operator assembly
and evaluation. These levels include full matrix assembly to CSR format,
partial assembly (where data at quadrature points is computed and stored during
the assembly and later used in the operator evaluation), and an ``unassembled''
approach where assembly is a no-op and all computations are performed on the fly
during the operator evaluation.

A recent addition to MFEM is the introduction of CPU SIMD intrinsics to the
HPC extension classes where vectorization is explicitly applied across a fixed
number of elements; for example, on CPUs with 256-bit SIMD instructions, four elements
are processed simultaneously by using double-precision arithmetic.

In this paper, we present results based on the MFEM HPC extensions both with and
without the use of SIMD intrinsics.

\subsubsection*{Implementation Details.}

In MFEM, the subdomain restriction, $P$, can be represented by two 
classes: (1) \code{ConformingProlongationOperator} when the finite element space has no
hanging nodes or (2) \code{HypreParMatrix} in all other cases. All tests in
this paper use the first class since we do not consider AMR meshes. This class allows for an
optimized implementation because all nonzero entries of $P$ are equal to $1$.

In the HPC extension of MFEM, the element restriction matrix $G$ is represented
by an implicit template class concept that allows flexible optimized
implementations. The specific class used in the numerical tests is class
\code{H1\_FiniteElementSpace}, which represents $G$ through an indirection array
of integer offsets: every row of $G$ has exactly one nonzero entry that is
equal to $1$, and therefore, only its column index has to be stored. The class
concept assumes, and \code{H1\_FiniteElementSpace} implements, two main methods,
\code{VectorExtract} and \code{VectorAssemble}, which implement the action of
$G$ and $G^T$ for one element or a small batch of elements.

The basis evaluator operator $B$ also is represented in MFEM by a template class
concept. In the tests considered here, the specific class template we used is
\code{TProductShapeEvaluator} instantiated in 3D (hex element) for a fixed
number of DOFs and quadrature points in each spatial dimension. The main methods
in the class are \code{Calc} and \code{CalcT}, for the case of a mass matrix
operator, and \code{CalcGrad} and \code{CalcGradT}, for the case of a diffusion
operator.

The combined action of $G$ and $B$ for one element, or a small batch of elements,
is represented in the template class \code{FieldEvaluator}, which
defines the methods \code{Eval} and \code{Assemble} for the action of $B G$ and
$G^T B^T$, respectively.

The discretization operator is represented by the template class concept of
a physics``kernel.'' The kernel is responsible for defining the specific form
of $B$ that is required (e.g., $B$ is a different operator for mass and
diffusion) and the specific operations for defining (assembling) and using the
quadrature point data, $D$. In the numerical tests, we used the kernel
class templates \code{TMassKernel} and \code{TDiffusionKernel}.

Putting all these components together, MFEM defines the processor-local version
of the discretization operator, $A_L = G^T B^T D B G$, as the template class
\code{TBilinearForm}. The main class methods are \code{Assemble}, for partial
assembly, and \code{Mult}, for operator evaluation.

As an example of the implementation, the method \code{Mult} of
\code{TBilinearForm} can be written roughly (ignoring the template parameters
and some of the method parameters) as follows.
\begin{verbatim}
void TBilinearForm::Mult(const Vector &x, Vector &y)
{
   y = 0.0;
   FieldEvaluator solFEval(...,
                           x.GetData(), y.GetData());
   const int NE = mesh.GetNE(); // num. elements
   for (int el = 0; el < NE; el++)
   {
      // declare 'R' with appropriate type
      solFEval.Eval(el, R);
      kernel_t::MultAssembled(0, D[el], R);
      solFEval.Assemble(R);
   }
}
\end{verbatim}
Clearly, a lot of the details are hidden in the calls \code{solFEval.Eval()}
and \code{solFEval.Assemble()}. Note that the compiled method \code{Mult} has
no function calls because all methods inside its call tree are force inlined by
adding the \code{\_\_attribute\_\_((always\_inline))} specification to their
definition.

\subsection*{Appendix A3: deal.II}

The deal.II finite element library \citep{dealII85}, written in the C++
programming language, aims to enable rapid development of modern finite
element codes. The library provides support for adaptive meshes with hanging
nodes, a large number of finite elements, and delivers functions and classes
implementing code patterns in typical matrix assembly tasks with backends to
PETSc and Trilinos; see \citet{dealII85} and references therein. For
high-order computations on elements with tensor product shape functions, there
is a separate ``matrix-free'' module \citep{Kronbichler12} optimized
for cache-based architectures, exemplified in the step-37, step-48, and
step-59 tutorial programs of the library.

The matrix-free module lets the user specify operations on quadrature points
in several ways. The most
generic interfaces are convenience access operations, organized through a class
called \texttt{FEEvaluation}, in terms of the values and gradients of the
tentative solution and test functions, respectively. The geometric factors are
precomputed for a given mesh configuration---which can be deformed through an
arbitrary high-order mapping---and get implicitly applied when accessing a
field, such as the gradient of the input vector via
\texttt{FEEvaluation::get\_gradient()}. For the case of the 3D Laplacian, this
approach would access 10 data fields for quadrature points, namely, $9=3^2$ for
the inverse of the Jacobian $(\mathcal J^e)^{-1}$ and one field for the
determinant of the Jacobian times the quadrature weight $\rho$. Depending on
the operators called at the user level, only the necessary data is loaded from
memory. For specialized operators, such as the Laplacian, an alternative entry
point is to access gradients and values in the reference coordinates and apply
geometry and the operator by an effective tensor $\rho (\mathcal
J^e)^{-1}(\mathcal J^e)^{-T}$, which reduces the data access to 6 fields for
the symmetric rank-2 tensor. This effective tensor format is used for the
BP3--BP6 problems with deal.II.

The global input and output vectors in the matrix-free operator evaluation
connect to the tentative solution and test function slot on the element level,
respectively. The vectors are natively implemented in deal.II and stored and
treated in fully distributed format outside the integration tasks. 
Their local arrays contain some space beyond the end of the
locally owned range in order to support an in-place transformation 
via \texttt{MPI\_Isend} and \texttt{MPI\_Irecv}
without a deep copy of all vector entries. Furthermore, the extraction of the
local vector for the cell operations is done immediately prior to the
interpolation operations $B$ and $B^T$ to ensure that the data remains in caches of
the processor.

The actions of $G$, $B$, $B^T$, and $G^T$ 
are provided by the \texttt{read\_dof\_values}, \texttt{evaluate},
\texttt{integrate}, and \texttt{distribute\_local\_to\_global} functions of
the class \texttt{FEEvaluation}. Since the interpolation operations $B$ and
$B^T$ are arithmetically heavy also when using optimal sum factorization
algorithms (\cite{dfm02}), deal.II uses SIMD
instructions on supported architectures with vectorization over several cells
\citep{Kronbichler12}. The SIMD arrays ``leaks'' into the user code in the
form of a tensor of SIMD type called \texttt{VectorizedArray<double>} in
\texttt{FEEvaluation::get\_gradient()}, with each lane of the SIMD vector
returning data to a different cell. However, typical application codes are
templated on the return type and become almost transparent of this fact. The
C++ template arguments also enable auxiliary operators,
such as multigrid V-cycles for preconditioning, to be easily implemented in single precision. The code
automatically detects the case when node points and quadrature points
coincide, as well as the more general case when one first must interpolate
from the nodal values into quadrature points. The arithmetic of
local operations is reduced almost half by a so-called even-odd
decomposition that speeds up higher-order computations
\citep{solomonoff92,Kopriva09}.

Inside the one-dimensional kernels of sum factorization, the deal.II
implementation applies a particular optimization to reduce the arithmetic
operations by the so-called even-odd decomposition \citep{solomonoff92,Kopriva09}.
It relies on the fact that the basis functions and quadrature points are
symmetric about the center $0$ of the one-dimensional reference element
$(-1,1)$. If we denote the basis functions of degree $p$ by $\varphi_i$,
$i=1,\ldots, p+1$, and quadrature points by $\xi_j$, $j=1,\ldots,q$, the
symmetry implies $\varphi_i(\xi_j) = \varphi_{p+2-i}(\xi_{q+1-j})$. For
example, the one-dimensional interpolation matrix $S_{ji} = \varphi_i(\xi_j)$
for the case $p=3$ and $q=5$ is of the form
\begin{equation}\label{eq:shape_symmetric}
  S = \begin{pmatrix}
    s_1 & s_2 & s_3 & s_4 \\
    s_5 & s_6 & s_7 & s_8 \\
    s_9 & s_{10} & s_{10} & s_9 \\
    s_8 & s_7 & s_6 & s_5 \\
    s_4 & s_3 & s_2 & s_1
    \end{pmatrix},
\end{equation}
for some numbers $s_1,\ldots,s_{10}$, representing only $10$ distinct entries
in this $(q)\times(p+1)  = 5\times 4$ matrix. The interpolation from the $p+1$
nodal values to the $q$ quadrature points (i.e., the matrix-vector product
$\underline{v} = S\underline{u}$) is then decomposed in the following three
steps: \begin{itemize}
\item Decompose the vector $\underline{u}$ into the even part $\underline{u}^+$
and the odd part $\underline{u}^-$. For the $5\times 4$ example, this yields
\begin{align*}
    &u^+_1 = u_1 + u_4, \quad u^+_2 = u_2 + u_3,\\
    &u^-_1=u_1-u_4,\quad u^-_2 = u_2-u_3.
  \end{align*}
\item Perform the matrix-vector product on the even and odd contributions separately,
  \begin{align*}
    \underline{v}^+ = S^+ \underline{u}^+, \quad \underline{v}^- = S^- \underline{u}^-,
  \end{align*}
  with the $\left\lceil\frac{q}{2}\right\rceil \times
\left\lceil\frac{p+1}{2}\right\rceil$ matrix $S^+$ with entries \begin{align*}
    S^+_{ij} = \frac{1}{2}\left(S_{i,j} + S_{i,p+1-i}\right)
  \end{align*}
  and the  $\left\lfloor\frac{q}{2}\right\rfloor \times
\left\lfloor\frac{p+1}{2}\right\rfloor$ matrix $S^-$ with entries
\begin{align*}
    S^-_{ij} = \frac{1}{2}\left(S_{i,j} - S_{i,p+1-i}\right).
  \end{align*}
\item Form the final vector $\underline{v}$ by concatenating the even and odd
parts of $\underline{v}^+$ and $\underline{v}^{-}$, respectively. For the
$5\times 4$ example, the result is \begin{align*}
    &v_1 = v^+_1 + v^-_1, \quad v_2 = v^+_2+v^-_2,\quad v_3 = v_3^+,\\
    &v_4 = v^+_2 - v^-_2, \quad v_5 = v^+_1-v^-_1.
  \end{align*}
\end{itemize}
Note that the entries of the matrices $S^+$ and $S^-$ internally contain an
addition and a multiplication, but these contributions are precomputed. As a
consequence, the $(p+1)q$ additions and multiplications (fused
multiply-accumulates, FMAs) of the naive matrix-vector product are turned into
two matrix-vector multiplications of size $\frac{p+1}{2} \times \frac{q}{2}$,
thus reducing the number of FMAs to $\frac{(p+1)q}{2}$. Since the number of
additions and subtractions $(p+1)+q$ is of lower complexity, the even-odd
decomposition approximately halves the operation count for the 1D kernel.
Similar techniques are possible for the reference-cell derivatives with
symmetry relation $\varphi'_i(\xi_j) = -\varphi'_{p+2-i}(\xi_{q+1-j})$.

An important ingredient to make the even-odd decomposition competitive is the
way the intermediate results $\underline{u}^{\pm}$ and $\underline{v}^\pm$ are
handled by the implementation. Since the small matrix multiplication kernels
are often limited by the load/store (to L1 cache) rather than actual
arithmetics, all temporary results should ideally be kept in registers. In
deal.II, the loop kernels are written using C++ templates on both $p$ and $q$,
giving loop bounds and vector sizes that are compile-time constants. Up to
degrees of approximately $p=9$, register allocation policies of compilers such
as gcc keep the values $u^+_1,\ldots,u_5^+$ and $u^-_1,\ldots,u_5^-$ in 10
registers. With two additional registers for the two sums $v_j^+$ and $v_j^-$
that are eventually written into the respective positions $v_j$ and
$v_{q+1-j}$, and some spare registers to hold some entries of $S^+$ and $S^-$,
no extra load/store operations are necessary for architectures with 16 visible
floating-point registers, assuming the nowadays usual out-of-order execution
and register renaming capabilities to hide the latency of the two FMAs. As
shown by the analysis of \cite{Kronbichler19}, performance is slightly lower
for $p=1$ on architectures with FMA support, similar for $p=2,3$, but by
$1.3\times$ to $1.7\times$ for $p\geq 4$, when compared with optimized dense
matrix multiplication kernels for small matrices analogous to what is done in
libxsmm \citep{libxsmm}.

\subsection*{Appendix A4: libParanumal}

The GPU results of Section 8 were based on the libParanumal library,
which is an experimental testbed for exploring high-performance kernels that
can be integrated into existing FEM/SEM codes as accelerator modules.
The kernel code is written in OCCA (\cite{occa}), which can produce
performant CUDA, OpenMP, or OpenCL code.  All the current cases
were run on OLCF Summit's Nvidia V100s using CUDA.

Here, we briefly describe the libParanumal development for BK5
through the ten successive kernels indicated in Figure
\ref{fig:bk5_perform}(a).  Each kernel $K$ includes the optimizations in kernel
$K-1$ unless otherwise indicated.  These kernels are described in detail by
\cite{warburton2019}.
\begin{itemize}
\item
Kernel 1 corresponds to a straightforward implementation of (\ref{eq:bk5x}).
Each spectral element is assigned to a separate thread block on the
GPU with a 2D thread structure.  Threads within a block are assigned to
an $i$-$j$ column in the $i$-$j$-$k$ spectral element data structure.
Entries of $\uu^e$ are fetched from global memory as needed, and the
partial sums in the last loop are added directly to the global memory
array to store the final result.  This kernel achieves 500 GFLOPS, which
is one-fifth of the predicted empirical roofline.
\item
In Kernel 2 all variables except for the final result are declared
by using {\tt const}.  The gains are marginal for $p > 7$.

\item
In Kernel 3 all of the $k$ loops are unrolled, leading to significant
gains for $p>7$.

\item
In Kernel 4 the $k$ loop is placed exterior to the $i$ and $j$ thread loops, making
$k$ the slowest running index.

\item
In Kernel 5 an auxiliary shared-memory array is replaced by two smaller ones.

\item In Kernel 6 the input vector $\uu^e$ is fetched into shared memory.

\item In Kernel 7 global memory transactions are reduced by caching
data at the beginning of the kernel and writing the output variable
only once.   This optimization improves performance by about 40\%.

\item In Kernel 8 arrays are padded for the cases $p=7$ and 15 to
avoid shared-memory bank conflicts.

\item In Kernel 9 $\uu^e$ is fetched to registers, $p_1$ entries at a time.

\item In Kernel 10 three two-dimensional shared-memory arrays are used for
partial results.
\end{itemize}
The results for Kernel 10 are aligned with the empirical roofline
model and achieve a peak of 2 TFLOPS for $p=15$.
Algorithm~\ref{fig:libp} gives the pseudocode.

\begin{algorithm}
\caption{libParanumal pseudocode for BK5, $\uw_L=A_L \uu_L$ corresponding
 to Kernel 10 of \ref{fig:bk5_perform}(a).
\label{fig:libp}}
\footnotesize
\begin{algorithmic}[1]
\STATE for $e \, \in \, \{1,2,\dots,E\}$ do
\STATE \hspace*{.1in} for $i,j,k \, \in \,  \{1,2,\dots,p_1\}$ do
\STATE \hspace*{.20in} If $k=0$, load $\Dh$ to shared memory
\STATE  \hspace*{.20in} Declare register variables $r_{qr}$, $r_{qs}$, $r_{qt}$.
\STATE  \hspace*{.1in} end for
\STATE  \hspace*{.1in} for $i,j,k \, \in \,  \{1,2,\dots,p_1\}$ do
\STATE  \hspace*{.20in} Load $J$ (Jacobian)
\STATE  \hspace*{.20in} Declare $q_r$,$q_s$,$q_t$:=0.
\STATE  \hspace*{.20in} $q_r = \sum_{n}\Dh_{in}u_{nji}^e$
\STATE  \hspace*{.20in} $q_s = \sum_{n}\Dh_{jn}u_{kni}^e$
\STATE  \hspace*{.20in} $q_t = \sum_{n}\Dh_{kn}u_{kjn}^e$
\STATE  \hspace*{.20in} Set $r_{qr}=q_r$, $r_{qs}=q_s$, $r_{qt}=q_t$.
\STATE  \hspace*{.20in} $\uw^e_{kji}=\beta J^e_{kji}*u^e_{kji}$
\STATE  \hspace*{.1in} end for (synchronize threads)
\STATE  \hspace*{.1in} for $i,j,k \, \in \,  \{1,2,\dots,p_1\}$ do
\STATE  \hspace*{.20in} Load $G_{00}, \, G_{01} \, G_{02}$.
\STATE  \hspace*{.20in} $\mbox{Sqtemp}_{kji}=G_{00}*r_{qr} + G_{01}*r_{qs} + G_{02}*r_{qt}$
\STATE  \hspace*{.1in} end for (synchronize threads)
\STATE  \hspace*{.1in} for $i,j,k \, \in \,  \{1,2,\dots,p_1\}$ do
\STATE  \hspace*{.20in} $\uw^e_{kji}=\uw^e_{kji}\,+\,\sum_{n} \Dh_{ni}*\mbox{Sqtemp}_{kjn}$
\STATE  \hspace*{.1in} end for
\STATE  \hspace*{.1in} for $i,j,k \, \in \,  \{1,2,\dots,p_1\}$ do
\STATE  \hspace*{.20in} Load $G_{01}, \, G_{11} \, G_{12}$.
\STATE  \hspace*{.20in} $\mbox{Sqtemp}_{kji}=G_{01}*r_{qr} + G_{11}*r_{qs} + G_{12}*r_{qt}$
\STATE  \hspace*{.1in} end for (synchronize threads)
\STATE  \hspace*{.1in} for $i,j,k \, \in \,  \{1,2,\dots,p_1\}$ do
\STATE  \hspace*{.20in} $\uw^e_{kji}=\uw^e_{kji}\,+\,\sum_{n} \Dh_{nj}*\mbox{Sqtemp}_{kni}$
\STATE  \hspace*{.1in} end for
\STATE  \hspace*{.1in} for $i,j,k \, \in \,  \{1,2,\dots,p_1\}$ do
\STATE  \hspace*{.20in} Load $G_{02}, \, G_{12} \, G_{22}$.
\STATE  \hspace*{.20in} $\mbox{Sqtemp}_{kji}=G_{02}*r_{qr} + G_{12}*r_{qs} + G_{22}*r_{qt}$
\STATE  \hspace*{.1in} end for (synchronize threads)
\STATE  \hspace*{.1in} for $i,j,k \, \in \,  \{1,2,\dots,p_1\}$ do
\STATE  \hspace*{.20in} $\uw^e_{kji}=\uw^e_{kji}\,+\,\sum_{n} \Dh_{nk}*\mbox{Sqtemp}_{nji}$
\STATE  \hspace*{.1in} end for
\STATE end for (end of $e$ loop)
\end{algorithmic}
\end{algorithm}

 \bibliographystyle{./SageH}
 \bibliography{./bibs/emmd,./bibs/dealii,./bibs/bakeoff}

\section{Author Biographies}

\medskip
\noindent
{\bf Paul Fischer}
is a Blue Waters Professor of Computer Science and Mechanical Science and
Engineering at the University of Illinois at Urbana-Champaign and an Argonne
senior scientist. He is a Fellow of the American Association for the
Advancement of Science (AAAS).  He is the chief architect of the fluid thermal
simulation code Nek5000, which scales to over a million processors and has been
recognized with the Gordon Bell Prize in high-performance computing. Nek5000 is
used by over 400 researchers worldwide in a variety of thermal
fluids applications. He was the deputy director for the DOE-ASCR Co-Design
Center for Exascale Simulation of Advanced Reactor (CESAR) and is currently the
deputy director for the DOE-ECP Co-Design Center for Efficient Exascale
Discretizations (CEED).

\medskip
\noindent
{\bf Misun Min}
is a computational scientist in the  Mathematics and Computer Science Division at
Argonne National Laboratory.  Her research focuses on developing scalable
high-order algorithms and software for solving electromagnetic,
drift-diffusion, and fluid systems on advanced high-performance computing
architectures.  She is the author of the spectral element discontinuous
Galerkin (SEDG) electromagnetics simulation code, NekCEM, which scales to over
million CPU cores and tens of thousands GPUs. She won an R\&D100 award on
``NekCEM/Nek5000: Scalable High-Order Simulation Codes.''  She is a PI on the
DOE Applied Mathematics Research project ``High-Order Methods for
High-Performance Multiphysics Simulations'' and the Argonne PI of the DOE-ECP
Co-Design project, CEED.

\medskip
\noindent
{\bf Thilina Rathnayake} is a Ph.D. candidate in the Department of Computer Science at the University of Illinois at Urbana-Champaign.
Thilina worked as an intern at Lawrence Livermore National Laboratory  in 2017 and a Givens Associate
in the Mathematics and Computer Science Division at Argonne National Laboratory in 2018--2019.
His research has focused on the library package libCEED for the DOE Co-Design Center for Efficient Exascale Discretizations,
the CFD solver Nek5000, and the communication library gslib.

\medskip
\noindent
{\bf Som Dutta} is currently an assistant professor in the Department of Mechanical and Aerospace Engineering at Utah State University (USU). 
He contributed to the paper during his post-doctoral training in the Department of Computer Science at the University of Illinois at 
Urbana-Champaign (UIUC), where he was part of the DOE-funded CEED project. Before joining USU, he was a  postdoctoral researcher in the Department of Mathematics at College of Staten Island, City University of New York (CUNY). Som received a Ph.D. from the Department of Civil and Environmental Engineering at UIUC. His research interests are in studying environmental and turbulent multiphase flows using high-order spectral element methods. 

\medskip
\noindent
{\bf Tzanio Kolev}
is a computational mathematician at the Center for Applied Scientific Computing (CASC) in Lawrence Livermore National Laboratory (LLNL), where he works on finite element discretizations and solvers for problems in compressible shock hydrodynamics, multi-material arbitrary Lagrangian Eulerian methods, radiation hydrodynamics, and computational electromagnetics. He won an R\&D100 award as a member of the {\em hypre} team. Tzanio is leading the high-order finite element discretization research and development efforts in the MFEM and BLAST projects in CASC and is the director of the Center for Efficient Exascale Discretization in DOE's Exascale Computing Project.

\medskip
\noindent
{\bf Veselin Dobrev} is a  computational mathematician in the numerical analysis and simulations group in the Center for Applied Scientific Computing. His research interests are in the areas of numerical methods for solving PDEs, which include finite element and discontinuous Galerkin methods, shock hydrodynamics simulations, and iterative and multigrid methods.
Veselin received his Ph.D. in mathematics from Texas A\&M University in 2007. He is currently working on high-order curvilinear finite elements for Lagrangian hydrodynamics (BLAST project).

\medskip
\noindent
{\bf Jean-Sylvain Camier} works in the Center for Applied Scientific Computing (CASC), Lawrence Livermore National Laboratory.
His current research focus is on computer architecture; parallel computing; and computing in mathematics, natural science, engineering and medicine.

\medskip
\noindent
{\bf Martin Kronbichler} is a senior researcher at the Institute for Computational Mechanics, Technical University of Munich, Germany. He is a principal developer of the deal.II finite element library and leads the high-order as well as high-performance activities in this project. His current research focus is on efficient high-order discontinuous Galerkin schemes and fast iterative solvers for flow problems as a PI in the exascale project ExaDG within the German priority program SPPEXA.

\medskip
\noindent
{\bf Tim Warburton}
is the John K. Costain Faculty Chair in the College of Science and a professor of mathematics at Virginia Tech. He also currently holds an appointment in the Department of Computational and Applied Mathematics at Rice University. He developed the first high-order nodal discontinuous Galerkin solver for time-domain electromagnetics on unstructured grids and led a decadal project to accelerate these methods by devising parallel local time-stepping methods, GPU acceleration, co-volume filtering techniques, novel rational bases for curvilinear elements, and numerous other innovations . He co-authored the first comprehensive book on discontinuous Galerkin methods. He  created the Open Concurrent Compute Abstraction (OCCA) as part of the CESAR co-design center at Argonne. The OCCA framework enables domain scientists and computational scientists to write portable threaded code. The OCCA library has been used as a foundational layer for higher-order finite element, spectral element, discontinuous Galerkin, and finite difference PDE solvers for industrial applications and lab miniapps.

\medskip
\noindent
{\bf Kasia \'Swirydowicz} is a postdoctoral researcher at the National Renewable Energy Laboratory.
She was formerly a postdoc researcher at Virginia Tech. Her scientific interests include 
GPU programming and code optimization.

\medskip
\noindent
{\bf Jed Brown}
is an assistant professor of computer science at the University of Colorado Boulder. He is a developer of PETSc and specializes in multiscale and high-order numerical methods for geoscience and engineering applications. His work has been recognized by the 2014 SIAG/SC Junior Scientist Prize and the 2014 IEEE TCSC Young Achiever award, and he was co-recipient of the 2015 SIAM/ACM Prize in Computational Science and Engineering.

 \newpage
The following paragraph should be deleted before the paper is published:

The submitted manuscript has been created by UChicago Argonne, LLC, Operator of
Argonne National Laboratory (``Argonne"). Argonne, a U.S. Department
of Energy Office of Science laboratory, is operated under Contract No. DE-AC02-06CH11357.
The U.S. Government retains for itself, and others acting on its behalf,
a paid-up nonexclusive, irrevocable worldwide license in said article to reproduce,
prepare derivative works, distribute copies to the public, and perform publicly and
display publicly, by or on behalf of the Government.
The Department of Energy will provide public access to these results of federally sponsored research in accordance with the DOE Public Access Plan. http://energy.gov/downloads/doe-public-access-plan.

Work performed under the auspices of the U.S. Department of Energy under
Contract DE-AC52-07NA27344 (LLNL-JRNL-782135).

LLNL Disclaimer: This document was prepared as an account of work sponsored by
an agency of the United States government. Neither the United States government
nor Lawrence Livermore National Security, LLC, nor any of their employees makes
any warranty, expressed or implied, or assumes any legal liability or
responsibility for the accuracy, completeness, or usefulness of any information,
apparatus, product, or process disclosed, or represents that its use would not
infringe privately owned rights. Reference herein to any specific commercial
product, process, or service by trade name, trademark, manufacturer, or
otherwise does not necessarily constitute or imply its endorsement,
recommendation, or favoring by the United States government or Lawrence
Livermore National Security, LLC. The views and opinions of authors expressed
herein do not necessarily state or reflect those of the United States government
or Lawrence Livermore National Security, LLC, and shall not be used for
advertising or product endorsement purposes.

ECP Disclaimer: This research is supported by the Exascale Computing Project
(17-SC-20-SC), a collaborative effort of two U.S. Department of Energy
organizations (Office of Science and the National Nuclear Security
Administration) responsible for the planning and preparation of a capable
exascale ecosystem, including software, applications, hardware, advanced system
engineering and early testbed platforms, in support of the nation’s exascale
computing imperative.


\end{document}